\documentclass[aps,prd,reprint,twocolumn,superscriptaddress,floatfix,nofootinbib]{revtex4-1}

\usepackage{epsfig,amsmath,amssymb,color,comment,physics}
\usepackage[makeroom]{cancel}
\usepackage[caption=false]{subfig}
\usepackage{mathrsfs}
\usepackage[countmax]{subfloat}
\usepackage[normalem]{ulem}
\usepackage[english]{babel}
\usepackage{dsfont}
\usepackage{braket}
\usepackage[bookmarks=true,colorlinks,linkcolor=OrangeRed,urlcolor=NavyBlue,citecolor=RoyalBlue]{hyperref}
\usepackage[dvipsnames]{xcolor}
\usepackage{graphicx}
\graphicspath{{./Figures/}}
\usepackage{amsmath, amssymb,latexsym,amsfonts}
\usepackage{cleveref}
\usepackage{multirow}

\newcommand{\Ztwo}{\mathbb{Z}_2}
\newcommand{\Uone}{U(1)}
\newcommand{\be}{\begin{equation}}
\newcommand{\ee}{\end{equation}}
\begin{document}
\title{Half-strings across the deconfinement transition of a $\Uone$ quantum link model}

\author{Indrajit Sau}
\affiliation{International Centre for Theoretical Sciences (ICTS-TIFR), Bengaluru 560089, India}

\author{Arnab Sen}
\affiliation{School of Physical Sciences, Indian Association for the Cultivation 
of Science, Kolkata 700032, India}

\author{Debasish Banerjee}
\affiliation{Department of Physics and Astronomy, University of Southampton, Southampton 1BJ, UK}

\begin{abstract}
  The phenomenon of confinement is arguably the key feature of gauge theories regularly studied in 
  high-energy and condensed matter physics. This causes hadrons to exist as stable particles in 
  quantum chromodynamics (QCD) as bound states of quarks (and gluons), and strings in pure gauge 
  theories. The promise of realizing such strings in quantum simulator setups have fueled 
  extensive studies of a plethora of simpler lattice gauge theories (LGTs). Quantum link models 
  (QLMs) generalize Wilsonian LGTs and yield richer scenarios of confinement. For example, the 
  string joining two static charges in the spin-$\frac{1}{2}$ $\Uone$ QLM fractionalize, giving rise to 
  confining phases that break the point group and internal symmetries spontaneously. 
  It was further shown in PRL 130, 071901 (2023) that adding annealed $Q = \pm 2$ charges to this 
  model qualitatively changes the confinement-deconfinement transition. In this paper, we examine 
  the finite temperature phase transition both in the presence and the absence of annealed charges 
  through detailed extraction of critical temperature, $T_c$, and exponents. The (effective) string tension 
  $\sigma$ at zero and finite-temperatures are compared between the theory with and without annealed 
  charges. The quantity $T_c/\sqrt{\sigma}$ for the QLM with annealed charges is similar to the usual 
  $\Ztwo$ gauge theory, while the theory without the charges provides a new universal value. 
\end{abstract}

\date{\today}
\maketitle
\tableofcontents

\section{Introduction} \label{sec:intro}
 The physics of Abelian gauge theories has received renewed attention in connection 
 with the development of novel experimental tools in the form of quantum simulators 
 and quantum computers \cite{Zohar2016,Banuls2020,Bauer2023}. While ambitious long-term 
 goals include the investigation of finite-density regimes and real-time dynamics of 
 quantum chromodynamics (QCD) using quantum technologies, significant milestones have 
 already been achieved in simulations of simpler Abelian gauge theories 
 \cite{Martinez2016,Mil2020,Yang2020}. 
 A first-principles, nonperturbative approach to strongly interacting gauge theories 
 is to regularize them on a spacetime lattice. A continuum limit, when it exists, is 
 obtained by tuning the bare parameters toward a critical point while sending the 
 lattice spacing to zero \cite{Wilson1974,Kogut1979}. In many condensed-matter systems, 
 including theories of doped Mott insulators relevant to high-temperature superconductivity 
 and quantum spin-ice materials, the low-energy physics can be described in terms of emergent 
 Abelian lattice gauge fields \cite{LeeNagaosaWen2006,Hermele2004,SavaryBalents2017,Castelnovo2008}. 
 Finite-temperature transitions in such theories can therefore provide valuable insights 
 into the phenomenology of these systems, particularly the confinement and deconfinement 
 of fractionalized excitations \cite{SavaryBalents2013}.

 The finite-temperature phase transitions in lattice gauge theories are particularly interesting,
 because they are often associated with the loss of confinement, and the emergence of a deconfined 
 thermal phase \cite{Polyakov1978,Susskind1979,McLerranSvetitsky1981,SvetitskyYaffe1982}.
 In the context of the $U(1)$ lattice gauge theory, recent results can be found in 
 \cite{Caselle:2025bgu,Caselle:2025vhx}. 
 Confinement is characterized by the free-energy cost of separating external test charges. A static 
 gauge charge--anti-charge pair in a confined phase at two spatially separated points, $x$ and $y$ 
 respectively, causes an electric flux 
 string between them. Confinement implies that the flux lines are collimated through a tube and cannot 
 spread out (unlike the Coulomb field of a photon) \cite{Wilson1974,Polyakov1977}. At nonzero temperature, 
 the corresponding diagnostic is the Polyakov loop. Its expectation value is related to the free energy 
 $F_q$ of an isolated static charge through $\braket{L} \sim e^{-F_q/T}$, while the Polyakov-loop 
 correlator determines the free energy of a static charge--anti-charge pair 
 \cite{Polyakov1978,McLerranSvetitsky1981}. In a pure gauge theory with an exact center symmetry, the 
 confined phase has $\braket{L}=0$, formally corresponding to an infinite free energy for an isolated 
 center-charged source, whereas the deconfined phase has $\braket{L}\neq0$ and finite $F_q$. 
 When this transition is continuous, its universal critical behavior is governed by an effective 
 theory with the center symmetry of the gauge group, as expressed by the Svetitsky--Yaffe conjecture 
 \cite{SvetitskyYaffe1982}.

 Renormalization-group (RG) and universality arguments have been extensively used to predict the 
 nature of finite-temperature deconfinement transitions. The central example is the Svetitsky-Yaffe 
 conjecture, which relates the critical behavior of a pure gauge theory in $d+1$ Euclidean dimensions 
 to that of a spin model in $d$ dimensions \cite{SvetitskyYaffe1982}. Pure gauge theories have an exact 
 global symmetry corresponding to the centre of the gauge group (e.g. $Z(N)$ centre for $SU(N)$ and 
 $U(1)$ centre for $U(1)$ LGTs). The Polyakov loop, $L(x)$, which is the path ordered product of 
 temporal gauge links transforms non-trivially under the centre symmetry and thus acts as an order 
 parameter for deconfinement \cite{Polyakov1978,McLerranSvetitsky1981}. Since it winds around the 
 compact Euclidean-time direction, $L(x)$, is naturally defined on the $d$-dimensional 
 spatial lattice. After all noncritical degrees of freedom are integrated out, one obtains a 
 short-ranged effective action for the Polyakov loops, schematically,
 \begin{equation*}
  S_{\rm spin} = -J \sum_{\braket{xy}} (L(x) L^\dagger(y) + \text{h.c.}) + \cdots,   
 \end{equation*}
 where $\braket{xy}$ denotes nearest neighbours and the omitted terms contain additional 
 center-symmetric local interactions. Since the Polyakov loop is 
 a product of the gauge links in the temporal direction, this operator is defined in one lower 
 spatial dimension; and so is the effective theory. The Svetitsky-Yaffe conjecture states that 
 if the effective theory has an continuous phase transition, its long-distance critical behavior 
 is governed by the 
 universality class of a $d$-dimensional spin model with global symmetry equal to the center of 
 the gauge group. This conjecture has been tested extensively using analytical methods and 
 nonperturbative Monte Carlo simulations \cite{CaselleHasenbusch1996,Coddington1986,Borisenko2015}.

 The present work explores this universality paradigm in a particularly extreme setting. In the 
 conventional Wilson-Kogut-Susskind formulation, the gauge field on each link is described by a 
 quantum rotor \cite{Wilson1974,KogutSusskind1975,Kogut1979} and the corresponding link Hilbert 
 space is infinite dimensional, making a direct implementation on quantum simulators or quantum 
 computers with finite-dimensional local degrees of freedom challenging. Considerable progress has 
 been made in constructing controlled truncations and digitizations of these Hilbert spaces 
 \cite{ZoharBurrello2015,Hackett2019}. Complementary approaches use quantum link gauge 
 theories that preserve exact local gauge invariance while employing finite-dimensional link Hilbert 
 spaces \cite{ChandrasekharanWiese1997,Wiese2022,LiuChandrasekharan2022}, or $q$-deformed Kogut-Susskind 
 theories \cite{Zache2023}. Here we investigate the finite-temperature transition of the 
 spin-$\frac12$ $\Uone$ quantum link model in $2+1$-d, for which each link contains only two states 
 and can therefore be represented by a single qubit.

 The Svetitsky-Yaffe argument suggests that the deconfinement transition of a $2+1$-d pure 
 $\Uone$ gauge theory should be governed by the two-dimensional XY model, since the center 
 symmetry is itself $\Uone$. The XY model undergoes a Berezinskii-Kosterlitz-Thouless (BKT) 
 transition driven by the binding and unbinding of topological vortex excitations 
 \cite{Berezinskii1971,KosterlitzThouless1973,Kosterlitz1974}. Below the transition, vortices 
 and antivortices occur predominantly in bound pairs and the spin-spin correlation function 
 decays algebraically, while above the transition, free vortices proliferate and correlations 
 decay exponentially. Approaching the critical point as $T \rightarrow T^+_{BKT}$, the correlation 
 length increases exponentially.

 The finite-temperature transition of the conventional compact $\Uone$ lattice gauge theory in 
 $2+1$-d has long been associated with this BKT universality class, in accordance with the 
 Svetitsky-Yaffe picture \cite{Coddington1986,Borisenko2015}. In the present work, we provide 
 extensive evidence that the deconfinement transition of the spin-$\frac{1}{2}$ $\Uone$ QLM 
 \emph{also} belongs to the same universality class. This result is nontrivial because the two 
 regularizations have radically different microscopic Hilbert spaces: Wilson's formulation employs an 
 infinite-dimensional rotor on every link, whereas the QLM contains only a two-state spin-$\frac{1}{2}$ 
 degree of freedom. Their common critical behavior therefore provides a particularly striking 
 illustration of the RG principle that universality is insensitive to microscopic details, 
 provided the relevant symmetries and long-distance degrees of freedom coincide.

 In Ref.~\cite{SauSenBanerjee2023}, we presented initial results extending this picture by 
 introducing thermally fluctuating charges $Q=\pm2$ into the theory, which modifies the gauge symmetry
 to $\Ztwo$. At the same time, the global center symmetry relevant to the Polyakov loop becomes a 
 $\Ztwo$. A straightforward application of the Svetitsky-Yaffe symmetry argument would then suggest 
 a transition in the universality class of the two-dimensional Ising model.

 Instead, we observed the phenomenon of \emph{weak universality}. This behavior is known in 2d 
 models such as Ashkin-Teller and eight-vertex models, which possess a $\Ztwo \times \Ztwo$ structure 
 and a line of critical points generated by an exactly or marginally varying coupling 
 \cite{Ashkin1943,Baxter1971,Suzuki1974,Jose1977}. Along such a critical line, individual 
 exponents such as $\nu$, $\gamma$, and $\beta$ can vary continuously, while exponent ratios such as
 $\gamma / \nu=2-\eta$ and $\beta / \nu = \eta / 2$ remain fixed at their 2d Ising 
 values. In Ref.~\cite{SauSenBanerjee2023}, we found $\eta\simeq1/4$, while $\nu$ and the associated 
 individual exponents varied with the microscopic coupling. This constituted the first observation of 
 weak universality at the finite-temperature deconfinement transition of a lattice gauge theory. 
 It also refines the conventional Svetitsky-Yaffe picture: the center symmetry strongly constrains 
 the critical theory, but in the presence of a marginal coupling it need not uniquely determine all 
 individual critical exponents.

 In this paper, we present detailed results about our determination of the phase transition 
 and the critical properties of the model, as well as results about flux tubes and string
 tensions (both with and without the thermally fluctuating $Q=\pm2$ charges).
 
 The organization of the rest of the paper is as follows: in \cref{sec:models}, we discuss the details of 
 the microscopic model with and without the annealed charges, in \cref{sec:phU1} we present
 detailed evidence about the BKT phase transition in the pure gauge theory, while in
 \cref{sec:phZ2} similar evidence about the weak universality of the theory with annealed
 charges is presented. In \cref{sec:OP} we show the the distribution of order parameter across 
 the phase transitions and in \cref{sec:phTr} we present an effective field theory for both the 
 theories. In \cref{sec:stringTen} we discuss the properties of string tensions for both the 
 theories in details, while in \cref{sec:TcSqrtSig} we discuss about the ratio of critical 
 temperature and squared root of the string tension that can be used to compare different LGTs. 
 Finally we conclude in \cref{sec:concl}.

\section{Model, Symmetries, Observables} \label{sec:models}
 We consider the spin-$\frac12$ $U(1)$ quantum link model on a square lattice. The gauge degrees 
 of freedom reside on the links $(x,\hat\mu)$ and are represented by quantum spins,
 \begin{equation}
   E_{x,\hat\mu}=S^3_{x,\hat\mu},~~U_{x,\hat\mu}=S^+_{x,\hat\mu},~~U^\dagger_{x,\hat\mu}=S^-_{x,\hat\mu}.
 \end{equation}
 Consequently, the electric flux on each link takes the two values $E_{x,\hat\mu}=\pm\frac12$. 
 The Hamiltonian is
 \begin{equation}
 \begin{split}
    H &= -J\sum_{\square} \left[ U_{\square}+U_{\square}^{\dagger} 
      - \lambda \left( U_{\square}+U_{\square}^{\dagger}\right)^2 \right], \\
    U_{\square} &= U_{x,\hat 1}U_{x+\hat 1,\hat 2}U^\dagger_{x+\hat 2,\hat 1}U^\dagger_{x,\hat 2}.
 \end{split}   
 \label{eq:H}
 \end{equation}
 The first term reverses the orientation of the electric flux around a flippable plaquette, 
 while the $\lambda$ term counts the number of flippable plaquettes. We set the lattice 
 spacing and the overall energy scale $J$ to unity unless stated otherwise. The local gauge 
 symmetry is generated by the Gauss-law operator
 \begin{equation}
   G_x = \sum_{\hat\mu} \left( E_{x,\hat\mu} - E_{x-\hat\mu,\hat\mu} \right).
   \label{eq:gauss}
 \end{equation}
 In the pure $U(1)$ gauge theory, the physical Hilbert space is restricted to states satisfying
 \begin{equation} \label{eq:GLUone}
    G_x\ket{\psi}=0
 \end{equation}
 at every lattice site. We also consider an enlarged thermal ensemble in which local charges 
 $Q_x=\pm2$ are allowed, such that
 \begin{equation} \label{eq:GLZtwo}
   G_x \ket{\psi}=Q_x\ket{\psi},~~Q_x=0,\pm2,~~\sum_x Q_x=0.
 \end{equation}
 The charges do not possess an independent kinetic-energy term, but are annealed variables 
 which fluctuate together with the gauge fields. Including these charges reduces the microscopic
 gauge invariance from $U(1)$ to $\mathbb{Z}_2$. Schematically, the partition function of the
 theory is influenced by the projector defined in eq.~\ref{eq:GLUone} or eq.~\ref{eq:GLZtwo}:
 \begin{equation}
     Z = \text{Tr}~\left[~\text{exp} (-\beta H) \cdot \prod_x \delta (G_x)~\right]
 \end{equation}
 In addition to the exact local gauge symmetry, the pure theory is invariant under lattice 
 translations and point-group transformations, as well as charge conjugation,
 \begin{equation}
  {\cal C}:~~U\longleftrightarrow U^\dagger,\qquad E\longrightarrow-E.
 \end{equation}
 On a periodic lattice there are also two global $U(1)$ centre symmetries associated with 
 the winding fluxes in the two spatial directions. Allowing charges $Q_x=\pm2$ reduces these 
 symmetries to their $Z_2$ subgroups. This reduction of the centre symmetry is central to 
 the different nature of the finite-temperature transitions in the two ensembles.

 \begin{figure}
     \centering
     \includegraphics[width=1.2\linewidth]{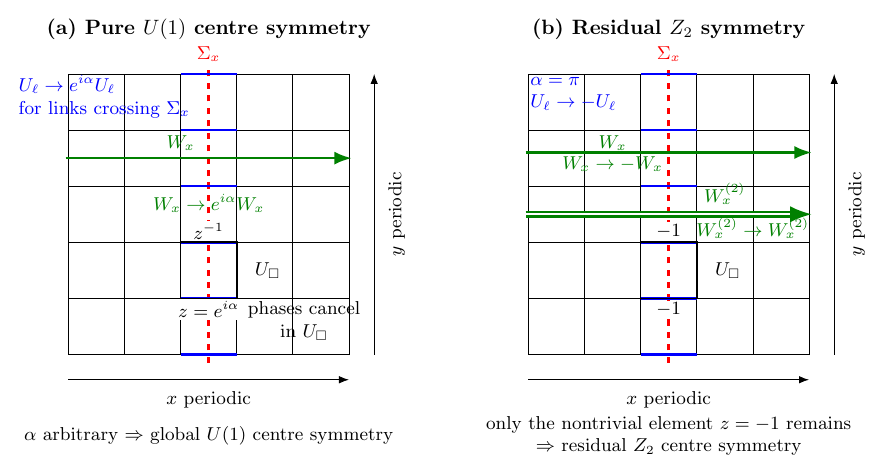}
     \caption{Centre symmetry on a spatial torus: a non-contractible dual cut $\Sigma_x$
     defines the generator ${\cal C}_x (\alpha)$. Links crossing the cut are multiplied 
     by $e^{i\alpha}$. Plaquettes are invariant under this transformation, but the Wilson
     loops transform under this transformation, both with and without annealed $\pm 2$ charges.}
     \label{fig:centreHam}
 \end{figure}

 The centre symmetry is clearest in the Euclidean finite-temperature formulation, which is 
 detailed in Sec.~\ref{sec:CenterS}. Here we will describe the symmetry through its action
 on the Hamiltonian. Let $\Sigma_i$ be a non-contractible cut through the spatial torus, 
 consisting of all links oriented in direction $\hat{i}$ which cross a fixed line perpendicular
 to $\hat{i}$, as schematically shown in Fig.~\ref{fig:centreHam}. 
 The corresponding centre transformation is generated by
 \begin{equation}
    {\cal C}_i(\alpha) = \exp\left[ i\alpha \sum_{\ell\in\Sigma_i} E_\ell \right], \qquad i=1,2.
 \end{equation}
 Using $[E_\ell,U_\ell]=U_\ell$, its action is
 \begin{equation}
  {\cal C}_i(\alpha)~U_\ell~{\cal C}_i^\dagger(\alpha) =
     \begin{cases}
              e^{i\alpha} U_\ell, & \ell\in\Sigma_i,\\ 
                          U_\ell, & \ell\notin\Sigma_i,
     \end{cases}
  \qquad E_\ell\longrightarrow E_\ell .
  \label{eq:hamiltonian-centre}
 \end{equation}
 Every plaquette intersects the cut either zero or two times. When the intersection is twice,
 then once $U_\ell$ and the other $U^\dagger_\ell$ is involved, and hence the phases cancel 
 in the Hamiltonian in all cases. A Wilson loop ${\cal W}_i$ winding once around direction 
 $\hat{i}$ intersects $\Sigma_i$ once and transforms as
 \begin{equation}
   {\cal W}_i \longrightarrow e^{i\alpha}{\cal W}_i.
 \end{equation}
 In the pure theory $\alpha$ is arbitrary, whereas in the theory containing $Q=\pm2$ charges 
 only $\alpha=0,\pi$ keeps the loop invariant. For the nontrivial $Z_2$ transformation,
 \begin{equation}
    U_\ell\longrightarrow -U_\ell \quad (\ell\in\Sigma_i), \qquad {\cal W}_i\longrightarrow-{\cal W}_i.
 \end{equation}
 Since there are two independent spatial directions on the torus, there are two independent 
 transformations, conventionally denoted by $[U(1)]^2$ in the pure theory and by $[Z_2]^2$ 
 when the $Q=\pm2$ charges are included.
 
 For the numerical simulations, the gauge theory is dualized to quantum height variables living at 
 the centres of the plaquettes (the dual lattice sites). Thus, for a site $x = (x_1,x_2)$, the 
 height variables are at $\tilde{x}=(x_1 + \frac{1}{2}, x_2 + \frac{1}{2})$. Moreover, the height 
 variables on different sublattices are distinct: $h_{\tilde{x}}^{\rm A}=0,1$ for even (A) sublattice 
 and $h_{\tilde{x}}^{\rm B}=\pm \frac{1}{2}$ for odd (B) sublattice. The divergence of the 
 height variables are the electric flux variables, 
 \begin{equation}\label{eq:flux_h}
  E_{x,\hat{i}}= \left[ h_{\tilde{x}}^X-h_{\tilde{x}+\tilde{i}-\tilde{1}-\tilde{2}}^{X^\prime} \right]
  ~~{\rm mod2} = \pm\frac{1}{2}; ~~~ {\rm X,X}^\prime \in \{{\rm A},{\rm B}\}    
 \end{equation}
 The corresponding magnetizations are \label{eq:magnetizations}
 \begin{equation}
   M_X = \frac{1}{L_t} \sum_{\tilde x\in X} \eta_{\tilde x}^{X}h_{\tilde x}^{X}, \qquad X=A,B,
 \end{equation}
 where the phases $\eta_{\tilde x}^{X}$ encode the appropriate ordering pattern. The two-component 
 quantity $(M_A,M_B)$ distinguishes the crystalline confined phases and also provides an order parameter 
 for the thermal transition.

 Defining $M^2=M_A^2+M_B^2$ and the spatial volume $V=L^2$, we measure the total and connected 
 susceptibilities,
 \begin{align}
   \chi_{\rm tot} = \frac{1}{V}\braket{M^2},~
   \chi_{\rm conn} = \frac{\beta}{V}\sum_{X=A,B} \left[ \braket{M_X^2} - \braket{|M_X|}^2 \right].
 \end{align}
 We further use the Binder ratios
 \begin{align}
   Q_1 &= \frac12\sum_{X=A,B} \frac{\braket{ |M_X| }^2}{\braket{ M_X^2 }},
   \qquad Q_{2a} = 2 - \frac{\braket{ M^4 }}{\braket{ M^2 }^2},\\
   Q_{2b} &= \frac{3}{2} -  \frac{1}{4} \sum_{X=A,B}\frac{\braket{ M_X^4 }}{\braket{ M_X^2 }^2}
 \end{align}
 which are particularly useful for locating the transition and carrying out the finite-size 
 scaling analysis. In the ensemble containing annealed charges, we additionally monitor the 
 charge density
 \begin{equation}
   \rho_Q = \frac{1}{4V} \sum_x \braket{ Q_x^2 },
 \end{equation}
 which measures the fraction of vertices carrying $Q_x=\pm2$.

 We note that the model in the absence of the annealed charges can be expressed
 as a quantum close-packed loop model on the square lattice. We refer to \cite{Penc2023}
 for the mapping, including the exposition of the point group symmetries. 

 The phase diagram of the dualized model in the height representation was studied 
 using a cluster quantum Monte Carlo algorithm. The details of the algorithm can be
 found in \cite{Banerjee2013,SauSenBanerjee2023}. 

\begin{figure}[!htpb]
    \centering
    \includegraphics[width=1.0\linewidth]{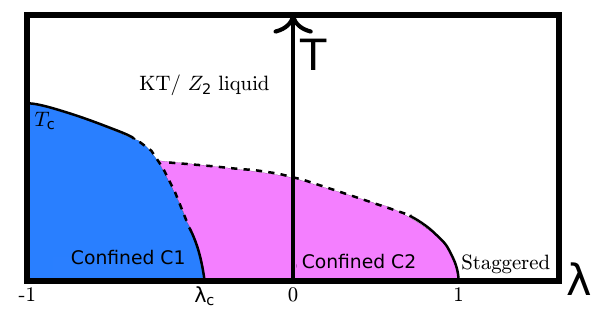}
    \caption{Phase diagram in $\lambda-T$ for the $U(1)$ and the $\mathbb{Z}_2$ LGTs.} 
    \label{fig:phDiagT}
\end{figure}

\section{Phase transition in the pure gauge theory} \label{sec:phU1}
  We first focus on the finite temperature phase transition in the pure $U(1)$
gauge theory. The schematic phase diagram for this theory was already conjectured
in \cite{SauSenBanerjee2023}, reproduced in Fig.~\ref{fig:phDiagT} for convenience. 
At $T=0$ two different phases are obtained by varying $\lambda$: both are confined, and
break different combinations of point-group and the internal charge conjugation
$\mathcal{C}$ symmetry and are separated by a weak first order phase transition at 
$\lambda_c \approx -0.359(5)$ \cite{Shannon2004,Banerjee2013,Penc2023}, before 
hitting the so-called Rokshar-Kivelson point at $\lambda = 1$, where all winding 
sectors meet the ground state. Beyond $\lambda > 1$, one runs into phases where 
system-spanning loops dominate the ground state. The global ground state belongs
to non-zero winding numbers.

 For the physics at $T=0$, the annealed charges are immaterial since they are only 
thermally activated. When the temperature is raised and is of order $J$, the 
ground state symmetry breaking gets restored and the annealed charges begin to 
influence the nature of the transition, leading to weak-universality, which we 
will address in the next section. 

\begin{figure}
    \centering
    \includegraphics[scale=0.3]{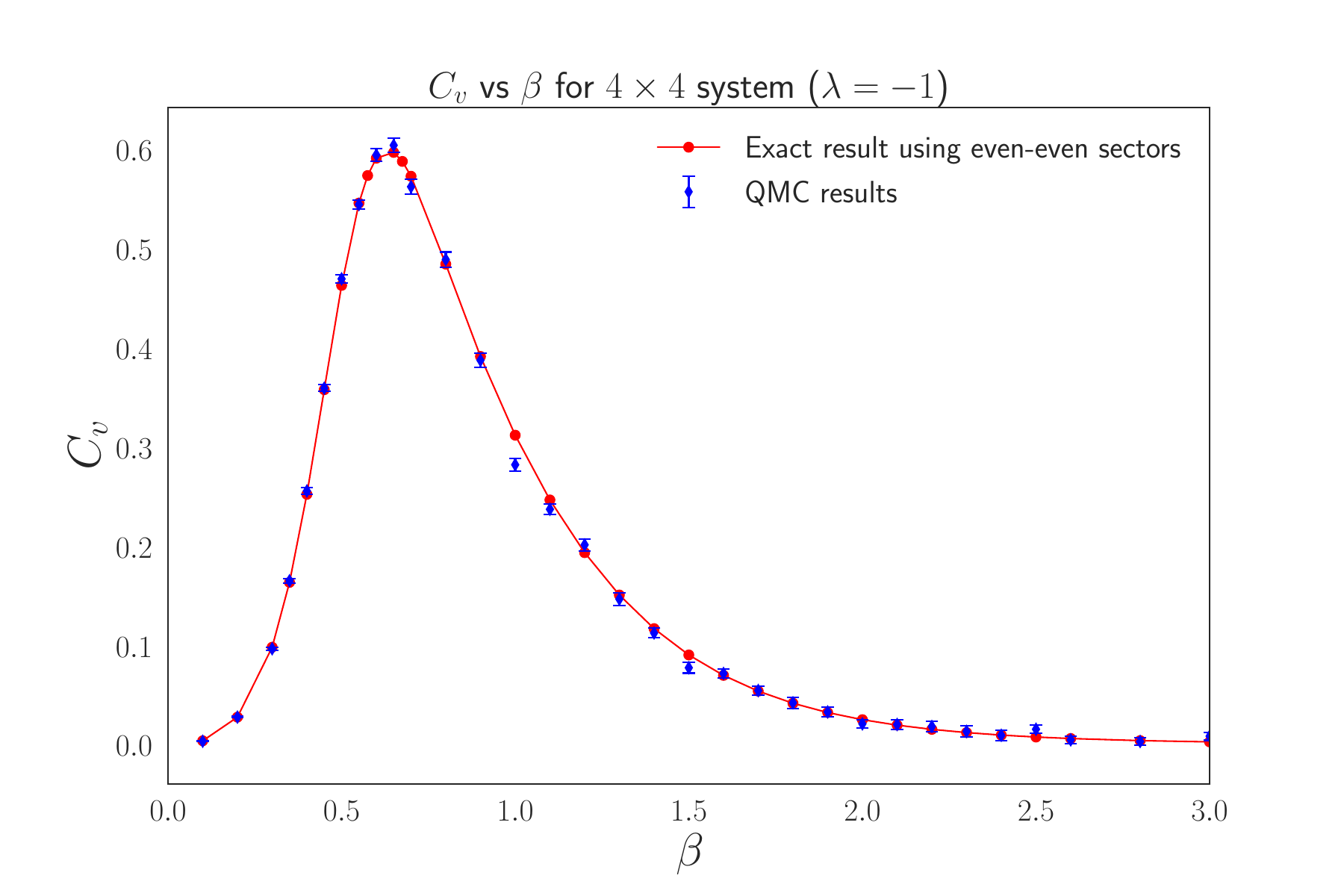}
    \includegraphics[scale=0.3]{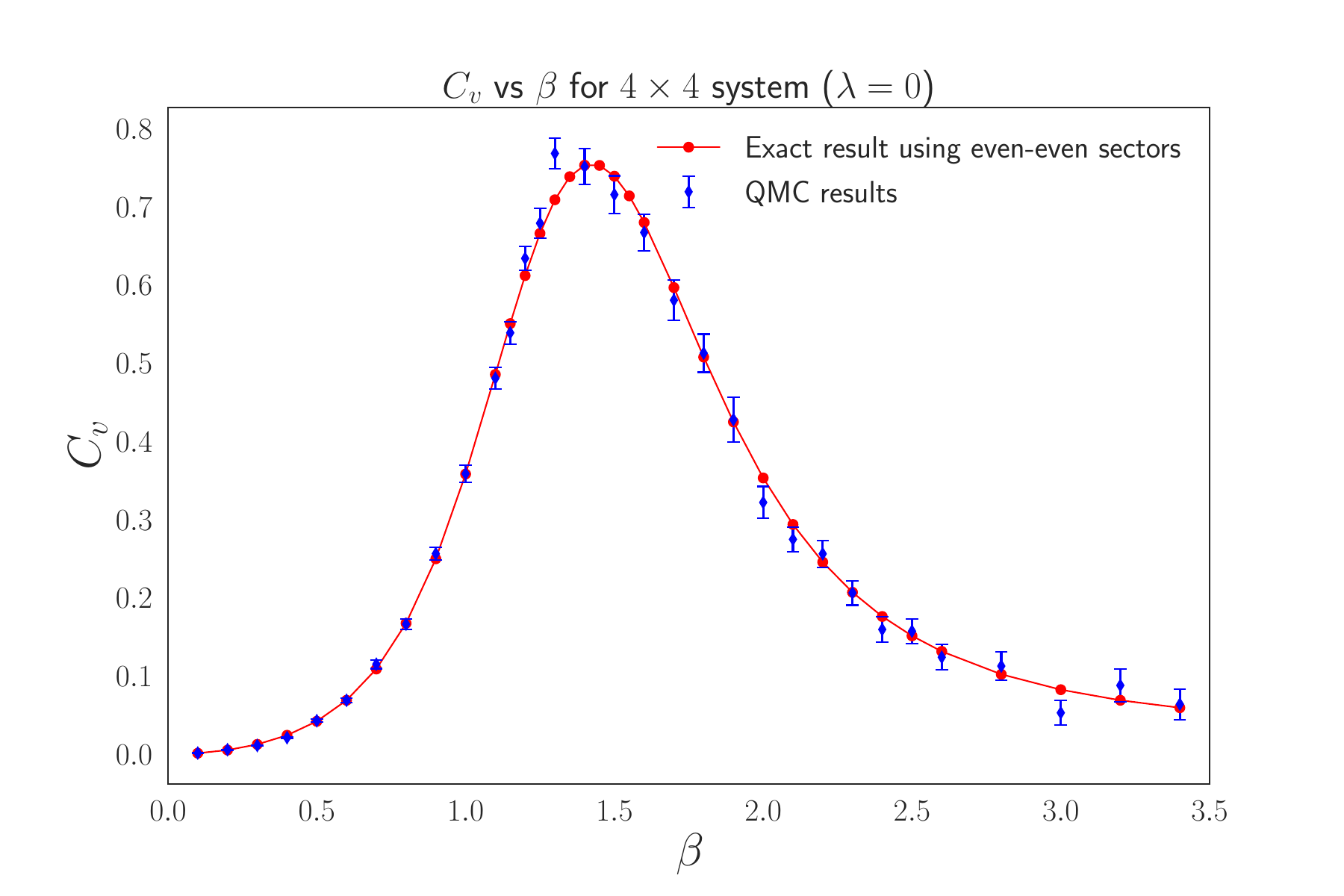}
    \caption{Specific heat $C_v$ computed on $4\times4$ lattice benchmarks the QMC
    algorithm and also provides an ordering of the (inverse) critical temperatures 
    $\beta_{1,2}$ for transitioning into the deconfined phase starting from the
    confined phase $C_{1,2}$ (corresponding to $\lambda=-1,0$).}
    \label{fig:CvU1}
\end{figure}

 Following the expectation from the Svetitsky-Yaffe conjecture, we expect the
finite temperature phase transition in the pure $U(1)$ theory to be in the BKT
universality class, as already motivated before. The zero temperature confined
phases $C1$ and $C2$ get deconfined at temperatues 
$\beta_c(\lambda_1) < \beta_c(\lambda_2)$, where $\beta = 1/T$, the inverse 
temperature, and the couplings are $\lambda_1 < \lambda_c < \lambda_2$, which
has influenced the schematic sketch in Fig.~\ref{fig:phDiagT}. We show results
from the calculation of the specific heat, $C_v$, in Fig.~\ref{fig:CvU1} on
a small $4 \times 4$ lattice benchmarking the quantum Monte Carlo algorithm
with exact diagonalization studies. The peak of $C_v$ identifies $\beta_{1,2}$
for $\lambda_{1,2}$ respectively, and while both values shift with the volume,
their relative ordering does not change.

\begin{figure}
     \centering
     \includegraphics[width=0.75\linewidth]{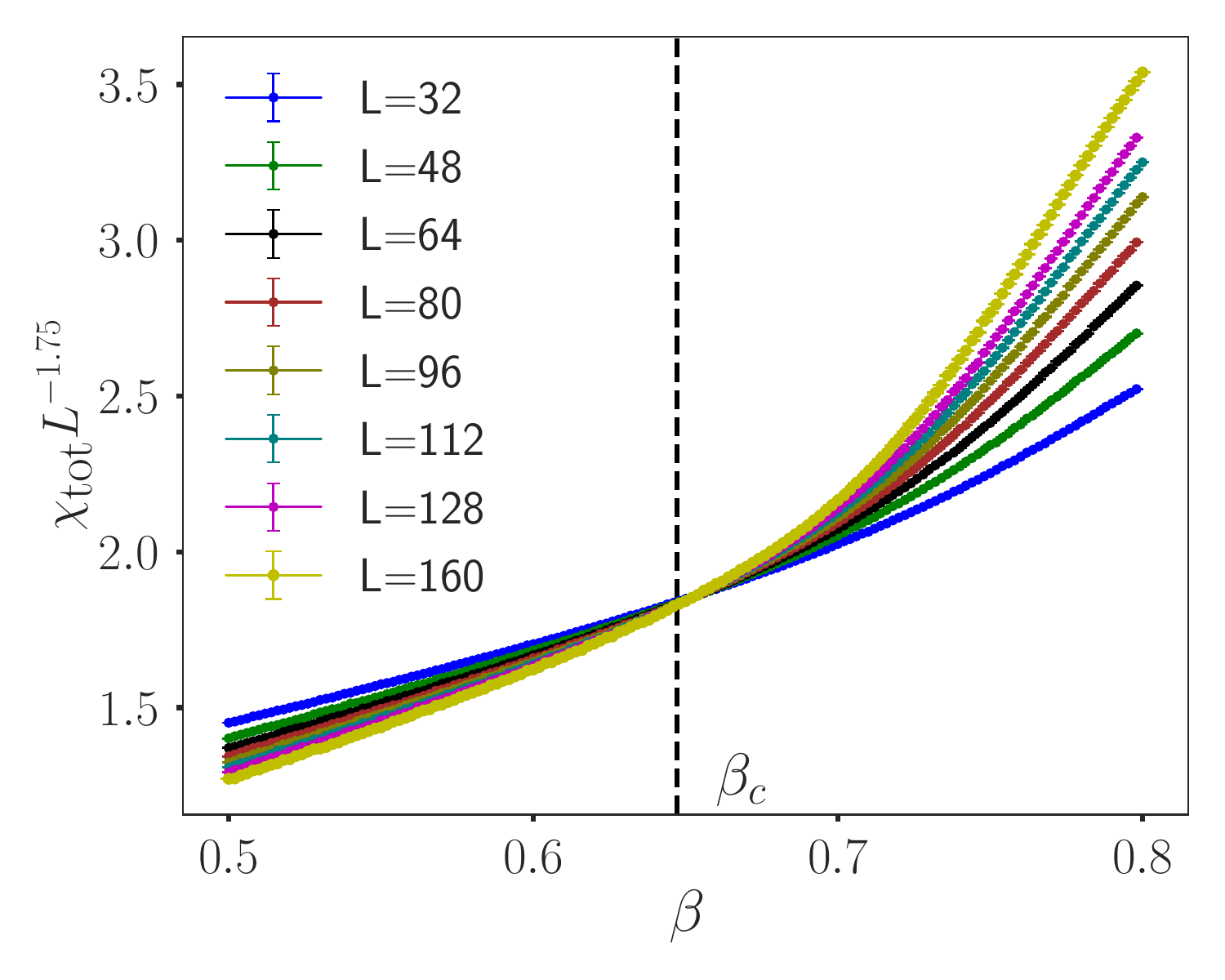}
     \includegraphics[width=0.75\linewidth]{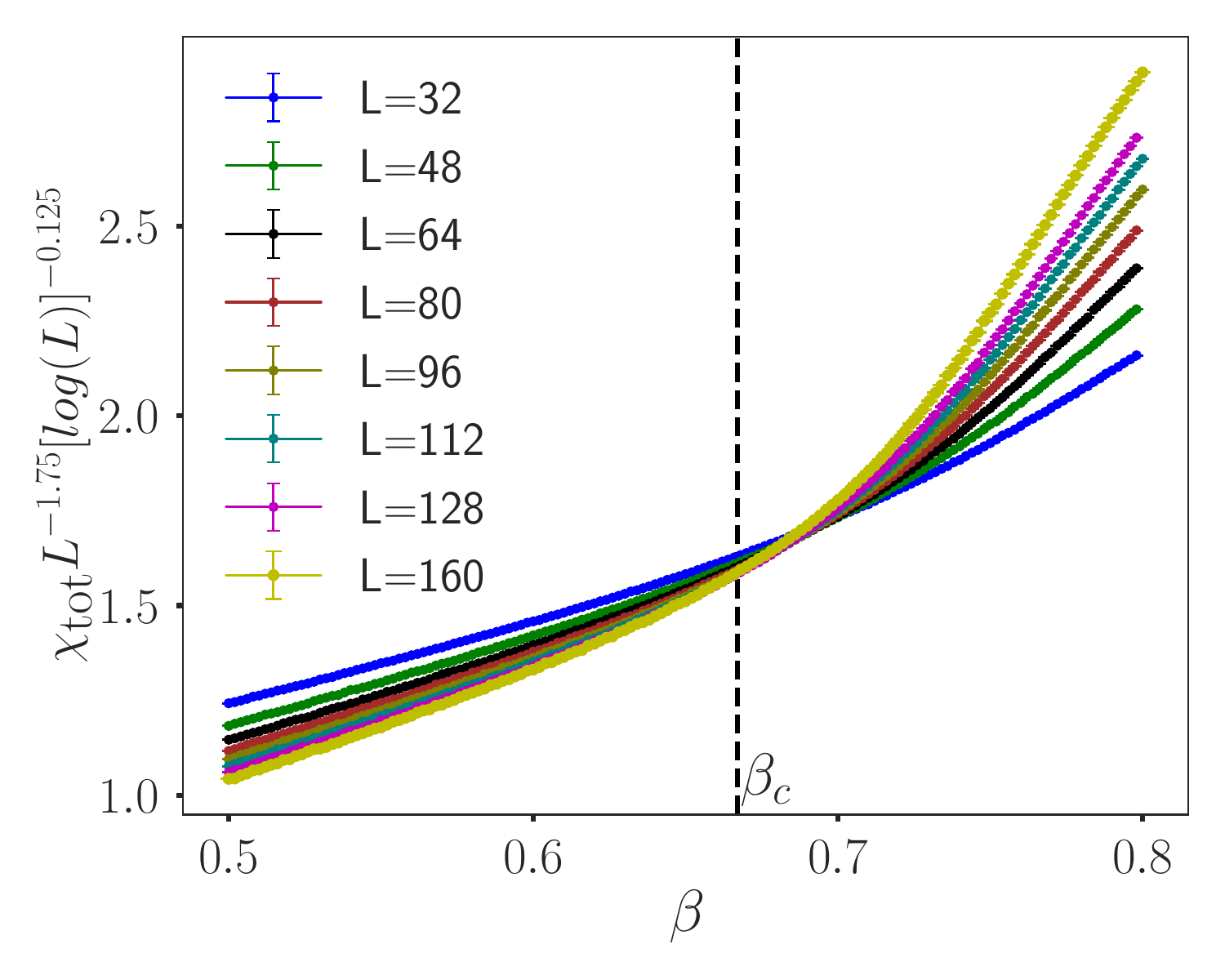}
     \caption{(Top panel) $\chi_{\textrm{tot}}L^{-1.75}$ vs $\beta$ for different system sizes. 
     The vertical dashed line indicates $\beta_c=0.6472(42)$ (Fig. \ref{fig:chi_crossBoth}). 
     (Bottom panel) $\chi_{\textrm{tot}}L^{-1.75}[{\textrm{log}}(L)]^{-0.125}$ vs $\beta$ for 
     different system sizes. The vertical dashed line indicates $\beta_c=0.6668(33)$ (Fig. 
     \ref{fig:chi_crossBoth}).}
     \label{fig:chiL}
\end{figure}

\begin{figure}
    \centering
    \includegraphics[width=1\linewidth]{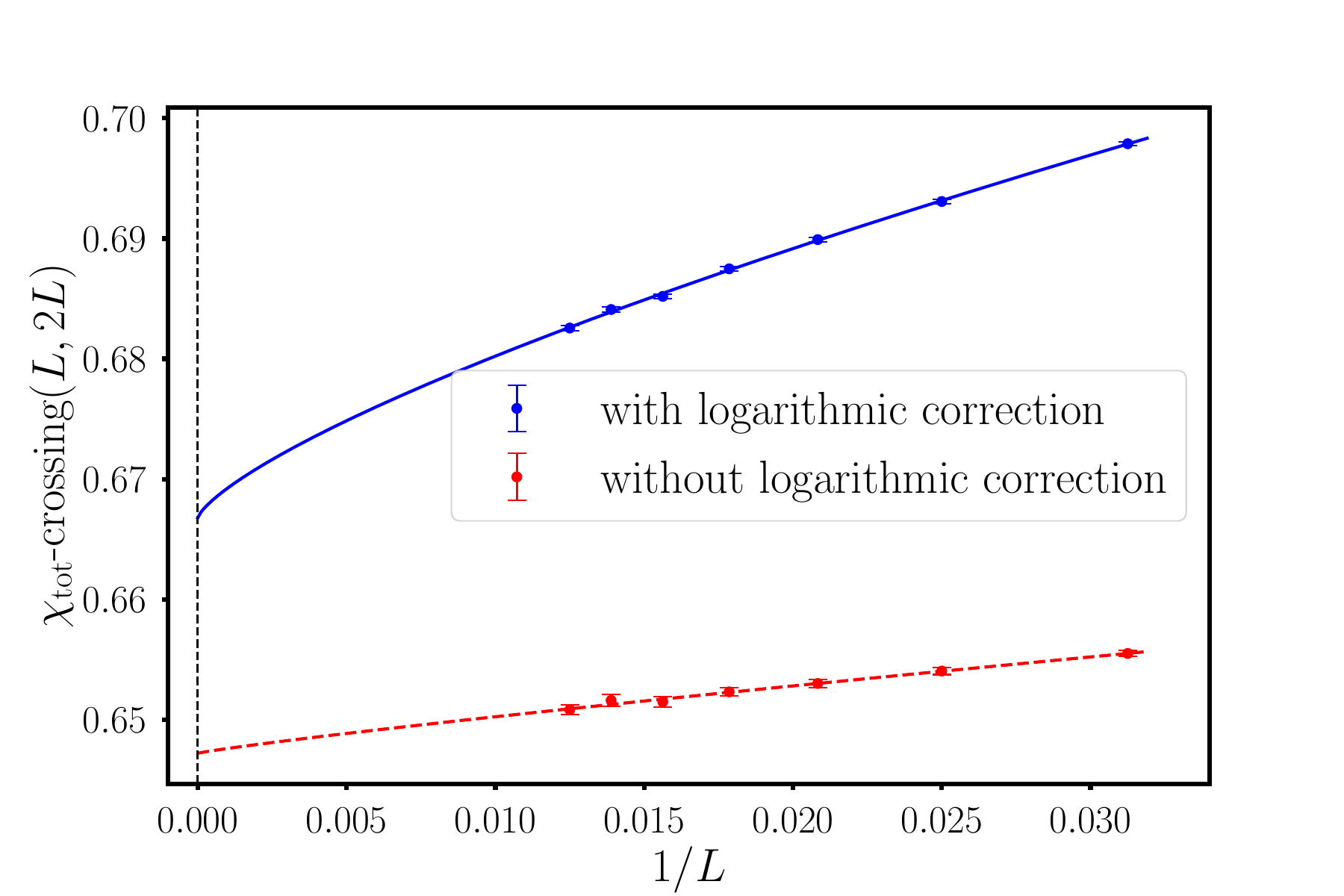}
    \caption{$\chi_{\rm{tot}}$-crossing points of  pair of system size $L$ and $2L$ as a 
    function of $1/L$ both with and without logarithmic correction. The data is fitted as 
    $a + b (\frac{1}{L})^c$. From the fitting, we extract the critical coupling 
    $\beta_c=0.6668(33)[0.6472(42)]$ with [without] logarithmic correction.}
    \label{fig:chi_crossBoth}
\end{figure}

 We now discuss our results for extracting the critical temperature $\beta_c$ at the BKT
phase transitions by using finite size scaling (FSS) of the thermodynamic observables. We 
will first consider the case when $\lambda = -1.0$, where we have analyzed different 
observables. It is known from the literature that at the BKT transition, the total 
susceptibility, $\chi_{\rm{tot}}$, scales as $L^{\frac{7}{4}}[\log(L)]^r$ at the critical 
point $\beta_c$ \cite{Balog:2000ra,Hasenbusch:2005xm}. The presence of the logarithmic 
corrections has been analytically computed in \cite{Balog:2000ra} to be $r=0.125$,
and was also tested phenomenologically by fitting the data to obtain $r$ in \cite{Hasenbusch:2005xm}. 
So, at $\beta_c$, $\chi_{\textrm{tot}}L^{-\frac{7}{4}}[\log(L)]^{-0.125}$ for different 
sizes, $L$, are expected to cross each other.  
Since the accurate extraction of logarithmic corrections are subtle, we have simply 
compared the systematics of including the expected log correction by
following the crossing points of the dimensionless susceptibility $\chi_{\textrm{tot}}L^{-\frac{7}{4}}$
for lattice sizes $L$ and $2L$ with and without the log-corrections as shown in Fig.~\ref{fig:chiL}. 
The second step was to then extrapolate the finite volume pseudo-critical couplings $\beta_c(L)$ with 
a power-law to the theromodynamic limit, as shown in Fig.~\ref{fig:chi_crossBoth}. The estimates
of the critical inverse temperature in the thermodynamic limit are $\beta_c=0.6668(33)$ with 
the logarithmic corrections, and $0.6472(42)$ without. These numbers differ at the 3.5-$\sigma$
indicating that the log corrections play a significant role at this BKT transition as well.

\begin{figure}
    \centering
    \includegraphics[width=0.95\linewidth]{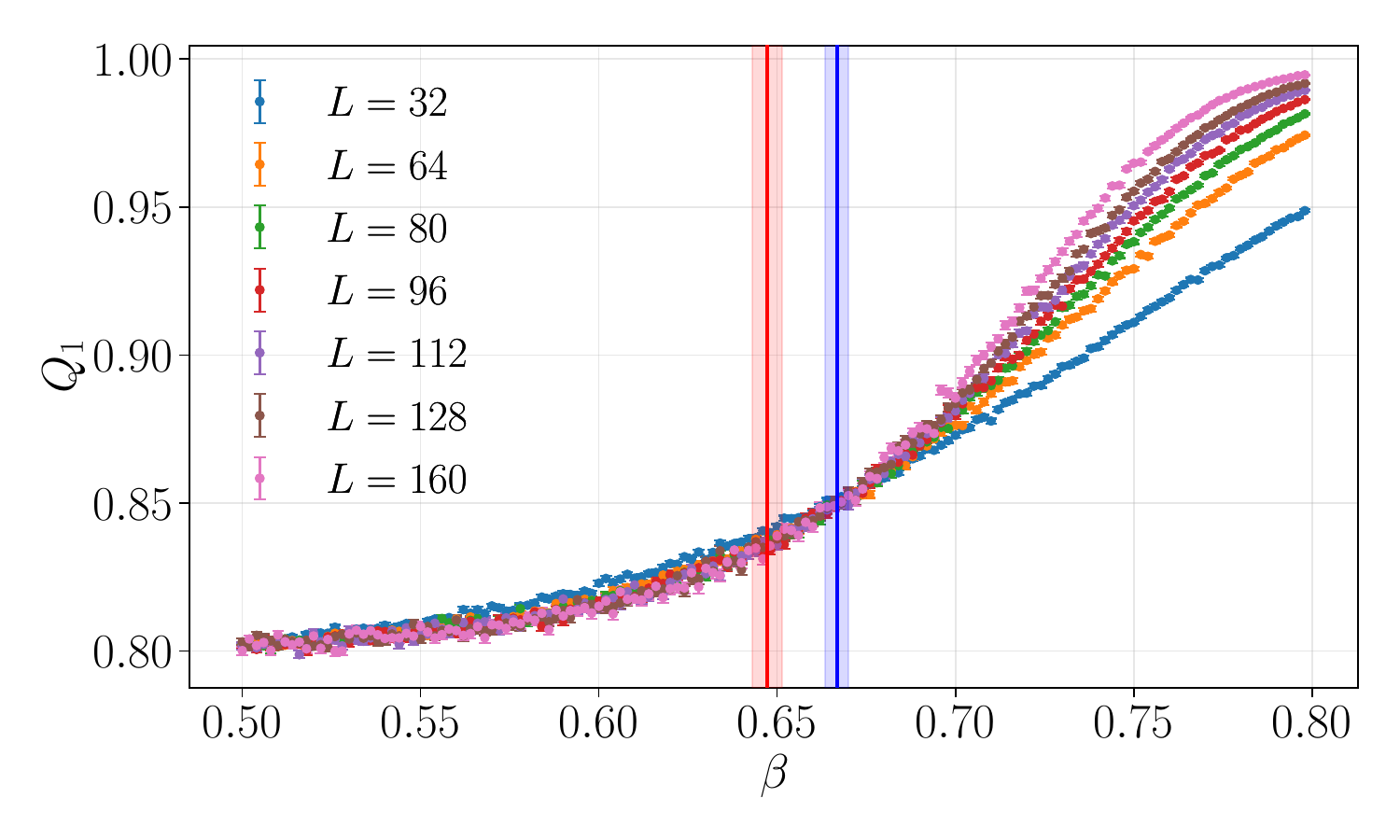}
    \includegraphics[width=0.95\linewidth]{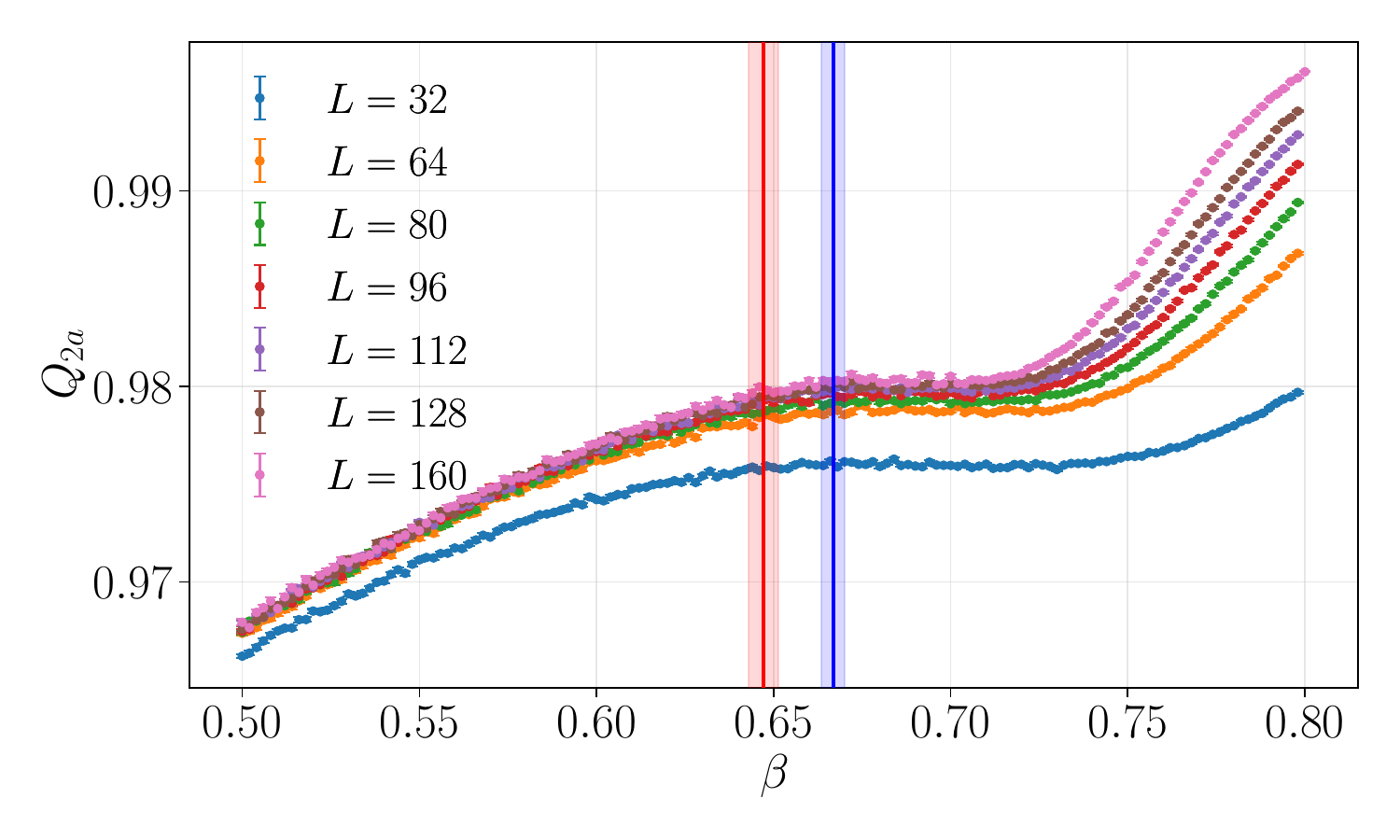}
    \includegraphics[width=0.95\linewidth]{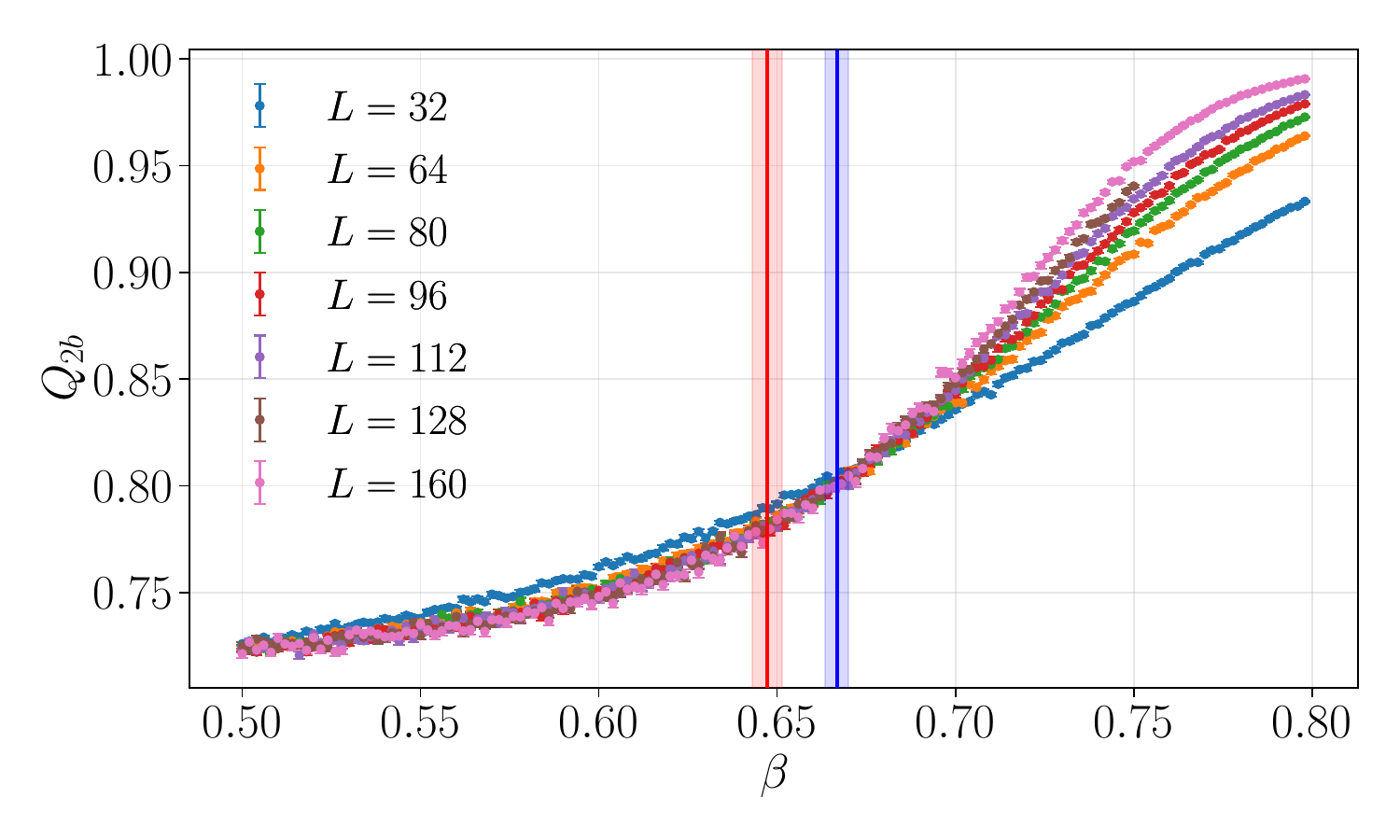}
    \caption{Different Binder ratios for pure gauge theory. The vertical blue (red) line shows the position of the critical temperature obtained from Fig.~\ref{fig:chi_crossBoth} (without) using logarithmic correction with the band indicating the error.}
    \label{fig:binders}
\end{figure}

 In addition to using the susceptibility, we have also explored the role of using the dimensionless
Binder ratios in order to extract $\beta_c$. In Fig.~\ref{fig:binders}, we show the behaviour of
the different Binder ratios $Q_1$, $Q_{2a}$, and $Q_{2b}$ respectively. All the Binder ratios 
have been normalized to lie in between 0 and 1. Qualitatively, the curves $Q_1$ and $Q_{2b}$ look
similar, at low-$\beta$ (high-temperatures) these curves almost fall on top of each other (except
for the smallest lattice, $L=32$), but appear to diverge around $\sim 0.67-0.68$ from each other
effectively placing the transition around this value. Note that this estimate of $\beta_c$ is
consistent with the estimate from dimensionless-$\chi$. Particularly, the estimate of $\beta_c$ 
using logarithmic correction coincides within this diverging region. We also notice plateaus 
in $Q_{2a}$ around $\sim0.67-0.68$ for all system sizes. It seems that the estimate of $\beta_c$ 
using the logarithmic correction is closer to the point where the cumulant splays out with 
increasing system sizes, particularly in the top and the bottom panels of Fig.~\ref{fig:binders}, 
supporting XY criticality.

 Since our principal aim behind the extraction of $\beta_c = \frac{1}{T_c}$ is to understand how the 
dimensionless ration $\frac{T_c}{\sqrt{\sigma}}$ behave as a function of the bare-coupling 
$\lambda$, the extraction at the two other values of $\lambda=-0.9,-1.1$ follow an even
simpler procedure, illustrated in Table.~\ref{tab:U1_betac}. 
The high-temperature phase in the gauge theory corresponds to the low-temperature phase of
the XY model where the susceptibility $\chi$ at a finite box of size $L$ behaves as $L^{2-\eta}$,
we perform this fit for $\chi$ as a function of decreasing the temperature (increasing $\beta$)
and note the $\beta$ at which this fit fails. This procedure yields a value of $\beta_c$, sufficient
for our purposes. From the table.~\ref{tab:U1_betac}, we identify $\beta_c(\lambda=-1.1)=0.59$,
$\beta_c(\lambda=-1.0) = 0.64$, and $\beta_c(\lambda=-0.9)=0.71$. Note that the value obtained
for $\lambda = -1.0$ is 0.64, consistent with the extrapolation without the log-correction 
as shown in Fig.~\ref{fig:chi_crossBoth}. We will use these values again in Sec.~\ref{sec:TcSqrtSig}.

\begin{table}
	\begin{tabular}{ | c | c | c | }
		\hline
            $\beta$ & $\eta$ & $\chi^2/dof$ \\
            \hline
            \multicolumn{3}{|c|}{$\lambda=-0.9$}\\
            \hline
            $0.65$ & $0.2794(4)$ & $0.43$\\
            \hline
            $0.68$ & $0.2655(3)$ & $0.91$\\
            \hline
            $0.7$ & $0.2557(3)$ & $1.1$\\
            \hline
            $0.71$ & $0.2505(3)$ & $0.61$\\
            \hline
            $0.72$ & $0.2445(3)$ & $2.64$\\
            \hline
            $0.73$ & $0.2388(2)$ & $2.78$\\
            \hline
            $0.74$ & $0.2322(2)$ & $5.99$\\
            \hline
		\multicolumn{3}{|c|}{$\lambda=-1.0$}\\
            \hline
            $0.6$ & $0.2803(4)$ & $0.73$ \\
            \hline
            $0.62$ & $0.2692(3)$ & $1.08$ \\
            \hline
            $0.63$ & $0.2643(3)$ & $2.32$ \\
            \hline
            $0.64$ & $0.2580(3)$ & $1.69$ \\
            \hline
            $0.65$ & $0.2520(3)$ & $2.17$ \\
            \hline
            $0.66$ & $0.2452(2)$ & $2.49$ \\
            \hline
            $0.67$ & $0.2380(2)$ & $2.67$ \\
            \hline
            $0.68$ & $0.2293(2)$ & $6.29$ \\
            \hline
            \multicolumn{3}{|c|}{$\lambda=-1.1$}\\
            \hline
            $0.55$ & $0.2837(4)$ & $1.29$\\
            \hline
            $0.58$ & $0.2657(3)$ & $0.74$\\
            \hline
            $0.59$ & $0.2597(3)$ & $1.65$\\
            \hline
            $0.594$ & $0.2573(3)$ & $2.51$\\
            \hline
            $0.6$ & $0.2526(3)$ & $3.38$\\
            \hline
            $0.61$ & $0.2449(2)$ & $10.65$\\
            \hline
    \end{tabular}
    \caption{Fitting of $\chi_{tot}(\beta)\sim L^{2-\eta}$ at different $\beta$ for 
       $\lambda=-0.9,\,-1.0$ and $-1.1$.}
	\label{tab:U1_betac}
\end{table}

\section{Order parameter distributions across phase transition} \label{sec:OP}
 Although the nature of the phase transition for the pure $U(1)$ theory is clear, it is insightful
 to consider the order parameter distributions both for the pure matter-free theory, and the one
 when we include the charge $\pm 2$ as defects, and can provide inputs for an effective field
 theory treatment. 

\begin{figure}[!htpb]
 \centering
 \includegraphics[scale=0.2]{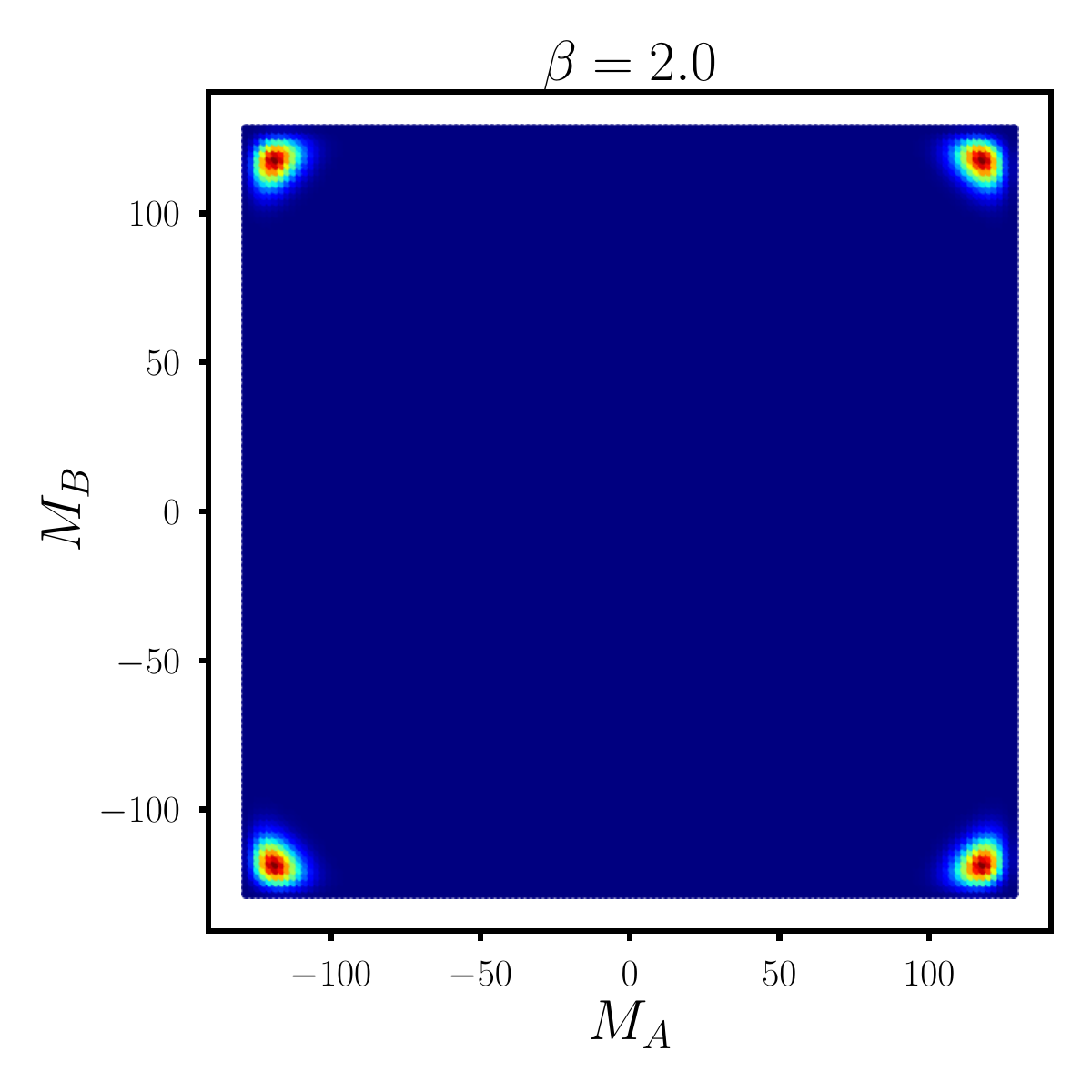}
 \includegraphics[scale=0.2]{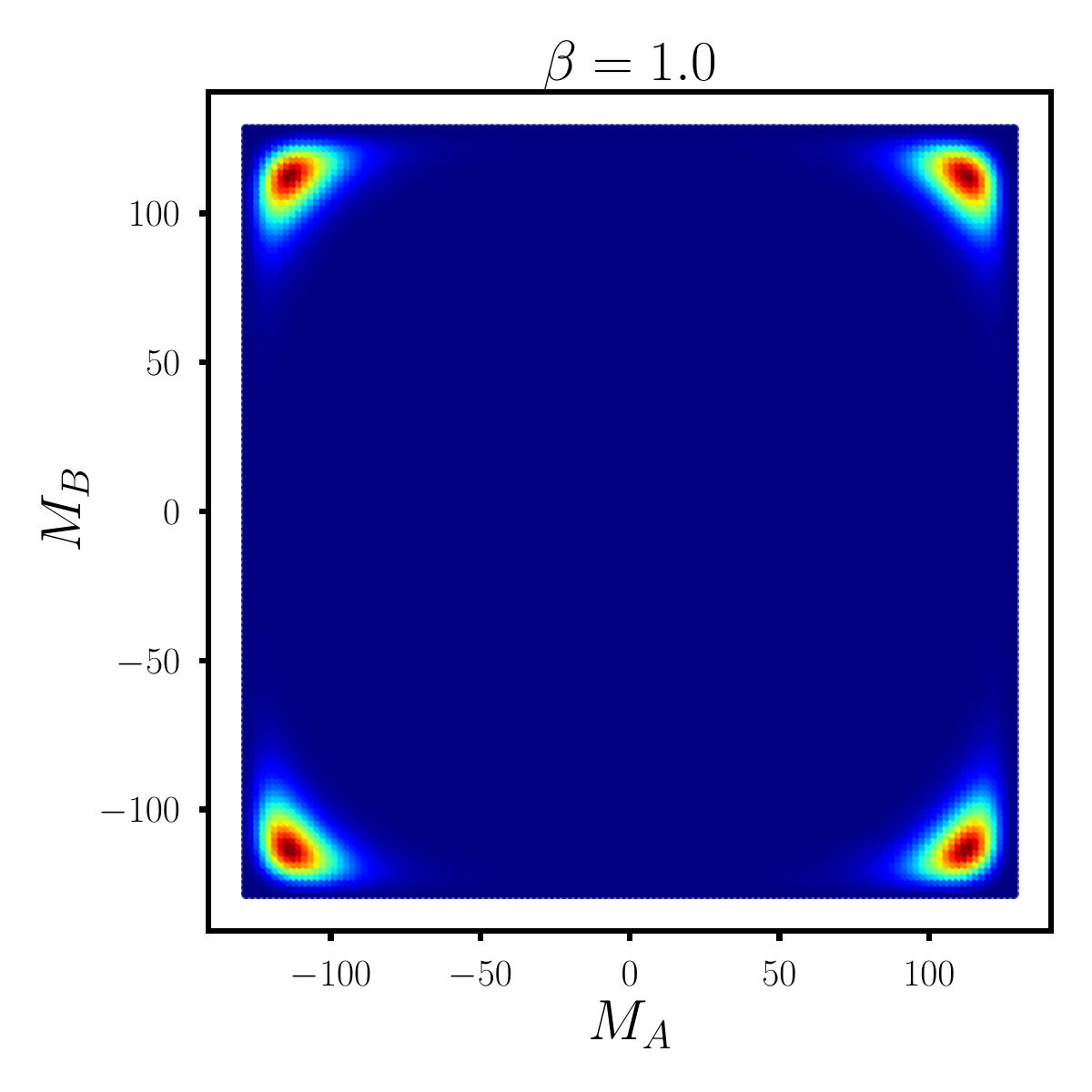}
 \includegraphics[scale=0.2]{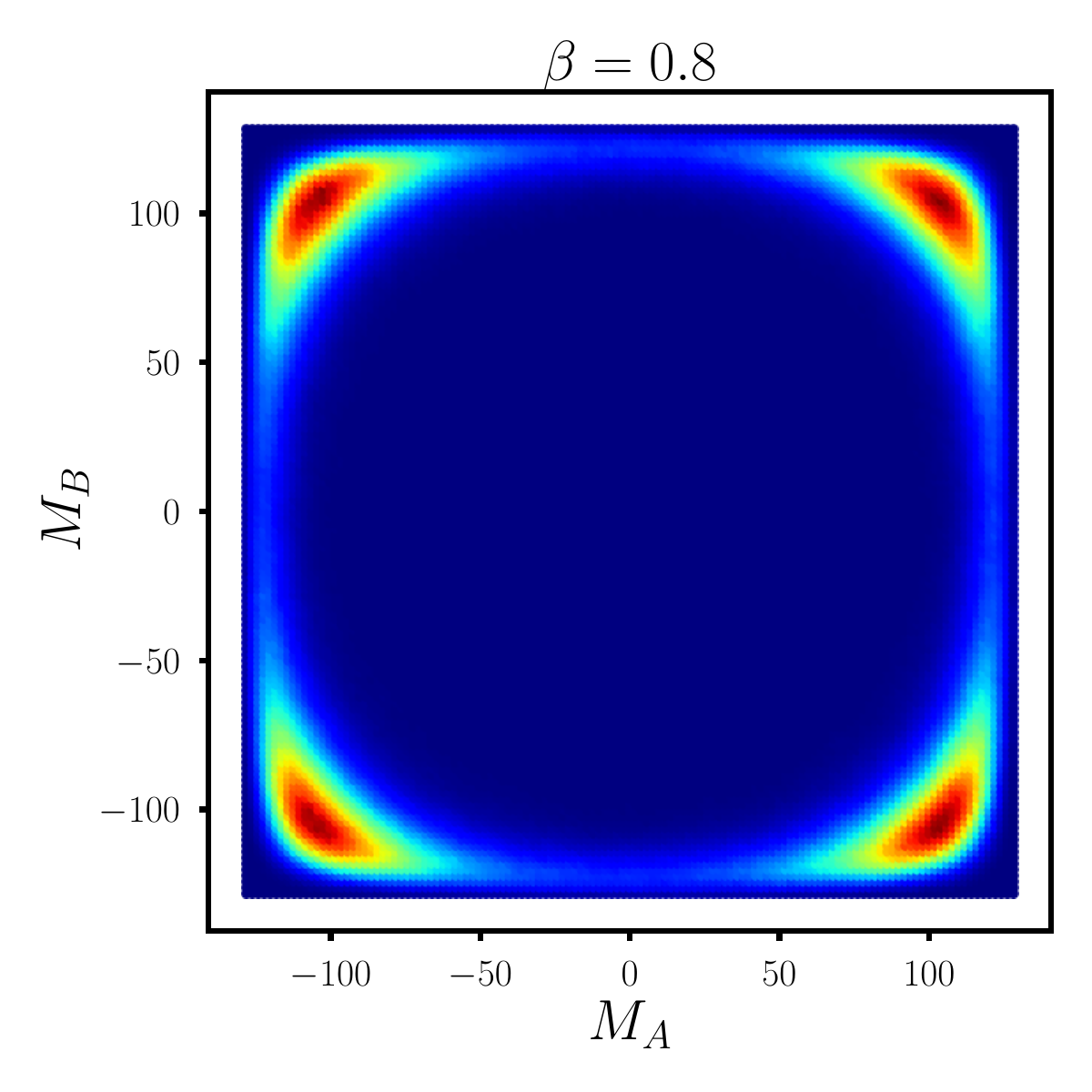}
 \includegraphics[scale=0.2]{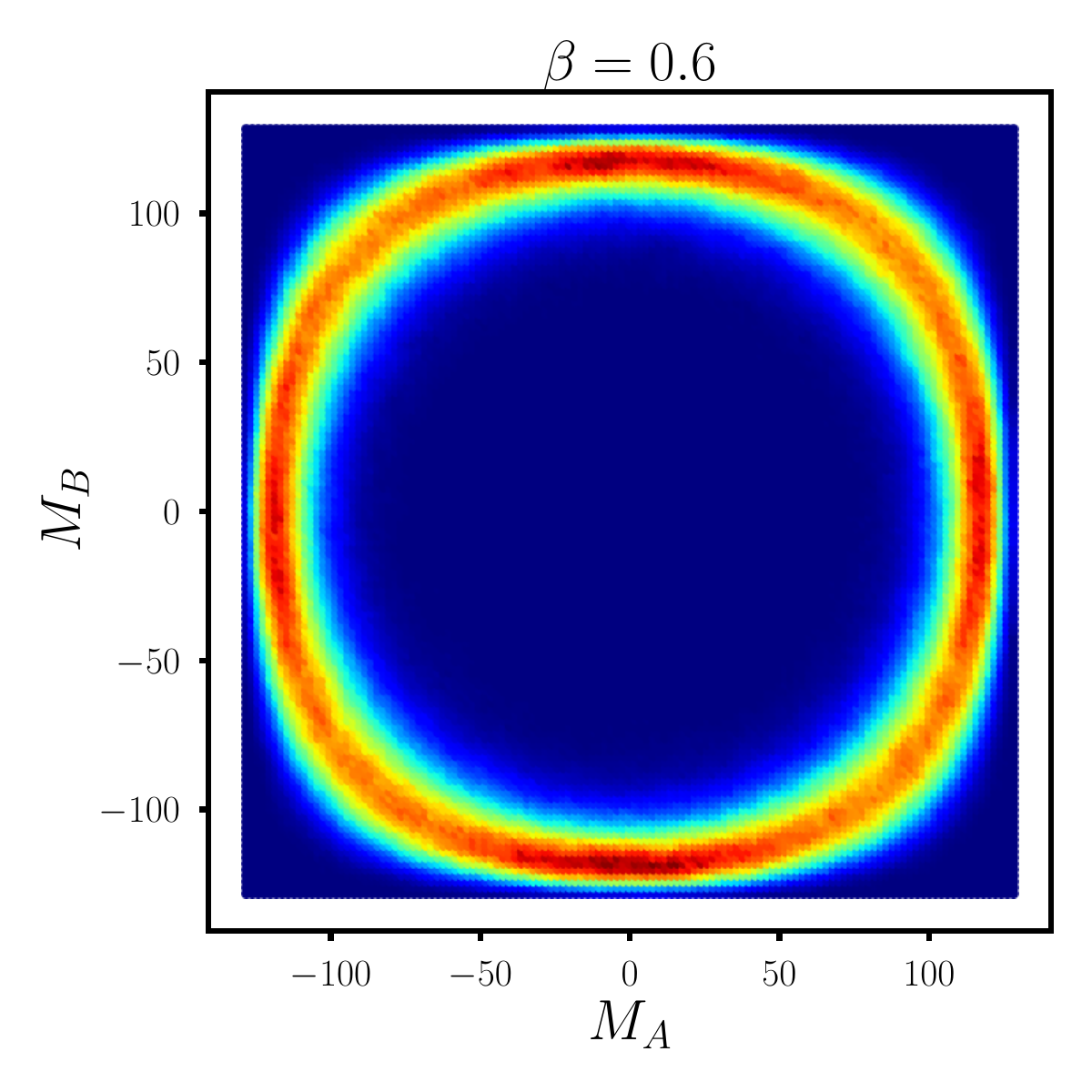}
 \caption{Distribution of ($M_A,M_B$) for pure $U(1)$ QLM at $\lambda=-1$ and 
 different $\beta$ ($L_x=L_y=16, L_t=24$) [no symmetrization is applied].}
 \label{fig:pmambU1lam-1}
 \end{figure}

 Let us first consider the finite temperature phase transition when we raise the temperature
 starting from the ground state at $\lambda = -1$. The probability distributions at four 
 successively higher temperatures of $\beta J = 2, 1, 0.8$, and $0.6$ are shown in 
 Fig.~\ref{fig:pmambU1lam-1}. At $T=0$ the symmetry breaking is manifest through the 
 peaks of the order parameters for which $\braket{M_A} = \braket{M_B} \approx v$. 
 Physically, this corresponds to the phase where the plaquettes on both the sublattice are
 equally flippable, but interlocked (i.e., one clockwise and the other anti-clockwise), 
 and thus both are ordered. 
 Note the efficiency of the cluster algorithm which causes it to tunnel through all the 
 vacuua with equal ease. 
 This is indeed possible since the cluster flip is sufficient to connect the 
 different vacuua, just as in the case of the Ising model. As the temperature is raised,
 the peaks gets diffused until the system undergoes a deconfined transition, beyond which
 the symmetry gets restored. In the deconfined phase, the order parameters $M_{A,B}$ can
 vary smoothly along a deformed circle, dictated by the $U(1)$ Gauss Law. 
 Another related point to note here is that the system is updated
 through the height variables which are themselves not gauge invariant. Indeed, 
 $(M_A,M_B)$ and $-(M_A,M_B)$ are gauge equivalent copies, and thus there are only two 
 physical vacuua. 

\begin{figure}
 \centering
 \includegraphics[scale=0.2]{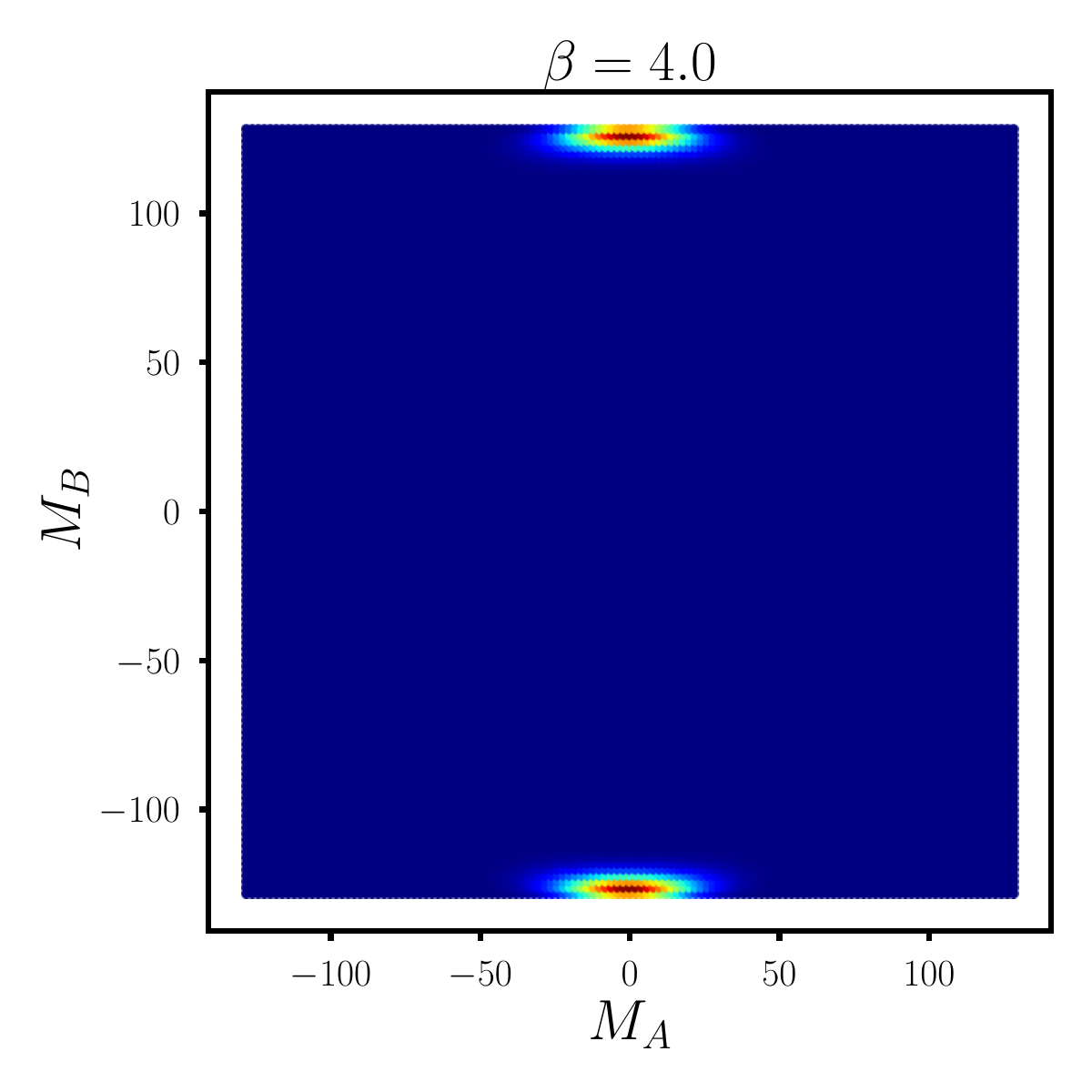}
 \includegraphics[scale=0.2]{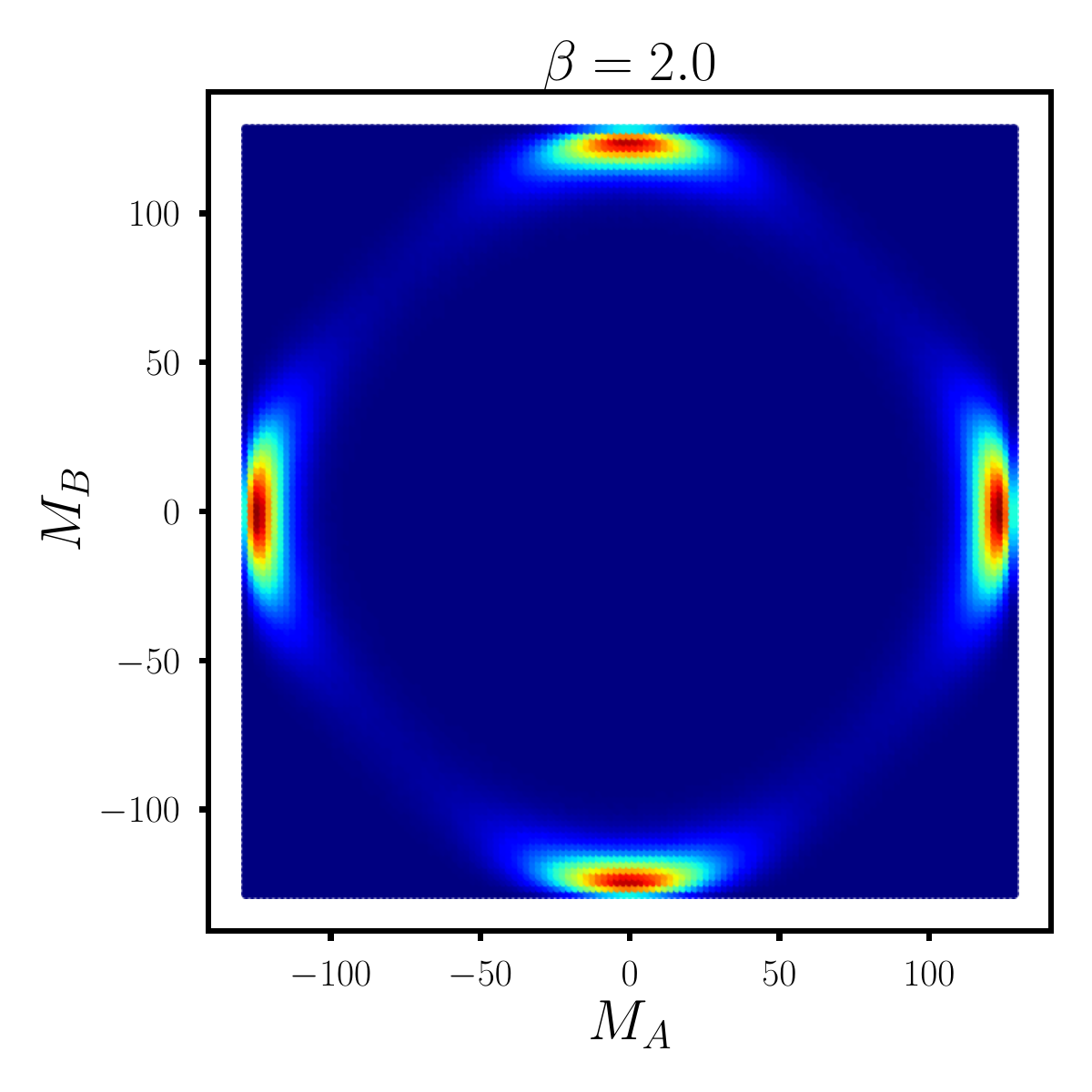}
 \includegraphics[scale=0.2]{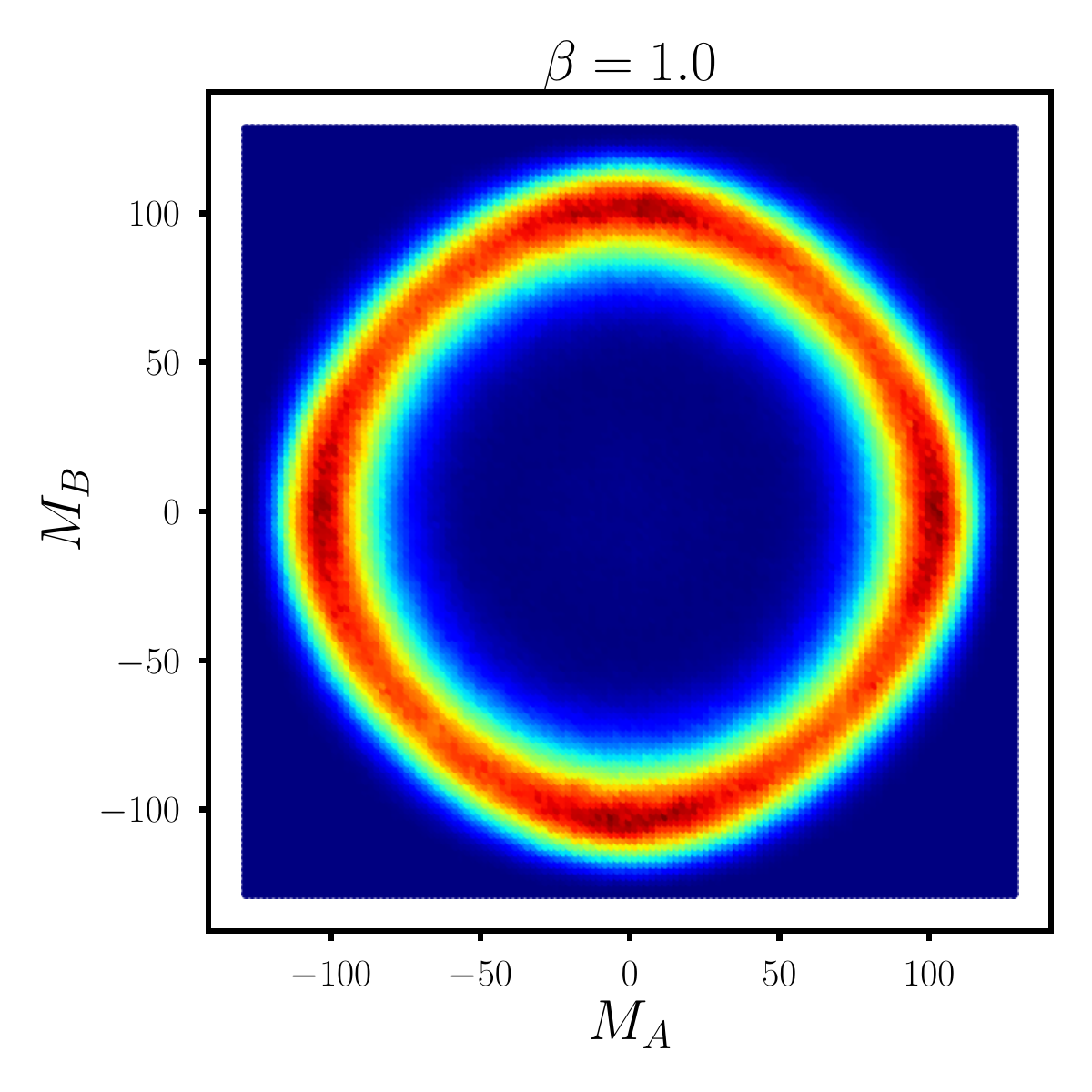}
 \includegraphics[scale=0.2]{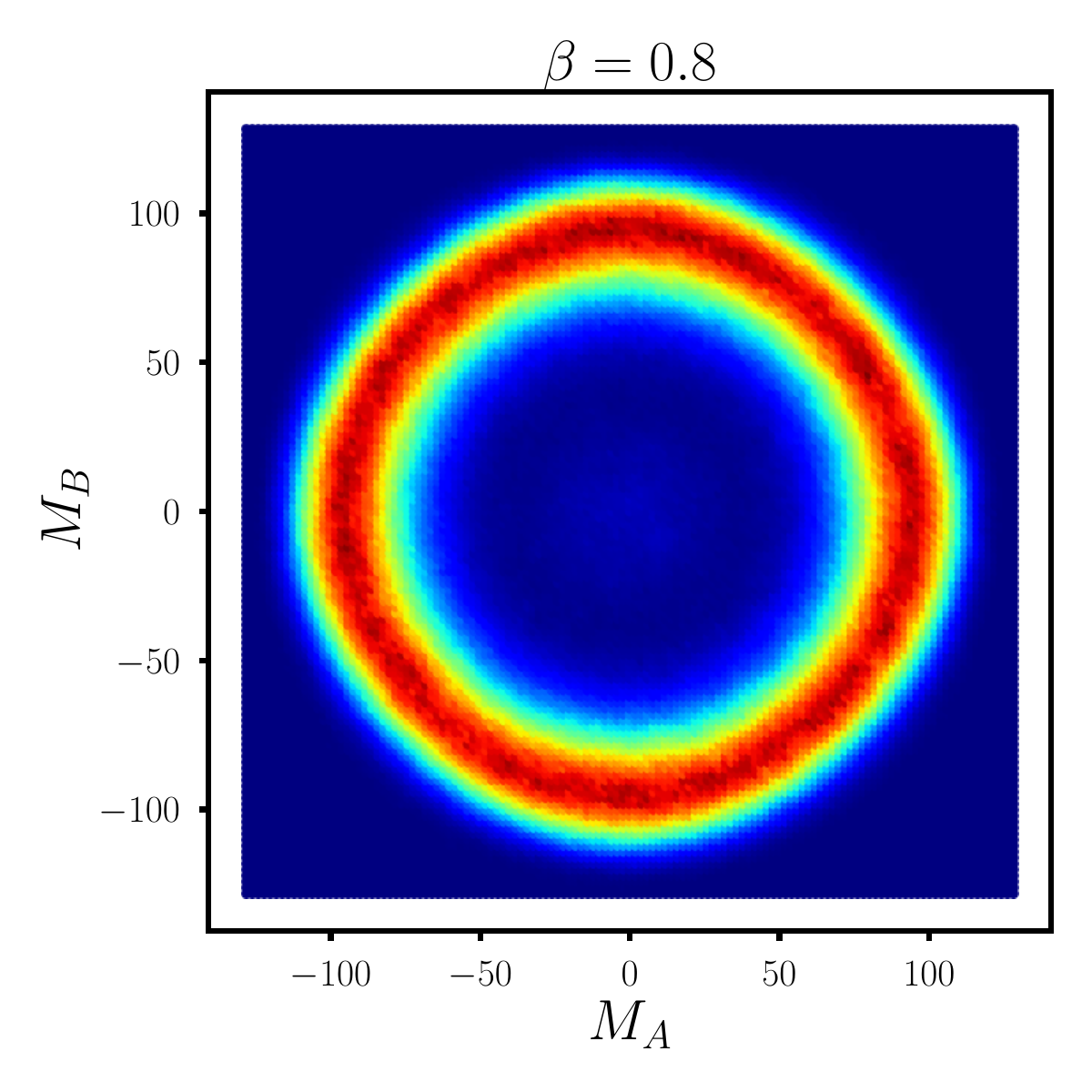}
 \caption{Distribution of ($M_A,M_B$) for pure $U(1)$ QLM at $\lambda=0$ and different 
 $\beta$ ($L_x=L_y=16, L_t=24$) [no symmetrization is applied].}
 \label{fig:pmambU1lam0}
 \end{figure}

  A related picture emerges if instead we study the deconfined transition while starting
  from $\lambda=0$. The ground state for this coupling breaks a different discrete symmetry,
  one that involves a single shift in the lattice spacing. In this region, \emph{only} one 
  of the sublattice is ordered, either A, or B; and this is spontaneously decided by the
  system. The physical picture is that either the A or the B sublattice is in a resonating
  valence bond, where the plaquettes are rapidly flipping, while the other sublattice is
  consequently rendered non-flippable. This is manifest by 
  $\braket{M_A} \approx 0, \braket{M_B} \approx \tilde{v}$ in the top left panel of 
  Fig.~\ref{fig:pmambU1lam0} which shows $p(M_A, M_B)$ at $\beta = 4.0$, a relatively
  low temperature. Note in particular that the algorithm does not tunnel to the other
  vacuum $\braket{M_A} \approx \tilde{v}, \braket{M_B} \approx 0$ at the lowest temperature
  since the native cluster flips do not involve a shift and cannot connect the two vacuua. 
  At low temperatures, therefore, the algorithm is stuck in one (physical) vacuum. The
  rise in temperatures is accompanied with the rise of fluctuations, which can then cause
  the system to tunnel to the other vacuum, as shown in the top right panel for $\beta = 2.0$.
  Increasing the temperature further, one transitions into a deconfined phase where the
  symmetry breaking disappears. However, the simulations for smaller values in $\lambda$
  are accompanied with large fluctuations, and since the cluster algorithm does not natively
  target the broken symmetry, the fluctuations are larger in this case. 
  
\begin{figure}
    \centering
    \includegraphics[scale=0.2]{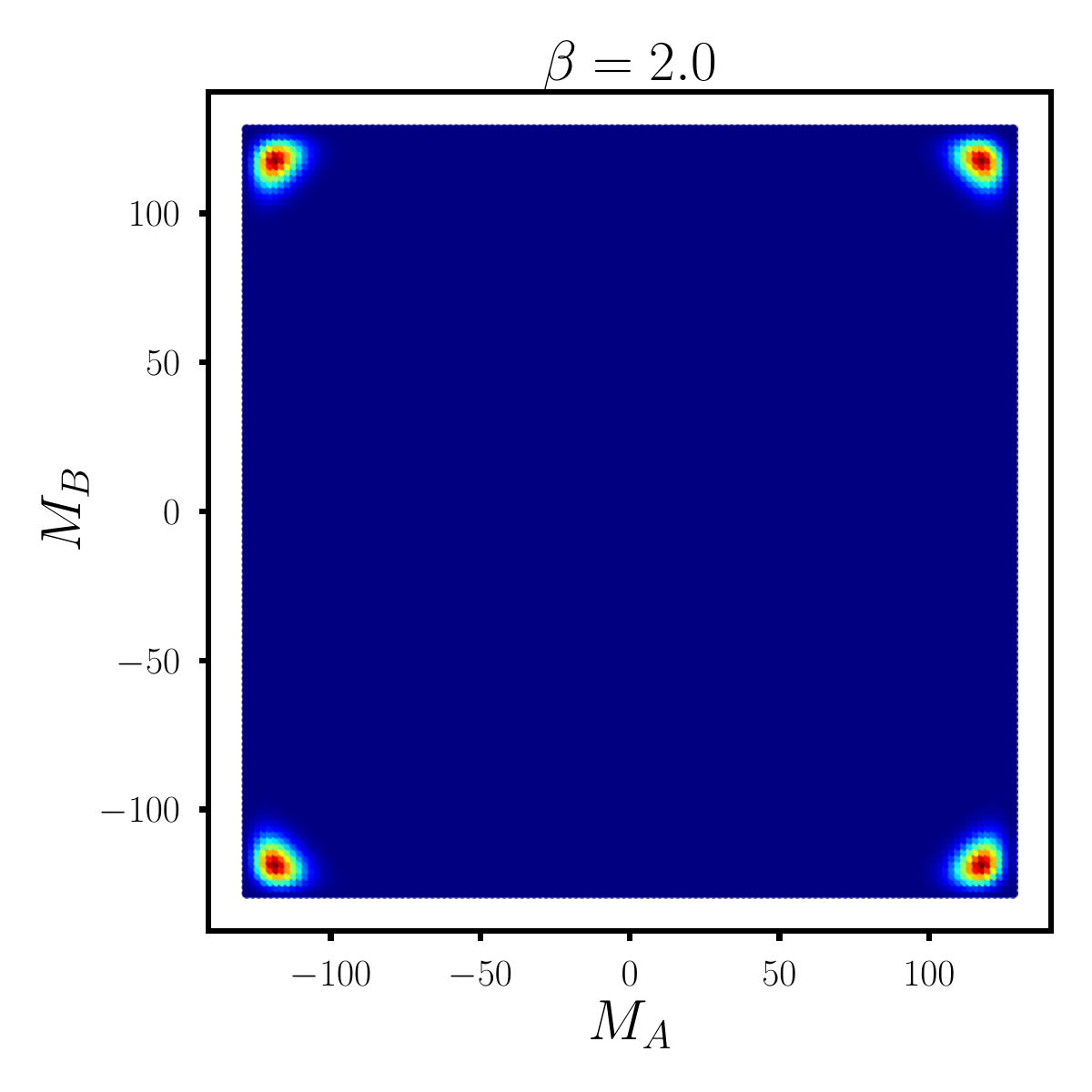}
    \includegraphics[scale=0.2]{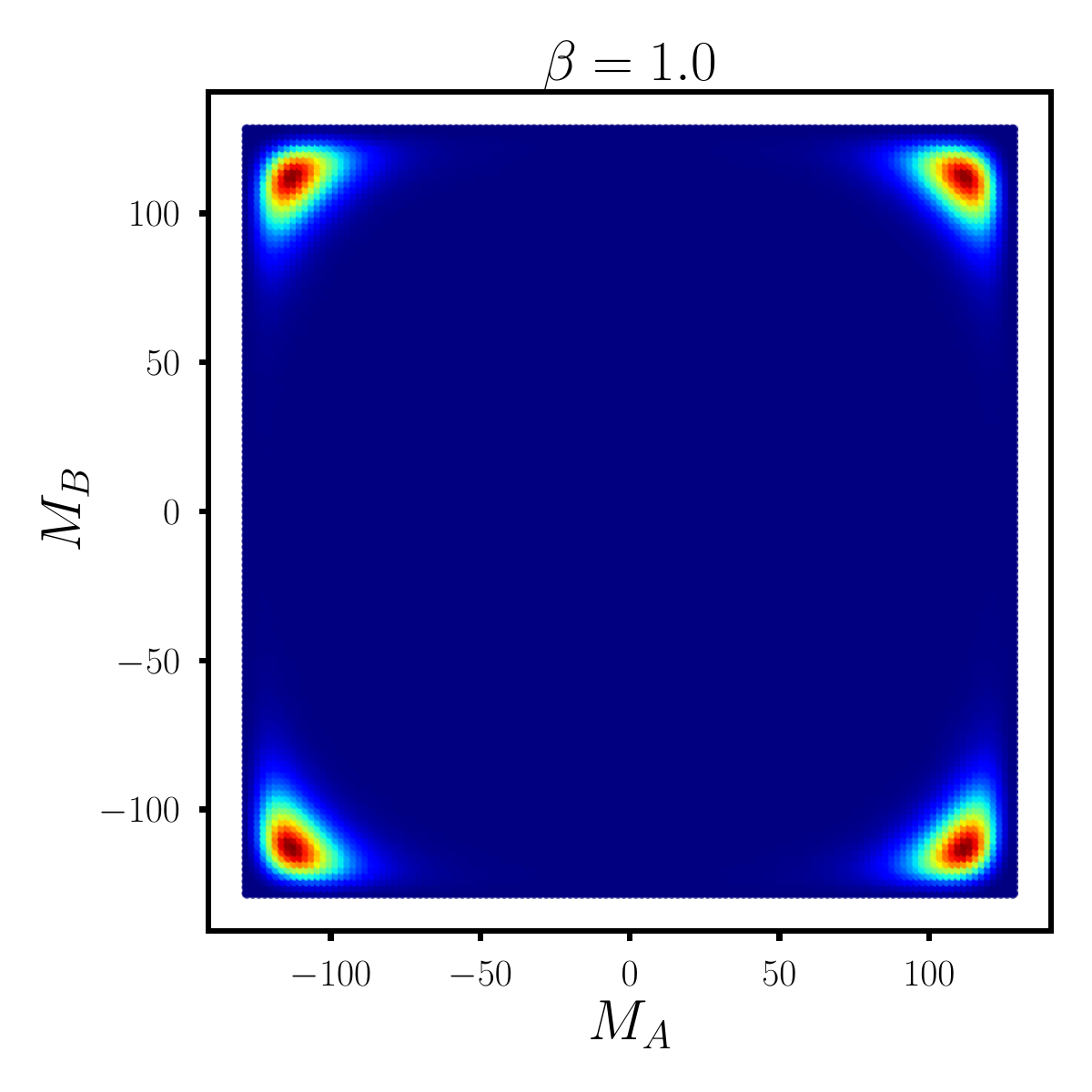}
    \includegraphics[scale=0.2]{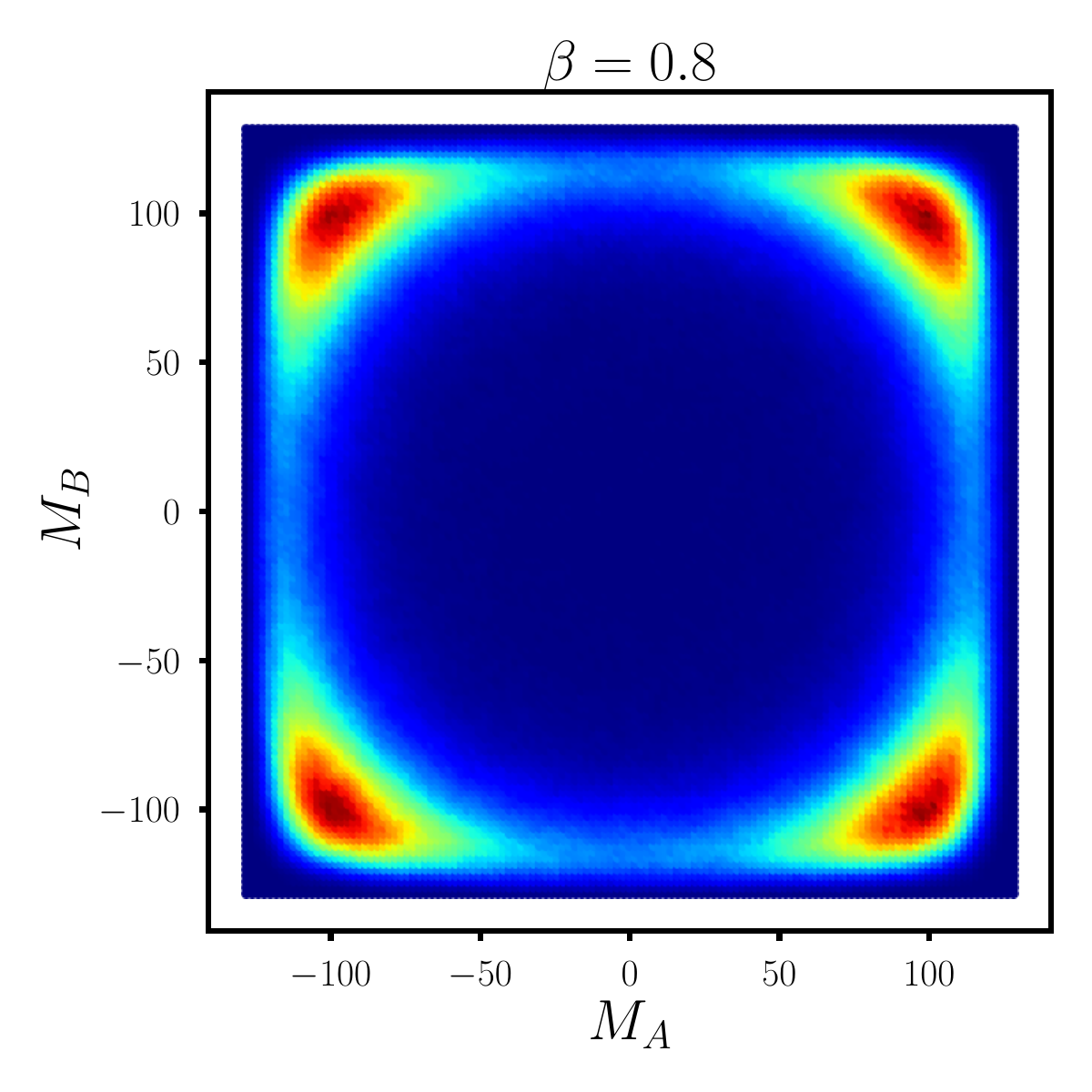}
    \includegraphics[scale=0.2]{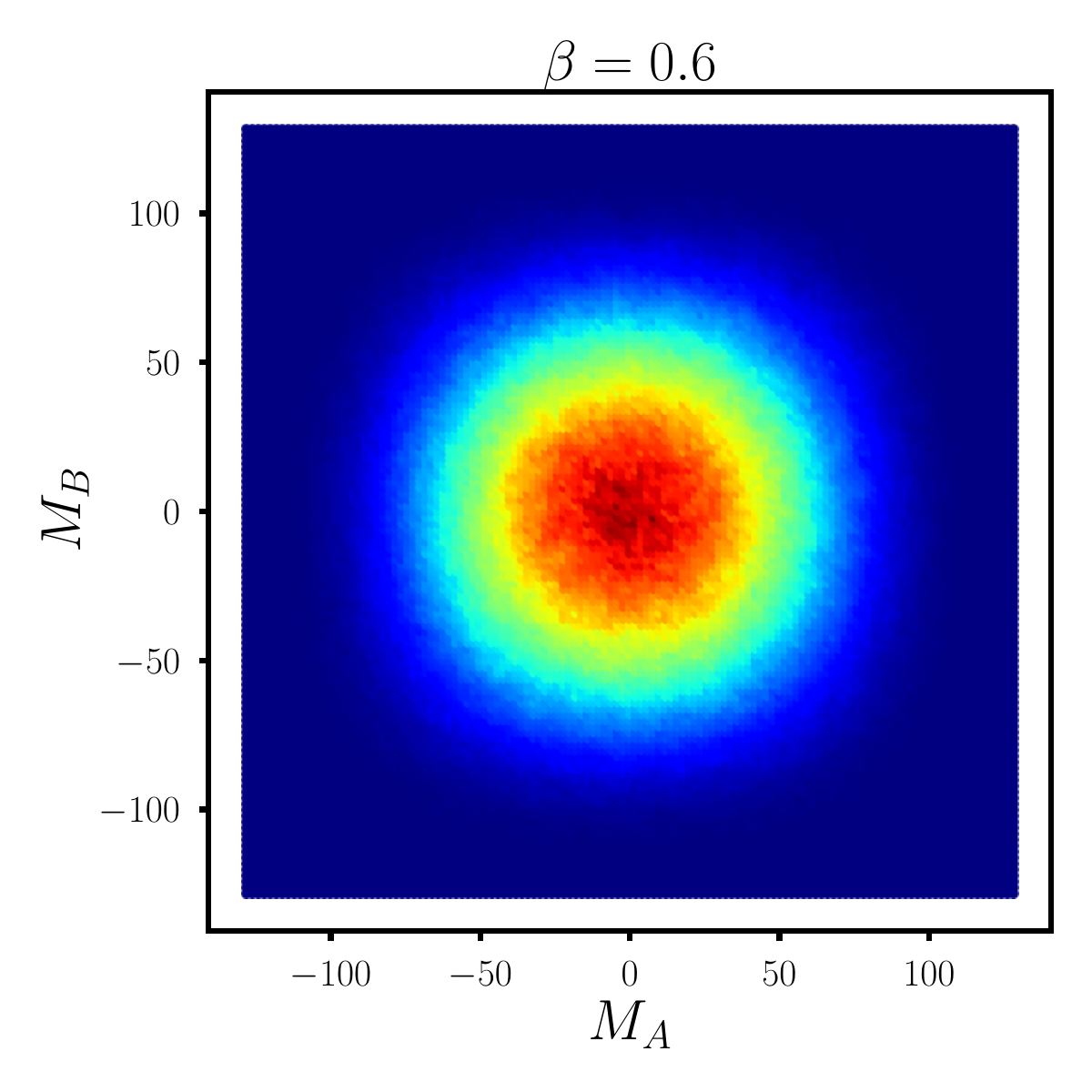}
    \caption{Distribution of ($M_A,M_B$) for $U(1)$ QLM with $\pm2$ charges at 
    $\lambda=-1$ and different $\beta$ ($L_x=L_y=16, L_t=24$) [no symmetrization is applied].}
    \label{fig:pmambZ2lam-1}
 \end{figure}

\begin{figure}
    \centering
    \includegraphics[scale=0.2]{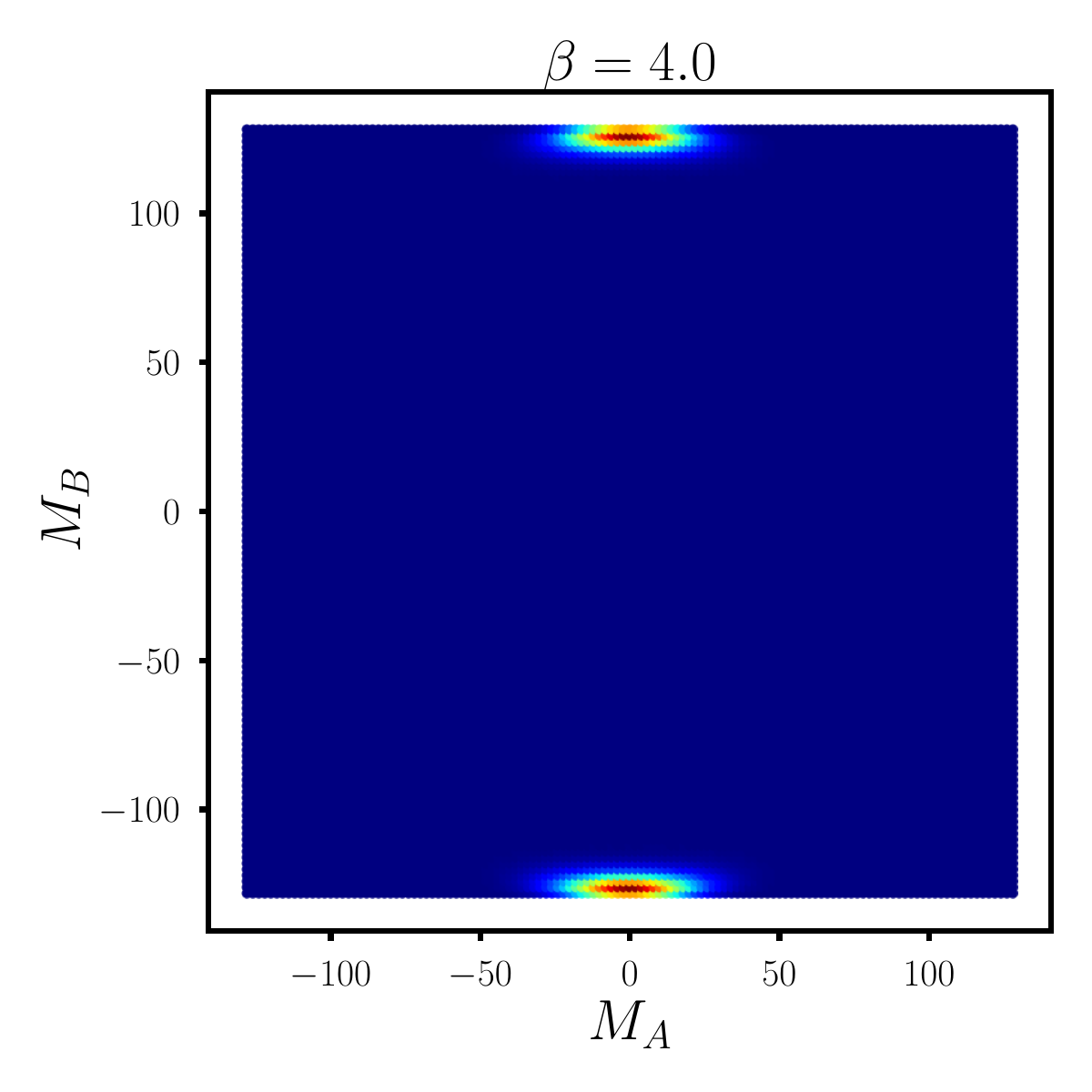}
    \includegraphics[scale=0.2]{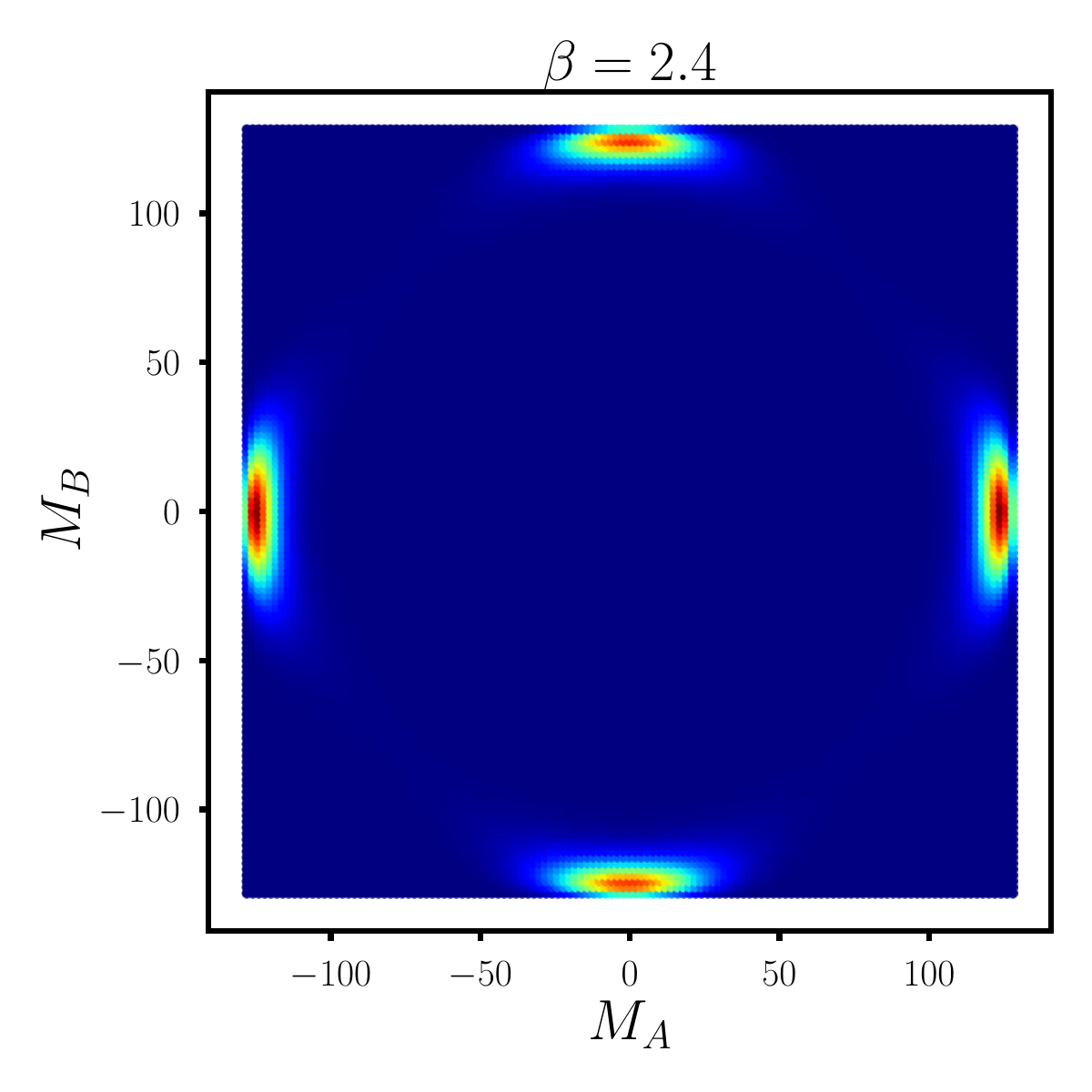}
    \includegraphics[scale=0.2]{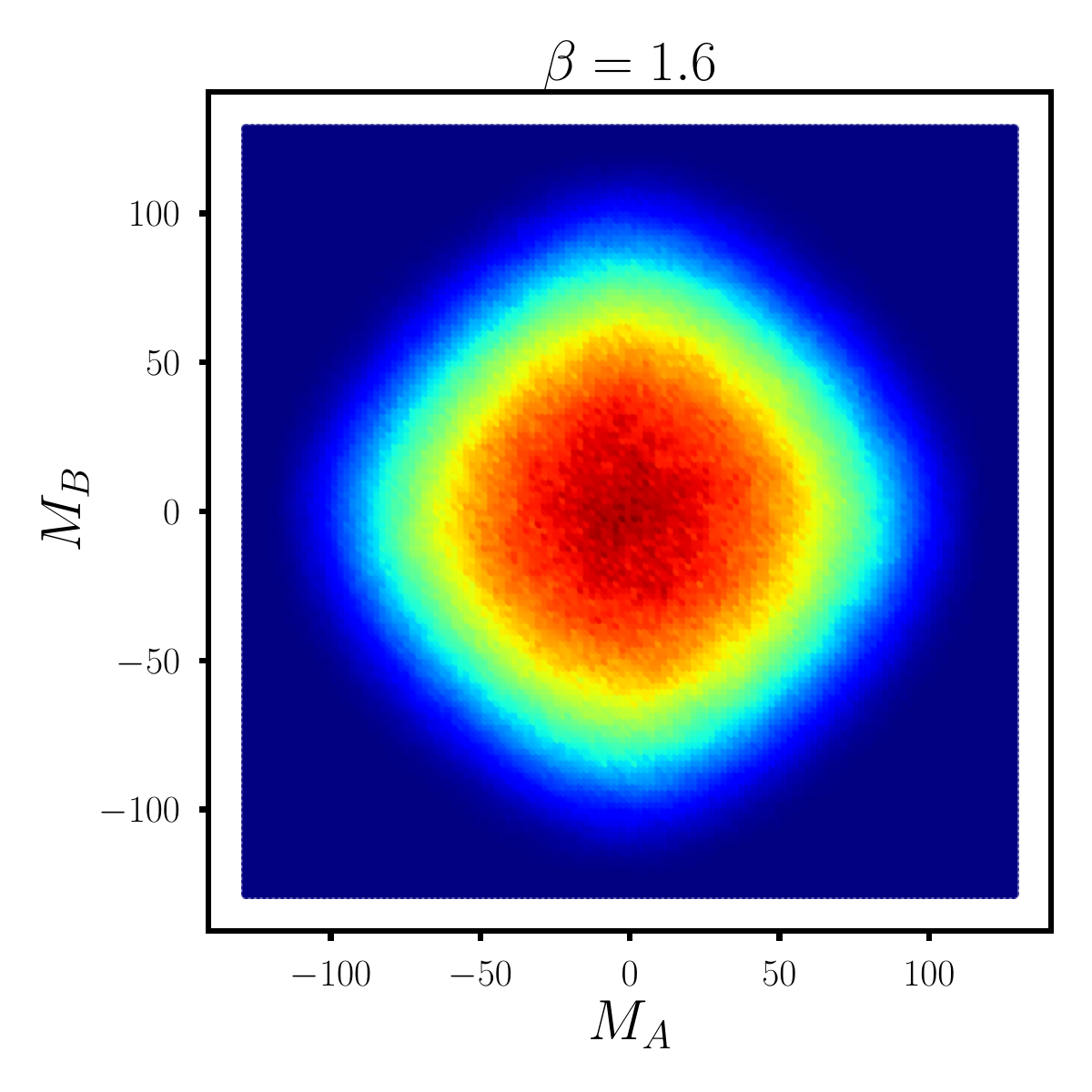}
    \includegraphics[scale=0.2]{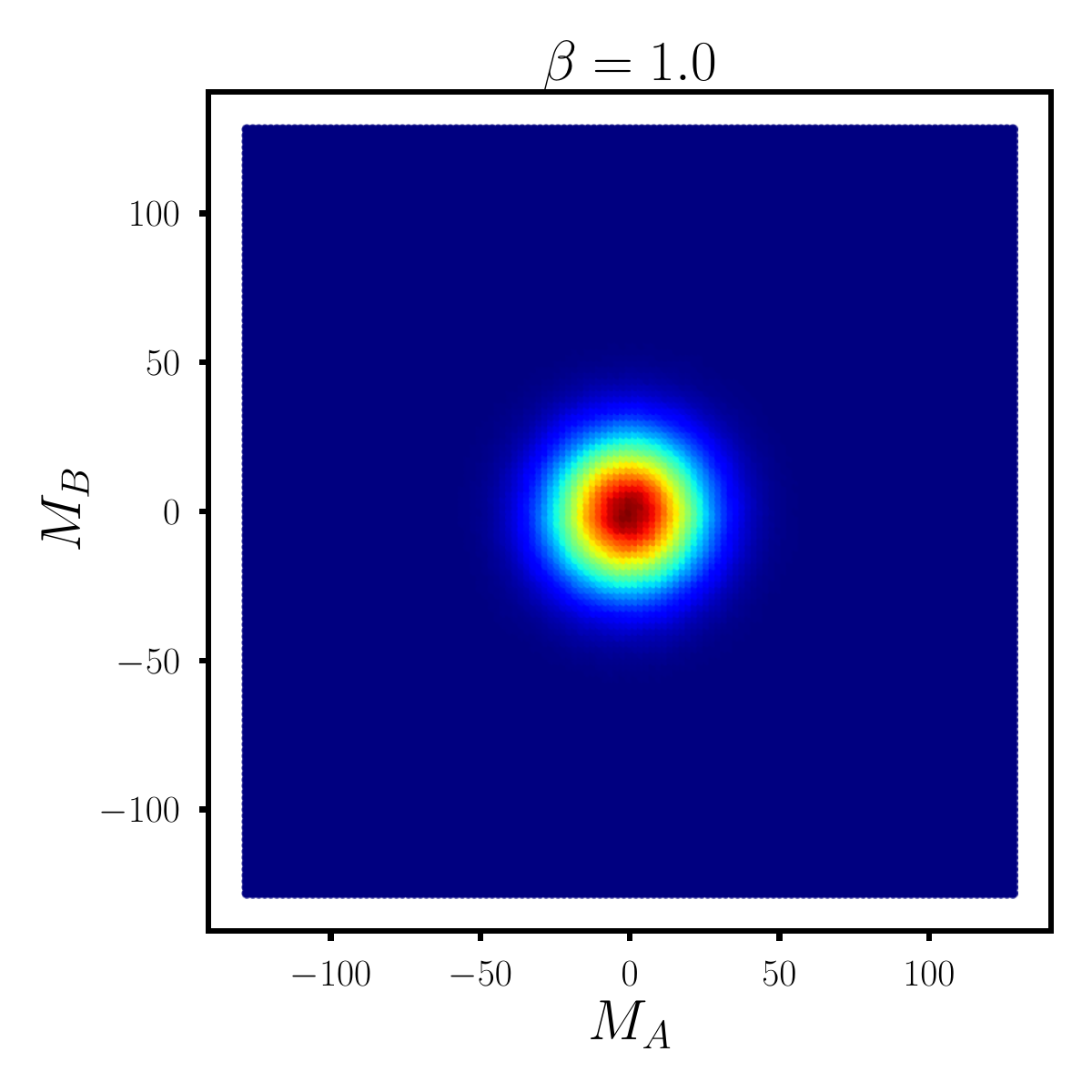}
    \caption{Distribution of ($M_A,M_B$) for $U(1)$ QLM with $\pm2$ charges at $\lambda=0$ 
    and different $\beta$ ($L_x=L_y=16, L_t=24$) [no symmetrization is applied].}
    \label{fig:pmambZ2lam0}
 \end{figure}

 The role played by the annealed charges becomes clear once we consider the 
 probability distribution $p(M_A,M_B)$ once again in the presence of the charges. 
 Figs.~\ref{fig:pmambZ2lam-1} and \ref{fig:pmambZ2lam0} show the effect of 
 starting at $T=0$ in the two different phases on either side of $\lambda_c$ and
 then raising the temperature. As can be expected, at low temperatures, the
 annealed charges do not play a role, since they are not created, and so 
 the ground state structure is unchanged. Although we do not explicitly show
 the data, we have checked that $\lambda_c$ is unchanged in the presence and
 the absence of the annealed charges. As the temperature is raised, annealed
 charges occur more and more frequently in the ensemble, and this causes the
 flippabilities to go down, while at the same time the broken symmetries get
 restored. This is evident from the figures because as the charges begin to 
 appear at the vertices, the flippabilities of all the plaquettes touching the 
 vertex get destroyed and one approaches $|\braket{M_A}| \sim |\braket{M_B}| \sim 0$, 
 i.e., the probability distribution becomes shrinks but remains roughly circular. 

  The remaining interesting \emph{unexplored} question is the shape of the probability
  distribution for the pure $U(1)$ LGT at sufficiently high temperatures. At 
  $\beta=0.8,~\lambda=-1$ and at $\beta=0.6,~\lambda=0$ (the lower right panels in
  Figs.~\ref{fig:pmambU1lam-1},~\ref{fig:pmambU1lam0}), the system is in the deconfined
  phase, and yet not only the radii of the distributions, but also the deformations are 
  different; indicating a clear influence of $\lambda$. Naively this is not to be expected, 
  since the symmetry breaking at zero temperatures are expected to be restored at such high 
  temperatures. In contrast, the corresponding plots for the $U(1)$ theory with annealed
  charges (Figs.~\ref{fig:pmambZ2lam-1},~\ref{fig:pmambZ2lam0}) show a perfect circle, with
  different radii, but no angular deformation. We leave this question to a future study. 

\section{EFT for the theory with and without the $Q = \pm 2$ defects}\label{sec:phTr}

  The above description in terms of the order parameters immediately motivates a description
 of the phase transition in terms of Landau-Ginzburg free energy analysis. In fact, this was
 performed in Ref.~\cite{Banerjee2013}, by defining a two-component order parameter as,
 \begin{equation}
    {\bf e} (x) = \bigl(\cos\varphi(x),\sin\varphi(x)\bigr), \qquad {\bf e} \sim - {\bf e}.
 \end{equation}
 Consequently, $\varphi \sim \varphi+\pi$, and the appropriate target space is $RP(1)$. This 
 ensures that only of the gauge equivalent copies are physical. Close to the zero-temperature 
 transition, the Euclidean effective action takes the form
 \begin{equation} \label{eq:zeroT-action}
    S[\varphi] = \int \dd^3x\, \left[ \frac{\rho}{2}(\partial_\mu\varphi)^2
        +\delta\cos^2(2\varphi) + \varepsilon\cos^4(2\varphi) \right],
 \end{equation}
 where we have ignored a renormalization constant between the space and time (i.e., the
 speed of light in this theory). Depending on the exact values of $\delta$ and $\epsilon$,
 this action can describe the phase transition across $\lambda_c$, and predicts a first
 order phase transition (unless there is a fine-tuning). In practice, the quantum Monte
 Carlo simulations \cite{Banerjee2013} report a weak first-order phase transition. 
 
 A useful aspect of this formulation is that electric charges appear as topological defects 
 of $\varphi$.  Following Eq.~(3) of Ref.~\cite{Banerjee2013}, the charge contained in a 
 spatial region $\Omega$ is
 \begin{equation}
     Q_\Omega  = \frac{1}{\pi}\oint_{\partial\Omega} \dd\ell_i\,\epsilon_{ij}\partial_j\varphi
    = \frac{\Delta\varphi}{\pi},
 \label{eq:charge-vorticity}
 \end{equation}
 where $\ell_i$ is the surface enclosing the area $\Omega$. A charge-$1$ defect therefore 
 corresponds to $\Delta\varphi=\pi$, an elementary vortex of the $RP(1)$ field, while a 
 charge-$2$ defect corresponds to, $\Delta\varphi=2\pi$, the double vortex of the $RP(1)$ field. 

  At a finite-temperature transition, the long-distance properties are sensitive to the two spatial
 dimensions. It is useful to introduce a compact field $\vartheta$, dual to $\varphi$, normalized 
 such that a charge-$Q$ defect is created by
 \begin{equation}
    {\cal V}_Q(x)=\exp~\bigl(i Q\vartheta(x)\bigr).
    \label{eq:defect-operator}
 \end{equation}
 Equivalently, $\exp(i\vartheta)$ has the quantum numbers of the fundamental Polyakov loop.  
 Under the global $U(1)$ centre transformation,
 \begin{equation}
    \vartheta \longrightarrow \vartheta+\alpha, \qquad {\cal V}_Q \longrightarrow e^{i Q\alpha} {\cal V}_Q .
    \label{eq:center-shift}
 \end{equation}
 In the pure gauge theory, charged sectors are absent from the thermal trace and the continuous 
 shift symmetry of $\vartheta$ is exact.  The finite-temperature transition is then governed by 
 the $U(1)$-symmetric BKT fixed point, in agreement with the Svetitsky-Yaffe argument and the 
 numerical results of Ref.~\cite{SauSenBanerjee2023}.

 When annealed $Q=\pm2$ sectors are included, a dilute gas of these defects contributes
 \begin{equation}
    y_2\left( {\cal V}_2+{\cal V}_{-2}\right) = 2 y_2 \cos(2\vartheta) \nonumber
 \end{equation}
 to the thermal ensemble.  Thus the leading charge-induced perturbation is
 \begin{equation}
    \delta S_{Q=2} = - 2 y_2 \int \dd^2x\,\cos(2\vartheta).
    \label{eq:charge-two-perturbation}
 \end{equation}
 At low charge density one expects schematically $y_2 \sim \exp(-m_2/T)$, where $m_2$ is the 
 free-energy cost of introducing a charge-$2$ defect.  The precise relation between $y_2$, the 
 microscopic charge density and the bare coupling $\lambda$ is non-universal.

 The perturbation in Eq.~\eqref{eq:charge-two-perturbation} is invariant only under 
 $\vartheta \rightarrow \vartheta+\pi$, and thus breaks the $U(1)_{\rm centre}$ to 
 $\mathbb{Z}_2{}_{\rm centre}$. In this language, the reduction of the winding symmetry 
 and the appearance of the charge-two operator are two descriptions of the same infrared perturbation.

 A schematic compact-boson description containing both the symmetry breaking perturbations and 
 the charge defects is
 \begin{align}
    S_{\rm eff} &= S_{\rm G}[\varphi] - u_4\int\dd^2x\,\cos(4\varphi) 
     - u_8\int\dd^2x\,\cos(8\varphi) \nonumber \\
     & - 2y_2\int\dd^2x\,\cos(2\vartheta) +\cdots .
    \label{eq:double-sine-gordon}
 \end{align}
 Here $S_{\rm G}$ denotes the Gaussian compact-boson action for $\varphi$.  
 Eq~\eqref{eq:double-sine-gordon} should be considered schematic: $\varphi$ and $\vartheta$ 
 are dual fields, and the two families of vertex operators are mutually non-local in a 
 formulation using only one elementary scalar field.  

 At the pure $U(1)$ BKT transition, the fundamental Polyakov-loop correlator behaves as
 \begin{equation}
    \braket{ e^{i\vartheta(x)} e^{-i\vartheta(0)} } \sim |x|^{-\eta}, \qquad \eta=\frac14.
    \label{eq:bkt-correlation}
 \end{equation}
 The scaling dimension of the fundamental vertex operator is therefore $x_1=\frac{\eta}{2}=\frac18$.
 At the Gaussian fixed point, the charge-$Q$ harmonic has $x_Q =Q^2 x_1$. Consequently, for the
 charge-$2$ operator we have,
 \begin{equation}
    x_2  = 4 x_1 = \frac12 .
    \label{eq:dimension-charge-two}
 \end{equation}
 The leading RG equation for its fugacity is therefore
 \begin{equation}
    \frac{\dd y_2}{\dd\ell} = \left(2-x_2\right)y_2+\cdots = \frac32\,y_2+\cdots .
    \label{eq:rg-y2}
 \end{equation}
 Thus, the charge-two operator is strongly relevant at the pure BKT fixed point.  Any non-zero 
 $y_2$ ultimately drives the system away from the $U(1)$-symmetric BKT theory.  Ignoring logarithmic 
 BKT corrections, the associated crossover scale behaves parametrically as
 \begin{equation}
    \xi_2 \sim a\,y_2^{-1/(2-x_2)} = a\,y_2^{-2/3}.
    \label{eq:crossover-length}
 \end{equation}
 This relevance argument explains why a small but non-zero critical density of annealed $Q=\pm2$ 
 charges can qualitatively change the transition. In \cite{SauSenBanerjee2023}, we have shown
 that $\braket{Q^2} \sim 0.05$ at the phase transition, showing that only a few percent is sufficient
 to trigger the instablility of the BKT fixed point. The perturbation changes the infrared theory 
 because it is relevant (unrelated to its microscopic coefficient).

\begin{figure}
    \centering
    \includegraphics[width=0.95\linewidth]{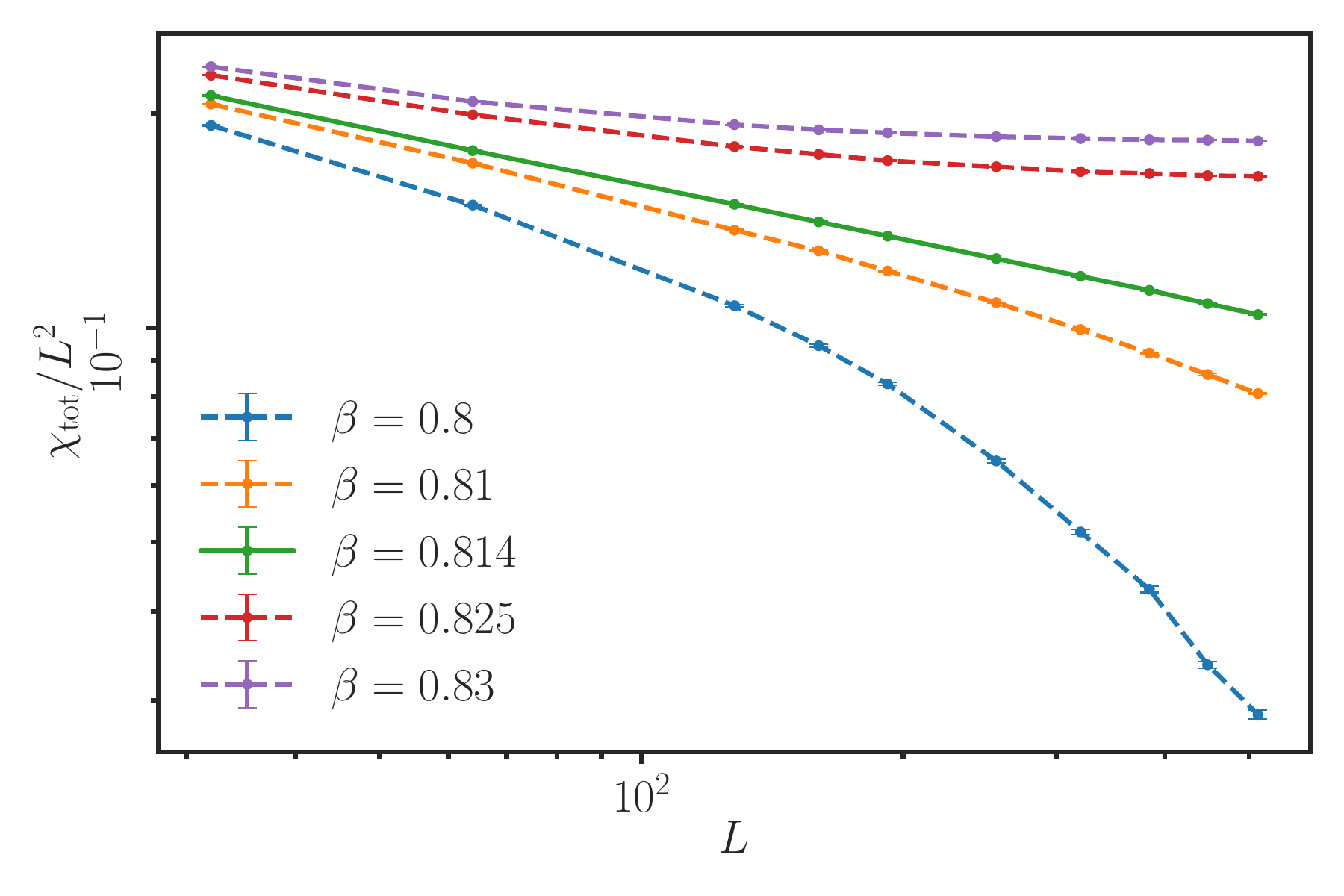}
    \includegraphics[width=0.95\linewidth]{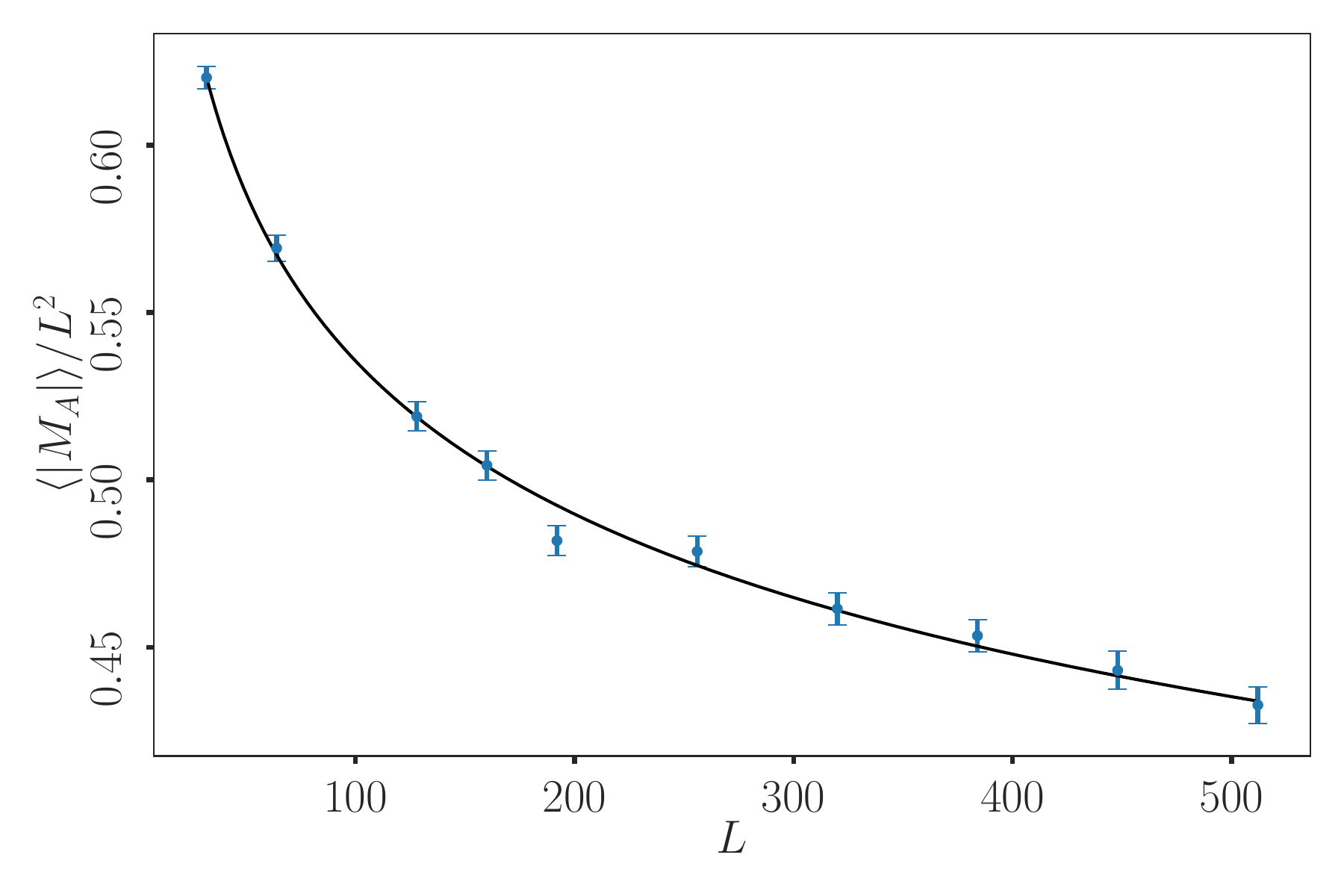}
    \caption{(Top) $\chi_{\rm{tot}}/L^2$ at $\beta_c\approx0.814$, below $\beta_c$ and above $\beta_c$. (Bottom) $\langle|M_A|\rangle/L^2$ as a function of system size $L$. The data is fitted according to Eq.~\ref{eq:mag_critical} (the solid line).}
    \label{fig:msqrL}
\end{figure}

  \begin{figure*}[t]
    \centering
    \includegraphics[width=0.32\linewidth]{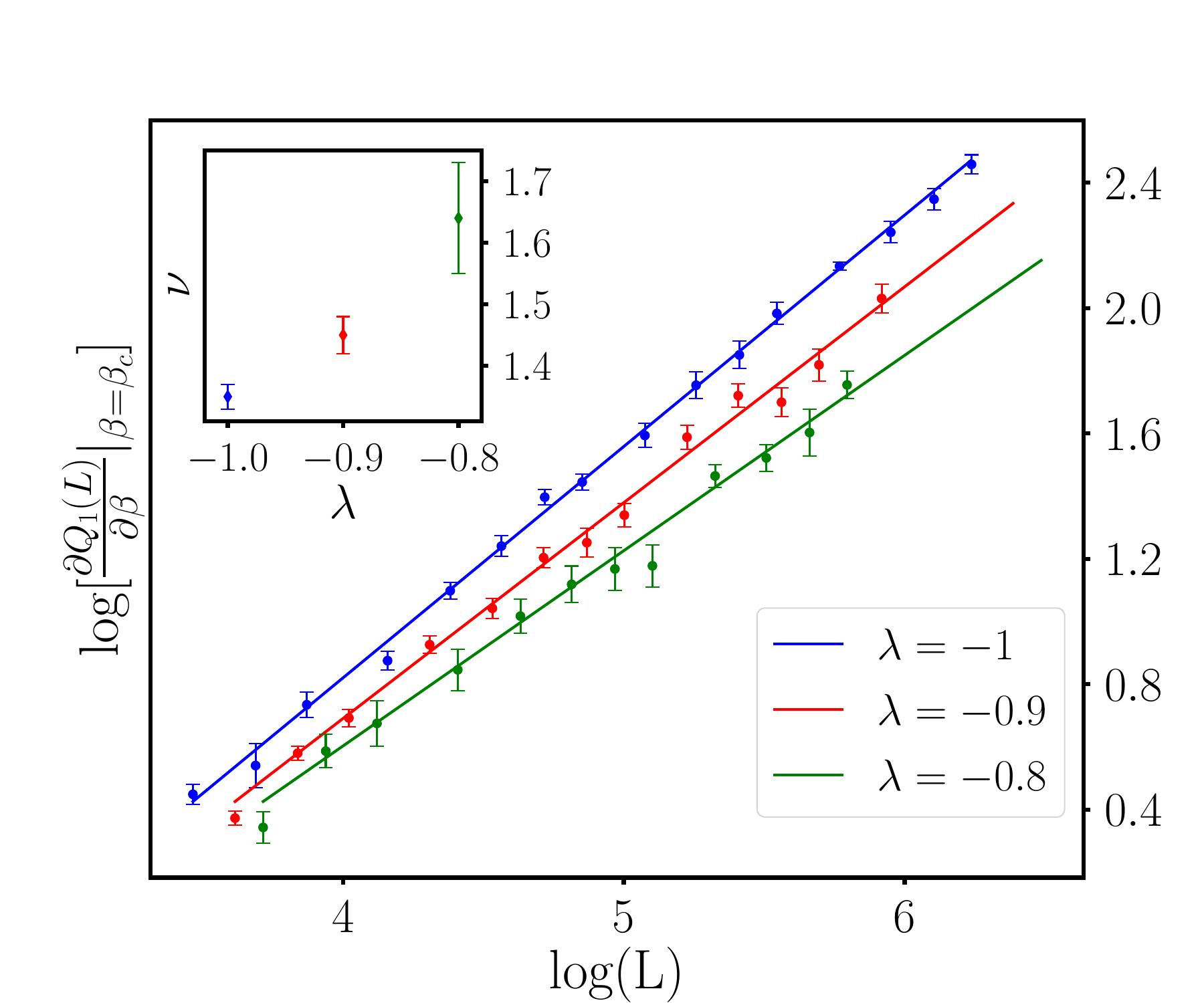}\hfill
    \includegraphics[width=0.32\linewidth]{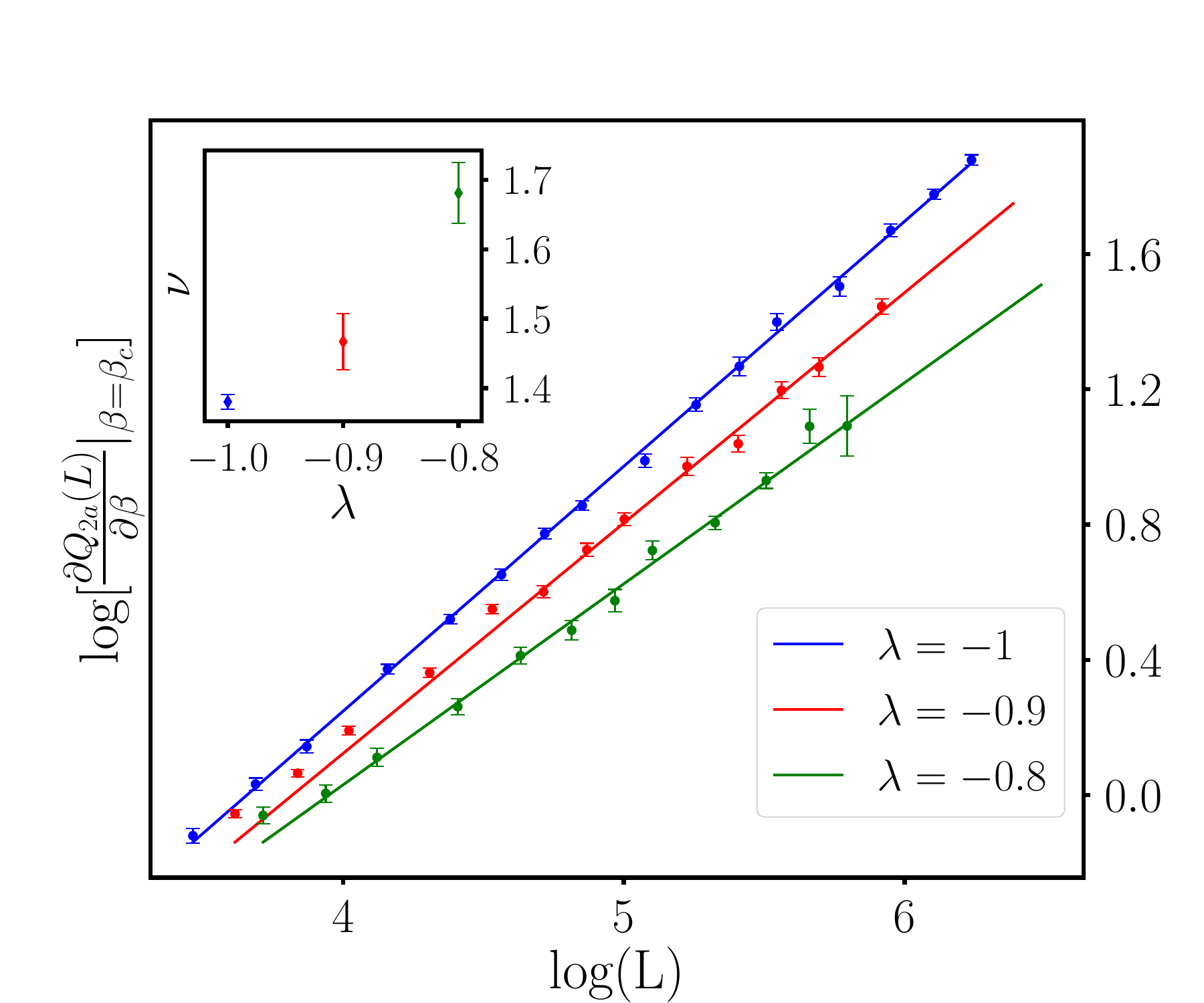}\hfill
    \includegraphics[width=0.32\linewidth]{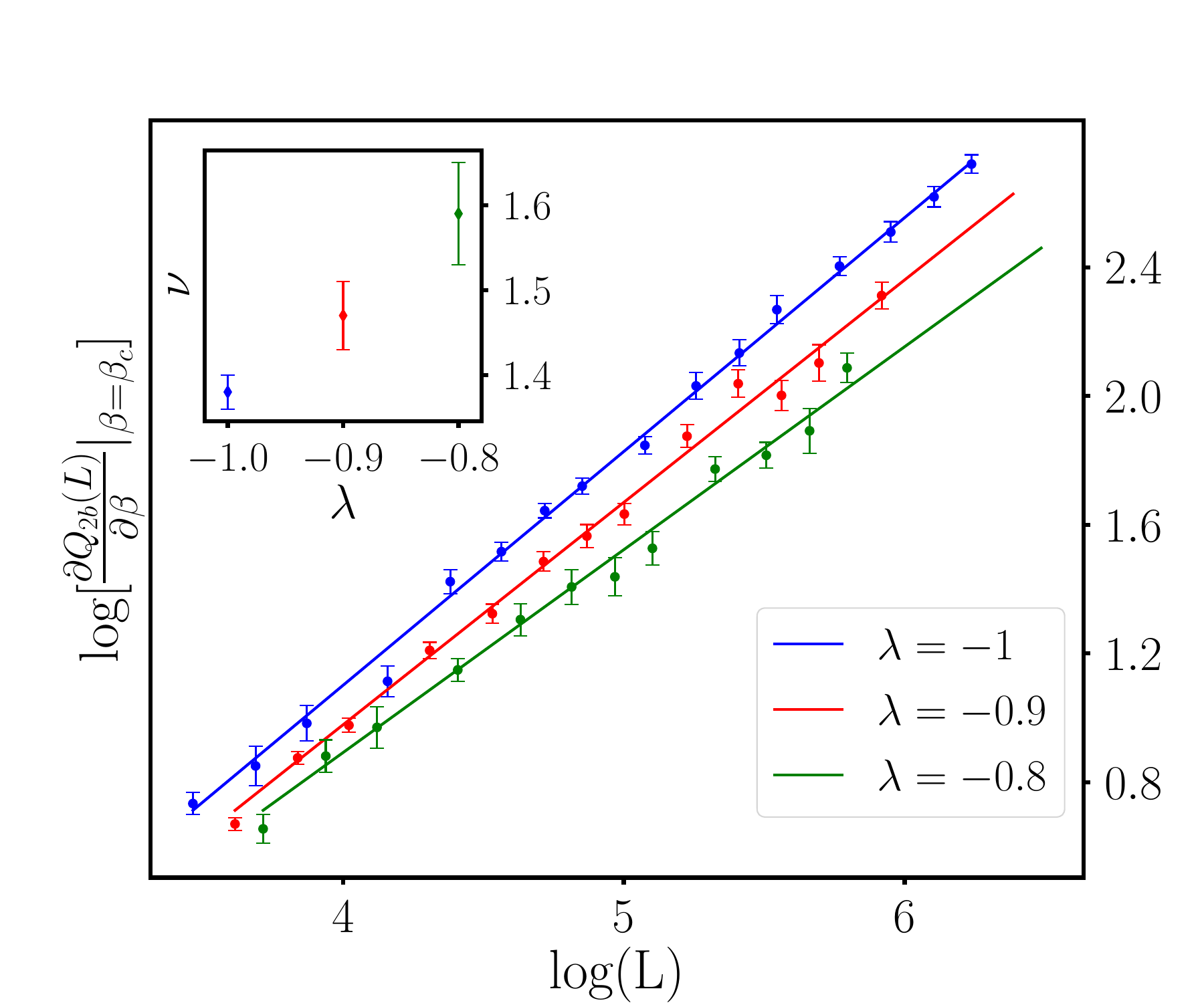}
    \caption{Using the slope of the derivative of the Binder cumulant in order to extract the 
    critical exponent $\nu$. Each plot shows the results for three different $\lambda = -1, -0.9, -0.8$.
    The three plots show the results for using the three different Binder ratios $Q_1,~Q_{2a},~Q_{2b}$.}
    \label{fig:hasenbusch}
\end{figure*}
\section{Phase transition with annealed charges}\label{sec:phZ2}
 In this section, we proceed to study the nature of the phase transition in the presence of
 the annealed $Q=\pm2$ charges, which was originally reported in \cite{SauSenBanerjee2023}.
 In the previous paper we identified the phenomenon of \emph{weak-universality}, once the 
 annealed charges starts to make their presence felt. This implies that certain critical
 exponents characterized like $\gamma, \nu, \beta$, change as a function of bare coupling,
 while others like $\eta, \delta$ defined directly at the critical point remain fixed at
 the values for the 2-d $Z_2$ Ising model. Further, the ratios of the critical exponents are 
 also fixed to that of the 2-d $Z_2$ Ising model. This has previously been observed in a 
 variety of systems \cite{Suzuki1974,Baxter1971,Ashkin1943,Pearce1987,Alet2005,
 Malakis2009,Queiroz2011,Jin2012,Jin2013,Suzuki2015}, but hardly yet for gauge theories. 
 In the previous work \cite{SauSenBanerjee2023}, we used the scaling property of the peak of 
 connected susceptibility, $\chi_{\rm{conn}}$, to show that the ratio $\frac{\gamma}{\nu}$ and 
 the critical exponent $\eta$ are same as that of the 2-d $Z_2$ Ising model. In this paper,
 using the scaling relation 
 \begin{equation}
     \frac{\braket{|M_A|}}{L^2}\sim L^{-\beta_{\rm{mag}}},
     \label{eq:mag_critical}
 \end{equation}
 at the critical temperature $\beta_c(\approx0.8143$ \cite{SauSenBanerjee2023}), we find
 the value of magnetic critical exponent, 
 $\beta_{\rm{mag}}=0.1289(29)$, which is same as the 2-d $Z_2$ Ising value of $\frac{1}{8}$ up to
 $2\sigma$ error bar (Fig.~\ref{fig:msqrL}). In Fig.~\ref{fig:msqrL}, we also show the behaviour of
 $\chi_{\rm{tot}}/L^2$ in the confined phase (above $\beta_c$), deconfined phase (below $\beta_c$) as well as at the critical temperature. While $\chi_{\rm{tot}}/L^2$ approaches a constant for $\beta > \beta_c$, it decreases as $\exp(-L)$ for $\beta < \beta_c$ and as $L^{-a}$ at $\beta\approx \beta_c$.
 We show detailed plots of the susceptibility and the different Binder ratios and the extraction
 of the critical exponents at different values of the bare coupling $\lambda$, which evidence
 this phenomenology.

 \subsection{Weak-universality from Binder ratios} \label{subsec:deriv}
 In the previous work \cite{SauSenBanerjee2023}, we demonstrated the emergence of weak-universality
 by the use of phenomenological couplings obtained using the Binder ratios. In particular, let 
 $R(\beta,L)$ be the dimensionless coupling, then the slope of $R$ at the critical point $\beta_c$
 directly yields the exponent $\nu$, 
 $\left. \frac{\partial R (L)}{\partial \beta} \right|_{\beta_c} = a L^{1/\nu} (1 + bL^{-\omega})$.
 For large lattices (or for large $\omega$), plotting the derivative (at $\beta_c$) vs $L$ in a 
 log-log scale enables us to compute $1/\nu$ from the slope. The requirement for the use of this 
 method is an accurate measurement of $\beta_c$, calculated through an extrapolation procedure of 
 different estimators. In \cite{SauSenBanerjee2023}, $\beta_c$ obtained from crossing-point analysis
 of $\chi_{\textrm{tot}}L^{-\gamma/\nu}$ was used for the above analysis. In this article, 
 we show the result of applying this procedure to extract $\nu$ in Fig.~\ref{fig:hasenbusch} and 
 mention their values obtained using different Binder ratios in Table~\ref{tab:nu_prl}.
 As evident from the plot and the table, there is a clear increase in $\nu$ as $\lambda$ increases. Moreover,
 three different Binder ratios provide a consistent value of $\nu$, indicating a robust extraction.

 \begin{table}
    \centering
    \begin{tabular}{|c|c|c|c|c|}
        \hline
        $\lambda$ & $\beta_c$ & $\nu(Q_{2a})$ & $\nu(Q_{2b})$ & $\nu(Q_1)$\\
        \hline
        $-1.0$ & $0.814279(14)$ & $1.38(1)$ & $1.38(2)$ & $1.35(2)$ \\
        $-0.9$ & $0.885292(17)$ & $1.47(4)$ & $1.45(3)$ & $1.45(3)$ \\
        $-0.8$ & $0.968196(26)$ & $1.68(4)$ & $1.64(8)$ & $1.64(9)$\\
        \hline
    \end{tabular}
    \caption{The values of $\beta_c$ obtained using crossing-point analysis of $\chi_{\rm{tot}}L^{-\gamma/\nu}$ and the values of $\nu$ for different Binder ratios at different $\lambda$.}
    \label{tab:nu_prl}
\end{table}

\begin{figure}
\center
\includegraphics[width=0.49\linewidth]{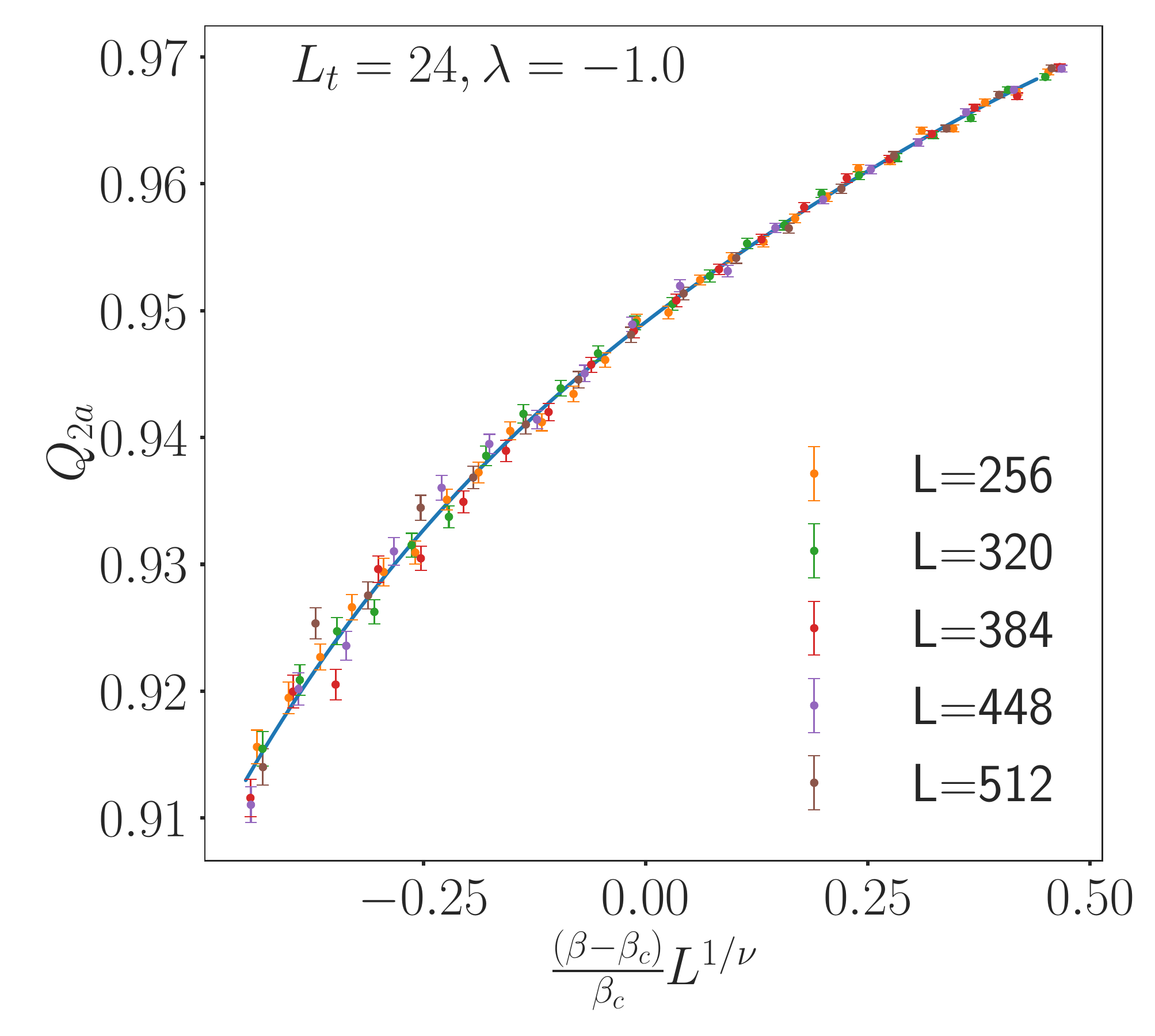}
\includegraphics[width=0.49\linewidth]{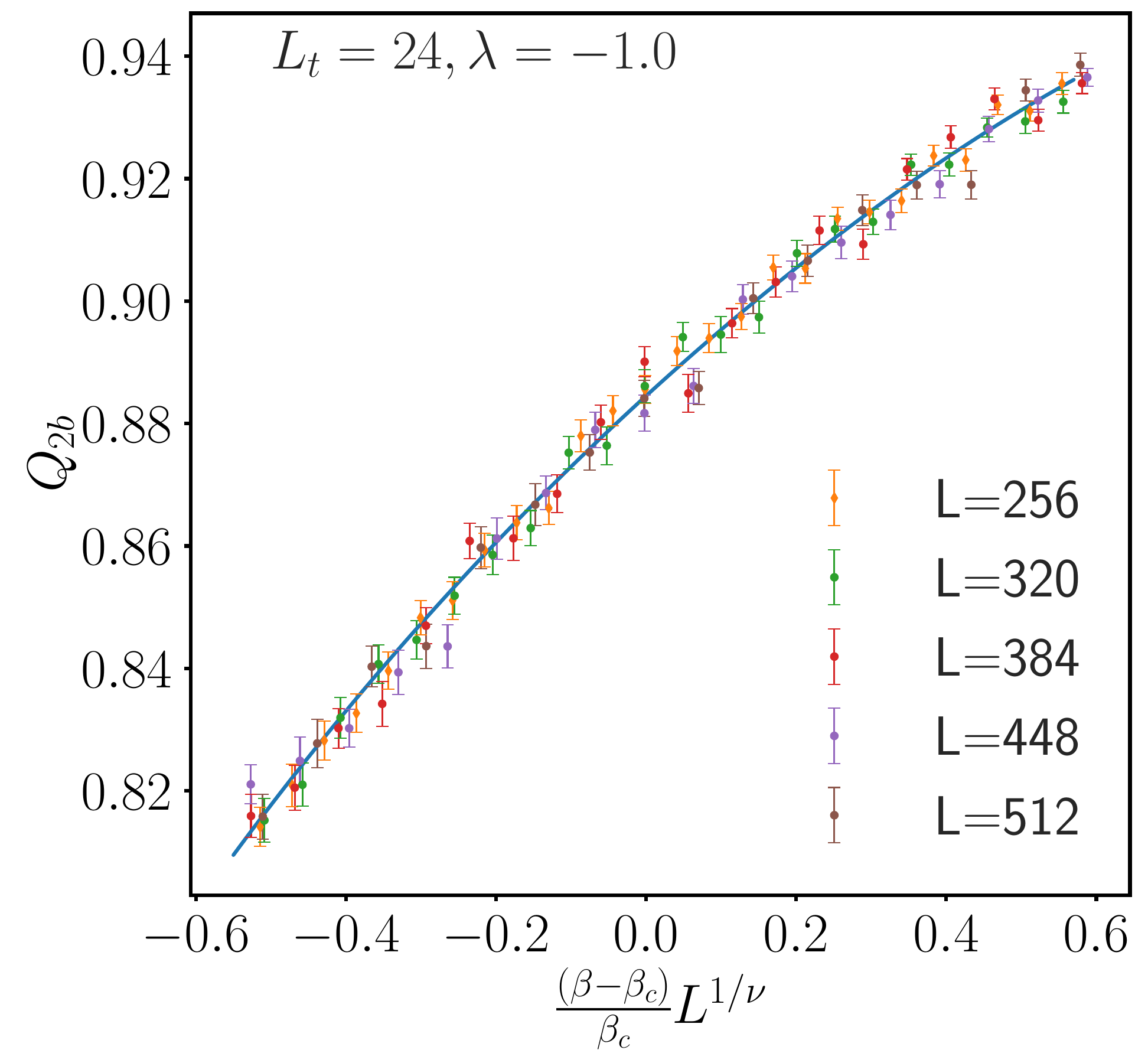}
\includegraphics[width=0.49\linewidth]{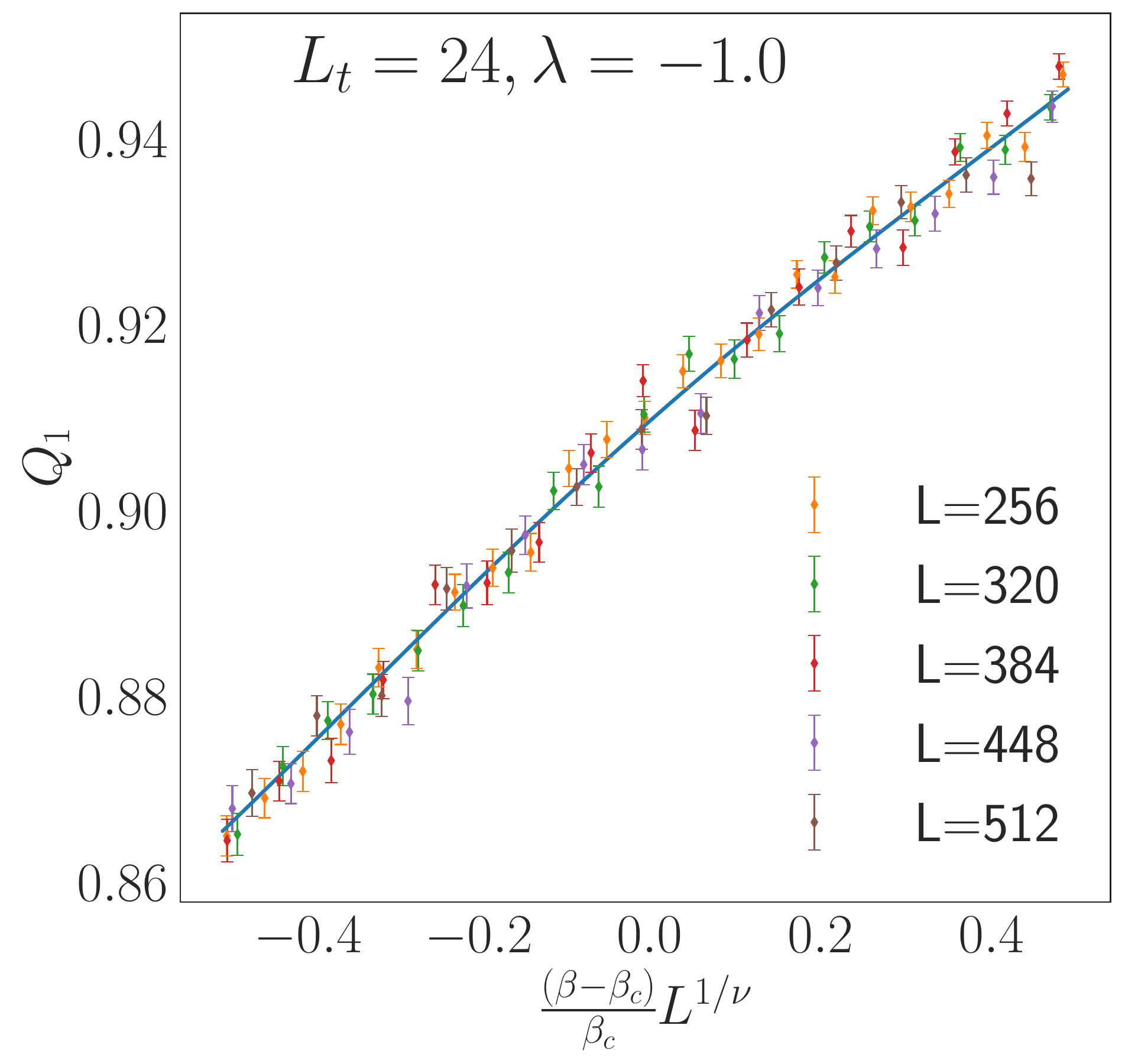}
\includegraphics[width=0.49\linewidth]{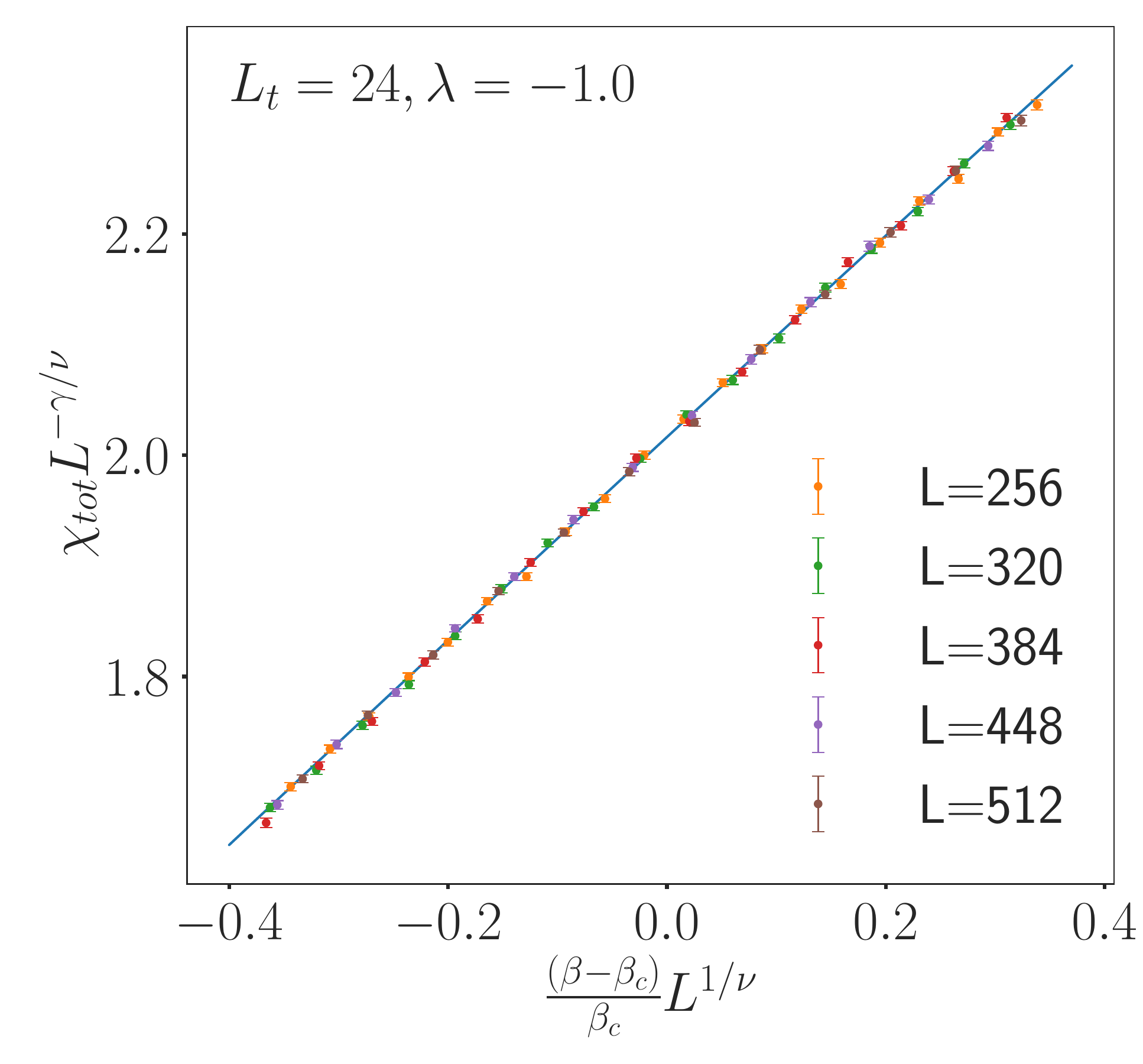}\\
\caption{Finite-size-scaling collapse of $Q_{2a}$, $Q_{2b}$, $Q_1$, and 
 $\chi_{\mathrm{tot}}L^{-7/4}$ for $L_t=24$, $\lambda=-1$, and $L=256,320,384,448,512$. 
 The horizontal axis is the scaling variable $(\beta-\beta_c)L^{1/\nu}/\beta_c$. For each 
 observable, $\beta_c$ and $\nu$ are determined from the corresponding collapse; their fitted 
 values are reported in Table~\ref{tab:collapseLts}.}
\label{fig:scaling-collapse-lamm1}
\end{figure}

\subsection{Scaling collapse analysis} \label{subsec:SCA}
 As a next step, we verify the extracted values of the critical coupling, $\beta_c$, and the exponent
 $\nu$, at different $\lambda$ through the scaling collapse. According to the theory of scaling collapse, 
 any physical observable, $A(\beta,L)$, near the critical point satisfies the scaling relation
 \begin{equation}
  A(\beta,L)=L^{\kappa/\nu}f_A(\hat{\beta}L^{1/\nu})
  \label{Eq:scaling}
 \end{equation} 
where $\hat{\beta}=\frac{\beta-\beta_c}{\beta_c}$ is the reduced inverse temperature and $\kappa$ 
is a critical exponent related to $A$, that is, $A(\hat{\beta},L\to\infty)\sim \hat{\beta}^{-\kappa}$. 
Thus, if $A$ is a susceptibility, then $\kappa = \gamma$, but $\kappa$ is zero if $A$ is Binder 
cumulant. $f_A$ is the corresponding scaling function for the observable $A$. If 
$A(\beta,L)L^{-\kappa/\nu}$ is plotted as a function of $x=\hat{\beta}L^{1/\nu}$, then the data for 
different system sizes collapse onto each other for the correct values of $\beta_c,\, \kappa$, and 
$\nu$. 

We have carried out the analysis for the total susceptibility, $\chi_{\rm{tot}}$, and the
different Binder cumulants, $Q_1,~Q_{2a}$, and $Q_{2b}$, for the different $\lambda$ values
at $L_T=24$. The analysis is carried out by noting that the scaling function $f_A$ can be 
expanded in Taylor series near critical point (very small $\hat{\beta}$)
\begin{equation}
f_A = A(\beta,L)L^{-\kappa/\nu} = q_0 + q_1 x + q_2 x^2 + q_3 x^3 + q_4 x^4 + ...
\label{Eq:fAtaylor}
\end{equation}
In the analysis, we fix $\kappa/\nu$ (i.e., $0$ for the Binder ratios and $\frac{7}{4}$ for 
$\chi_{\rm{tot}}$), and perform a fit of $\beta_c$, $\nu$, along with the parameters, $q_0$, 
$q_1$, $q_2$, $q_3$, $q_4$, of the polynomial in Eq.\,\ref{Eq:fAtaylor}. We have cross-checked
the robustness of the critical couplings and exponents by changing the order of the polynomial,
and verified that our quoted estimates are consistent with the systematics. 

 \Cref{tab:collapseLts} shows the estimates of $\beta_c$ and $\nu$ obtained 
using scaling collapse of $Q_{2a}, Q_{2b}$, $Q_1$, and $\chi_{\rm{tot}}$ for $\lambda=-1,-0.9$ 
and $-0.8$ at $L_T=24$.  For $\lambda=-1$, we use the system sizes $L=256,~320,~384,~448,~512$. 
The resulting collapses of $Q_{2a}$, $Q_{2b}$, $Q_1$, and $\chi_{\mathrm{tot}}$ are shown in 
Fig.~\ref{fig:scaling-collapse-lamm1}. Since the Binder ratios are dimensionless, they are plotted 
without an additional power of $L$, whereas the susceptibility is rescaled by 
$L^{-\gamma/\nu}=L^{-7/4}$. All four observables exhibit a common collapse within the critical 
scaling window within $2\sigma$, and with the estimates obtained from the slopes of the Binder 
ratios in Sec.~\ref{subsec:deriv}. This agreement is particularly significant because the Binder 
ratios and the susceptibility probe different moments of the order-parameter distribution. The 
scaling collapse for the two other $\lambda$ values, $\lambda = -0.9, -0.8$ are shown in 
Appendix~\ref{app:collapse}. 

  The scaling-collapse analysis therefore provides an independent confirmation of both the 
location of the transition and the anomalously large, coupling-dependent value of $\nu$. 
Together with the fixed ratios $\eta=1/4$ and $\gamma/\nu=7/4$, these results provide the 
characteristic signature of weak universality.

\begin{widetext}
    
\begin{table}
	\begin{tabular}{ | p{0.8cm} || l | l | l | l | l | l | l | l | l | l | }
		\hline
		\multirow{2}{*}{$\lambda$} &
		\multicolumn{2}{|c|}{Collapse of $Q_{2a}$} &
            \multicolumn{2}{|c|}{Collapse of $Q_{2b}$} &
            \multicolumn{2}{|c|}{Collapse of $Q_1$} &
		\multicolumn{2}{|c|}{Collapse of $\chi_{\rm{tot}}$} \\
		\cline{2-9}
		& $\beta_c$ & $\nu$ &
            $\beta_c$ & $\nu$ &
            $\beta_c$ & $\nu$ &
		  $\beta_c$ & $\nu$ \\
		\hline
		
		$-1.0$ & $0.814136(58)$ & $1.3673(244)$ & $0.814155(101)$ & $1.3513(493)$ & $0.814290(123)$ & $1.4003(727)$ & $0.814289(27)$ & $1.3634(176)$\\
		\hline
		
		$-0.9$ & $0.885207(44)$ & $1.4639(281)$ & $0.884846(76)$ &  $1.4492(490)$ & $0.884880(84)$ & $1.4557(542)$ & $0.885299(16)$ & $1.3832(100)$\\
		\hline
		
		$-0.8$ & $0.967738(100)$ & $1.6362(757)$ & $0.967918(191)$ & $1.6386(1499)$ & $0.967861(172)$ & $1.6187(1325)$ & $0.968159(26)$ & $1.4941(159)$\\
		\hline
	\end{tabular}
	\caption{$\beta_c$ and $\nu$ obtained using scaling collapse of $Q_{2a}$, $Q_{2b}$, $Q_1$ 
    and $\chi_{\rm{tot}}$ for different $\lambda$ at $L_t=24$. $\chi^2/\rm dof$ of scaling collapses 
    varies within $1.1$ to $1.6$.}
	\label{tab:collapseLts}
\end{table}
\end{widetext}

\section{Study of string tension} \label{sec:stringTen}

\subsection{Fractionalized flux tubes}
 A phenomenological signature of confinement is the formation of a flux tube joining 
a pair of static charges. At sufficiently large separation $r$, the energy stored in 
the tube grows linearly,
\begin{equation} \label{eq:pot}
  V_{Q\bar Q}(r) = V_0 + \sigma_Q r + \cdots,
\end{equation}
where $\sigma_Q$ is the string tension associated with a pair of external charges $(-Q,+Q)$. 
In the confined phase $\sigma_Q$ remains nonzero in the thermodynamic limit, whereas it 
vanishes upon crossing into the deconfined phase. Confining strings in the spin-$\frac{1}{2}$
U(1) QLM have a structure that is qualitatively different from that of conventional Wilsonian 
lattice gauge theories. As demonstrated in Refs.~\cite{Banerjee2013,Banerjee:2021tfo}, the 
electric flux carried by a string is fractionalized. A string connecting charges $(-1,+1)$ 
splits into two strands, each carrying one half-unit of electric flux, while a pair of charges 
$(-2,+2)$ is connected by four such strands.

 This fractionalization has a direct interpretation in terms of the broken symmetries of the 
two confined phases separated by the zero-temperature transition at $\lambda_c$. On one side 
of $\lambda_c$, there is a spontaneous breaking of $Z_2$ symmetry which breaks both translation 
and charge-conjugation symmetries, whereas on the other side breaks a different combination of
symmetries which in the gauge theory variables corresponds to a restoration of the charge 
conjugation, while translation remains broken. Thus, both sides show confinement by breaking
different combinations of the point group symmetries. An individual fractionalized strand is an 
interface between two distinct symmetry-breaking patterns. Consequently, the region within
the strands is locally in the confined phase that is stable on the opposite side of $\lambda_c$. 
For example, at $\lambda < \lambda_c$, the bulk is in the C1 phase, while the interior of the 
flux tube has the local ordering pattern of the C2 phase (see Fig.~\ref{fig:phDiagT}). The 
confining string should therefore be viewed as a collection of domain walls separating regions 
with different broken symmetries, rather than as a single featureless concentration of electric flux.

 The square lattice and the underlying symmetry-breaking patterns also make the strings anisotropic. 
Their transverse profiles retain a visible lattice-scale modulation, and the different strands 
need not have identical microscopic shapes. At low temperatures, the electric flux-$\frac{1}{2}$ is 
channeled through narrow tubes from the source to the sink. Thermal fluctuations broaden the 
interfaces and progressively smooth their microscopic structure.

\begin{figure}
    \centering
    \includegraphics[scale=0.14]{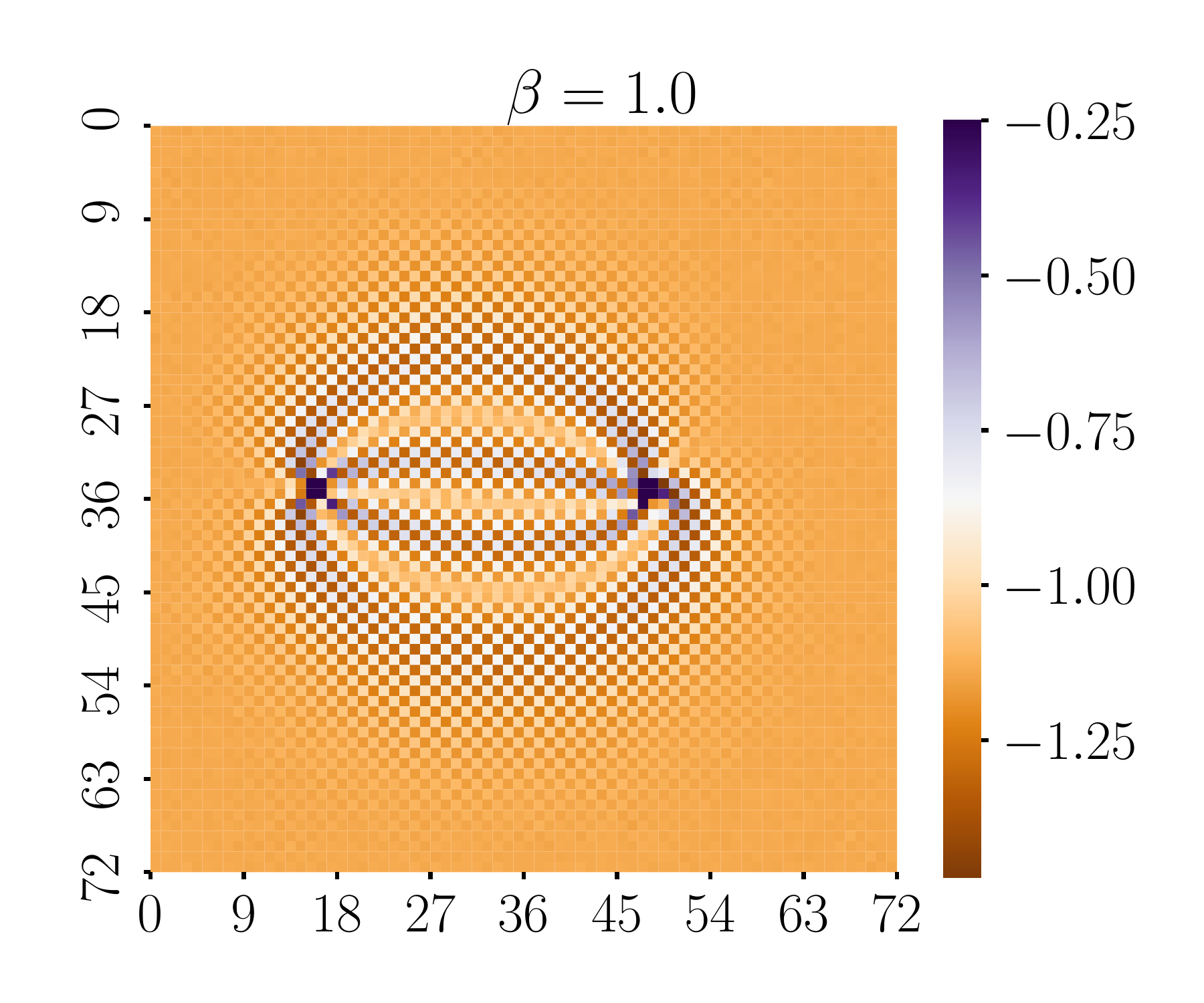}
    \includegraphics[scale=0.14]{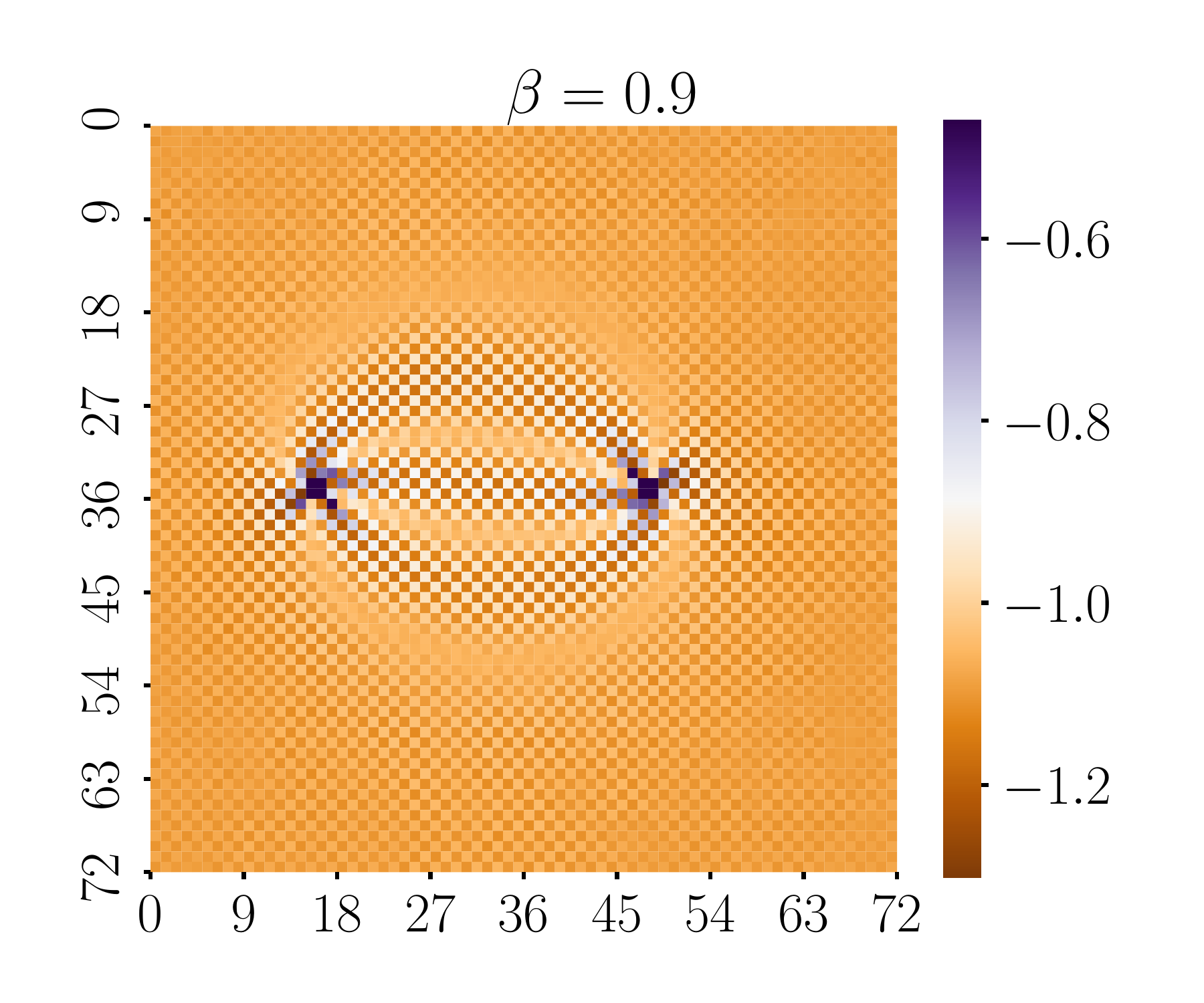}
    \includegraphics[scale=0.14]{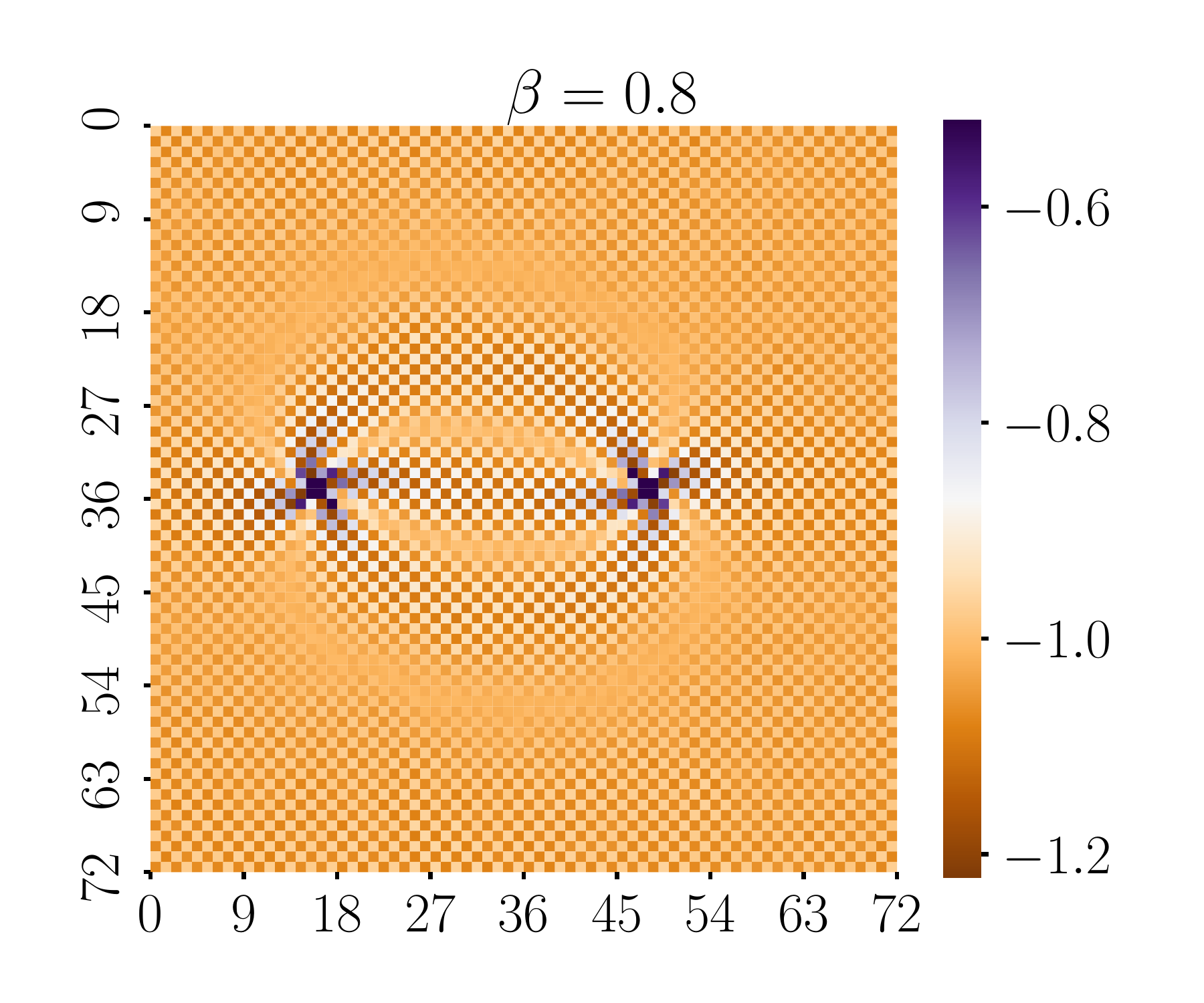}
    \includegraphics[scale=0.14]{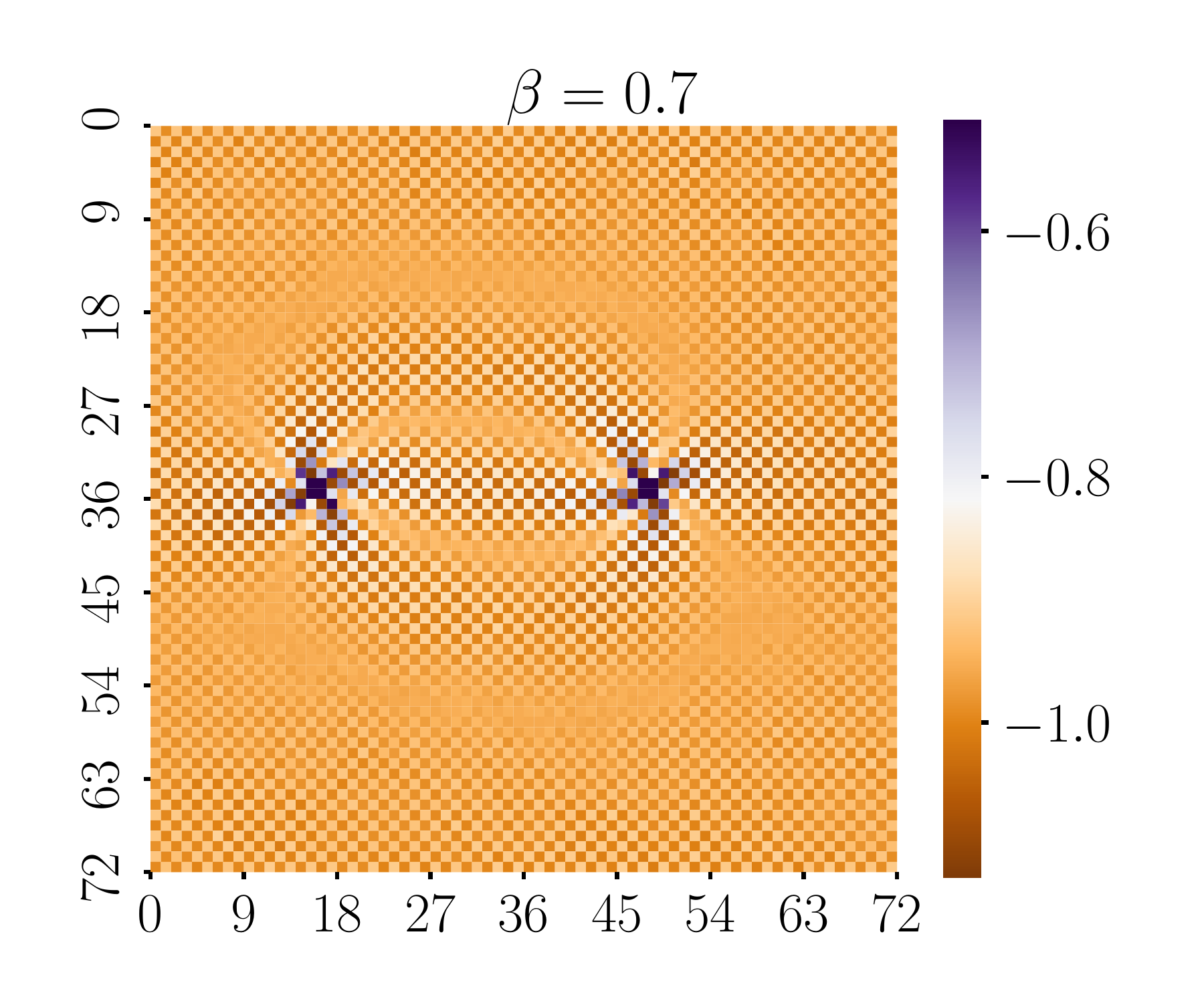}
    \caption{Total energy density with $(-2,+2)$ charge pair for pure $U(1)$ QLM at 
    $\lambda=-1$ ($L=72$). From left to right and top to bottom, the inverse temperatures 
    are $\beta=1.0,0.9,0.8,0.7$. $-2[+2]$ charge is placed at $(16,36)[(48,36)]$.
    Four fractionalized strands connect the sources. With increasing temperature the 
    strands broaden, interact more strongly, and eventually explore configurations 
    that wrap around the periodic spatial direction.}
    \label{fig:stringU1Q2}
\end{figure}

 To illustrate the above statements, Figure~\ref{fig:stringU1Q2} shows the energy-density 
distribution generated by external charges $(-2,+2)$ at $\lambda=-1$. The energy-density 
plots are computed using the expression, 
\begin{equation}
 {\cal E} = -\frac{1}{L^2} \frac{\partial Z}{\partial \beta} = -\frac{1}{L^2 N_t} \frac{\partial Z}{\partial \epsilon},
\end{equation}
where $L$ is the spatial extent, $\beta = N_t \epsilon$ the inverse temperature is given as
the product of the number of time-slices, $N_t$, and the Trotter step, $\epsilon$. 
This quantity can be computed
at each time-slice using the expression of the partition function, $Z$, in terms of the transfer
matrix elements, and then averaged over all time-slices. At the lowest temperature 
shown, $\beta=1$, four separate strands can be resolved between the charges. The patterns inside 
and between the strands correspond to different realizations of the broken lattice and 
charge-conjugation symmetries. This directly visualizes the interpretation of the strands as 
interfaces between the C1 bulk and domains with the local structure of the C2 phase. 
As the temperature is raised, corresponding to decreasing $\beta$, the four interfaces 
broaden. Their transverse fluctuations increase, the distinction between neighbouring strands 
becomes less sharp, and the region occupied by the competing phase grows. Because the spatial 
directions are periodic, a sufficiently broad fluctuating strand can lower its free energy by 
changing topology and wrapping around the lattice. The resulting configuration contains a 
system-spanning interface and an additional strand appearing from the opposite boundary. Upon 
approaching deconfinement, the interfaces spread over the system and the notion of a narrow 
flux tube eventually ceases to be applicable.

 \begin{figure}
    \centering
    \includegraphics[width=0.48\linewidth]{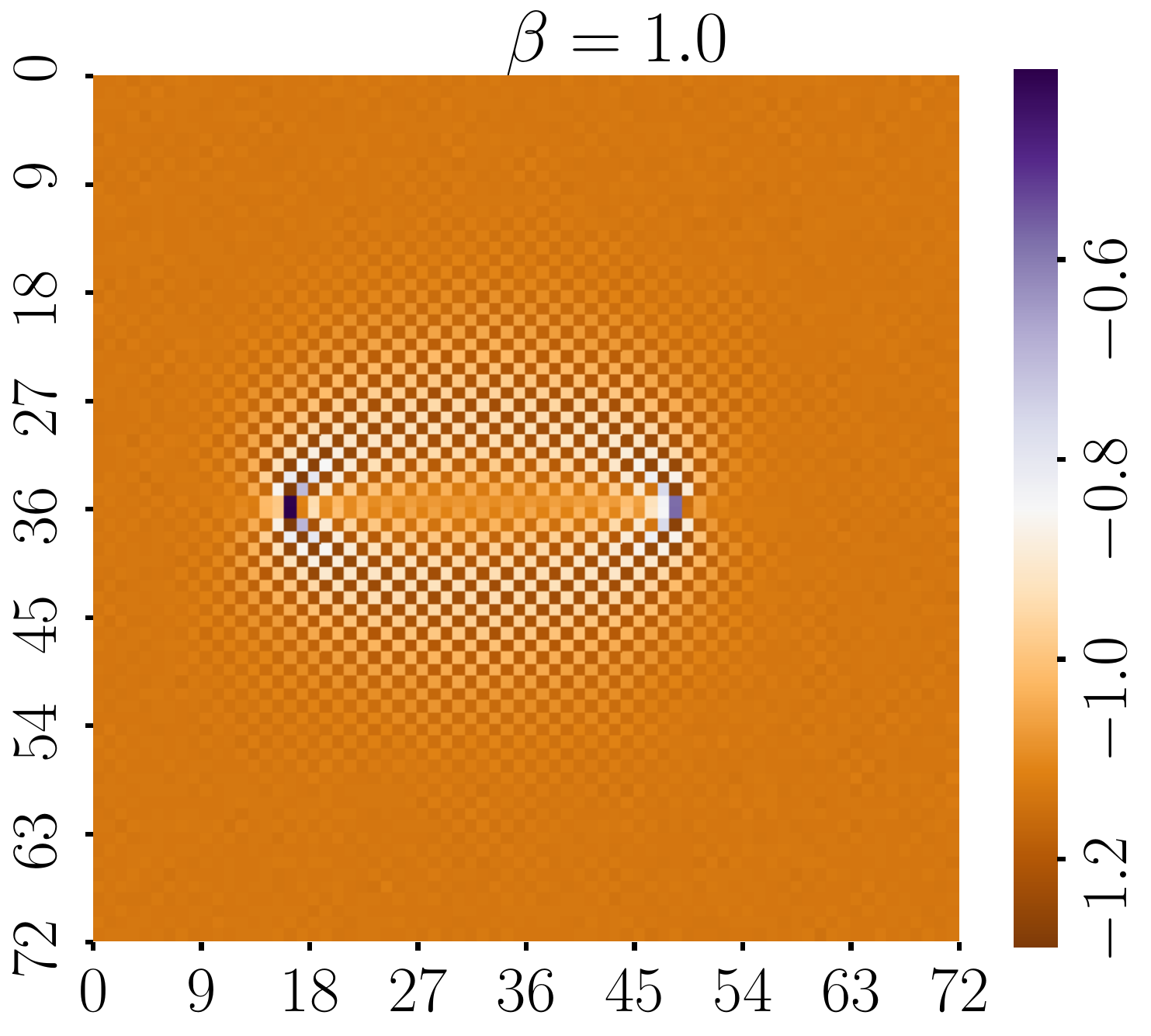}
    \includegraphics[width=0.48\linewidth]{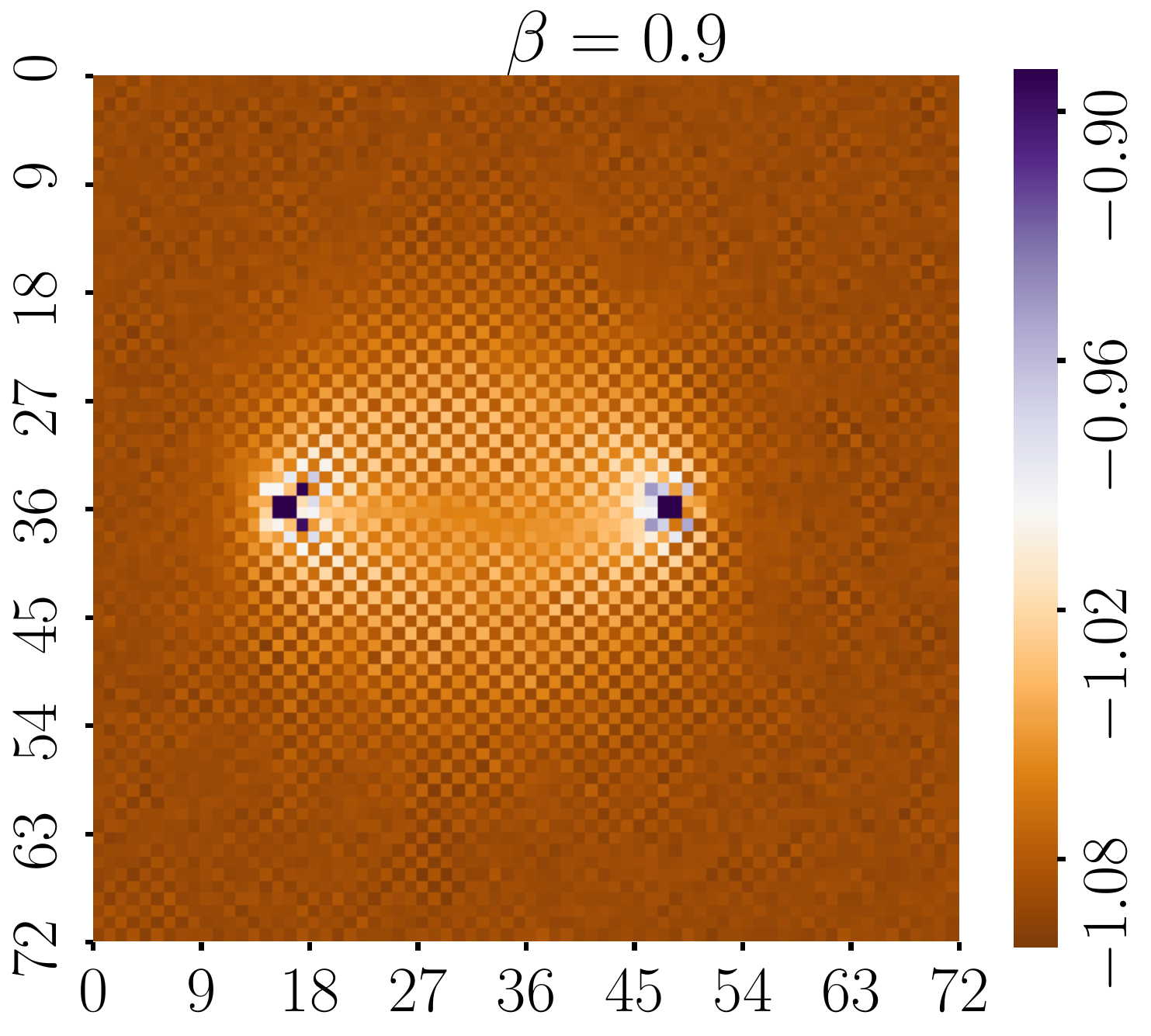}
    \includegraphics[width=0.48\linewidth]{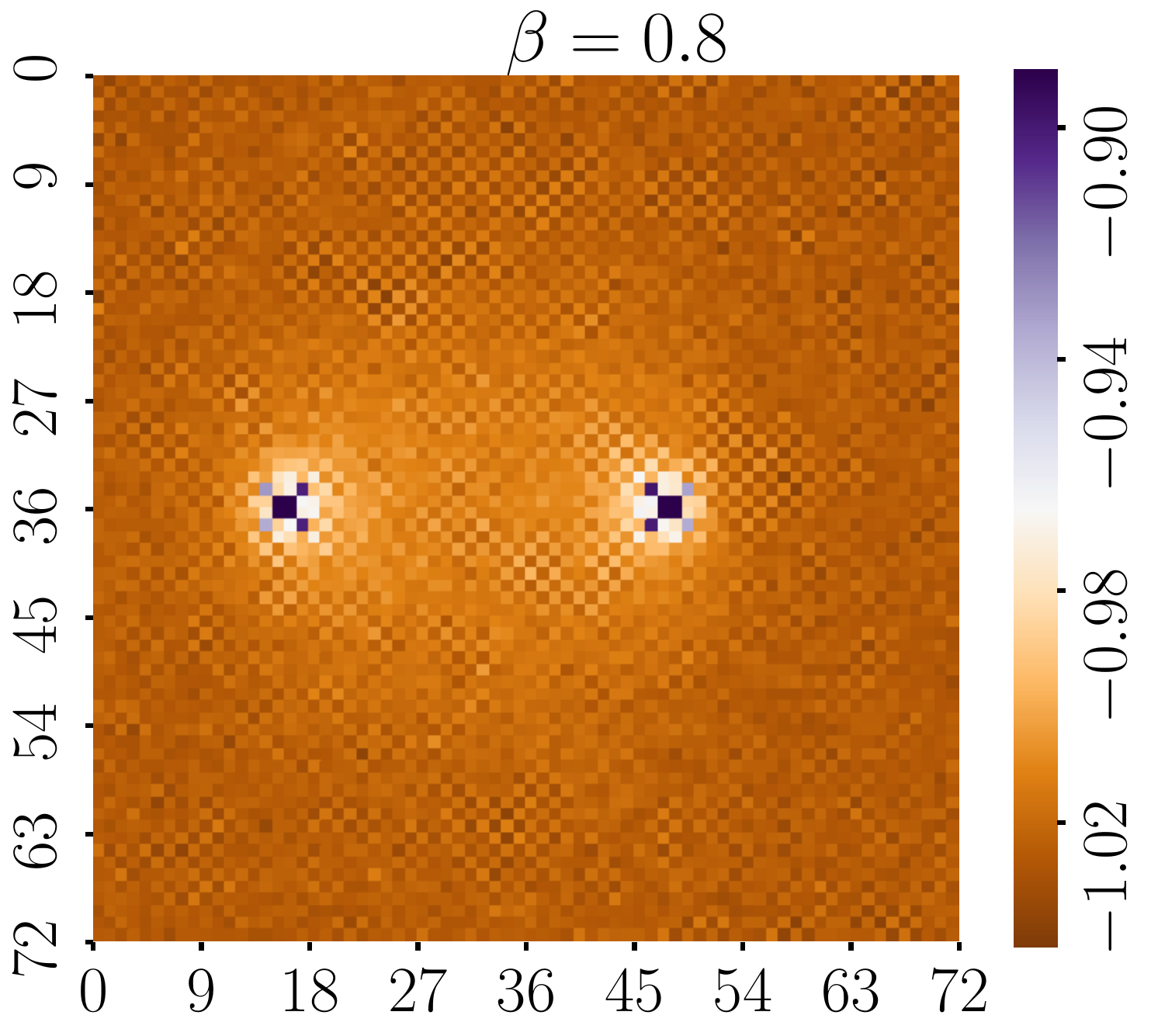}
    \includegraphics[width=0.48\linewidth]{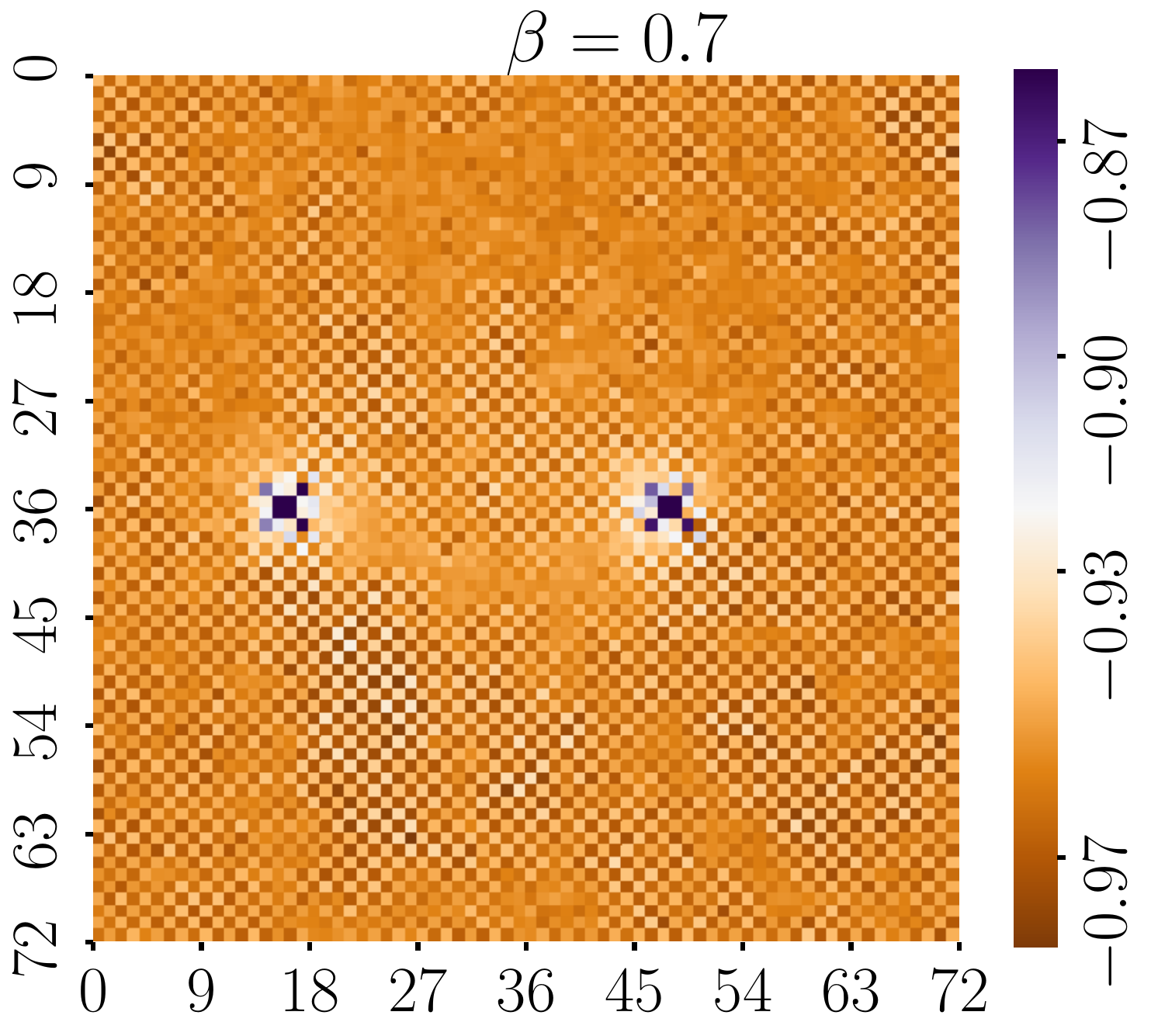}
    \caption{Total energy density with $(+1,-1)$ charge pair for pure $U(1)$ QLM at 
    $\lambda=-1$ ($L=72$). $-1[+1]$ charge is placed at $(16,36)[(48,36)]$. The same
    $\beta$ values as for the case of $Q =\pm 2$ is considered here. The fundamental 
    flux tube splits into two strands carrying one half-unit of electric flux each. 
    The strands broaden as the temperature is increased, as also for the $Q = \pm 2$
    case.}
    \label{fig:stringU1Q1}
\end{figure}

 A clearer picture demonstrating the fractionalization of the fundamental flux is obtained 
by inserting a fundamental charge pair $(-1,+1)$ into the pure theory. As shown in 
Fig.~\ref{fig:stringU1Q1}, the corresponding flux tube consists of two strands rather 
than four, as expected from EFT arguments in \cite{Banerjee2013}. Each strand again
carries flux-$\frac{1}{2}$ and forms an interface enclosing a region with the competing 
symmetry-breaking pattern. The strands become progressively broader as $\beta$ is lowered. 
As before, their thermal broadening is accompanied by an increasing overlap between the 
interfaces and by a visible deformation of the flux-tube profile. Comparing 
Figs.~\ref{fig:stringU1Q2} and \ref{fig:stringU1Q1} therefore gives a direct real-space 
demonstration that the number of strands is proportional to the magnitude of the 
external charge.

\begin{figure}
    \centering
    \includegraphics[scale=0.14]{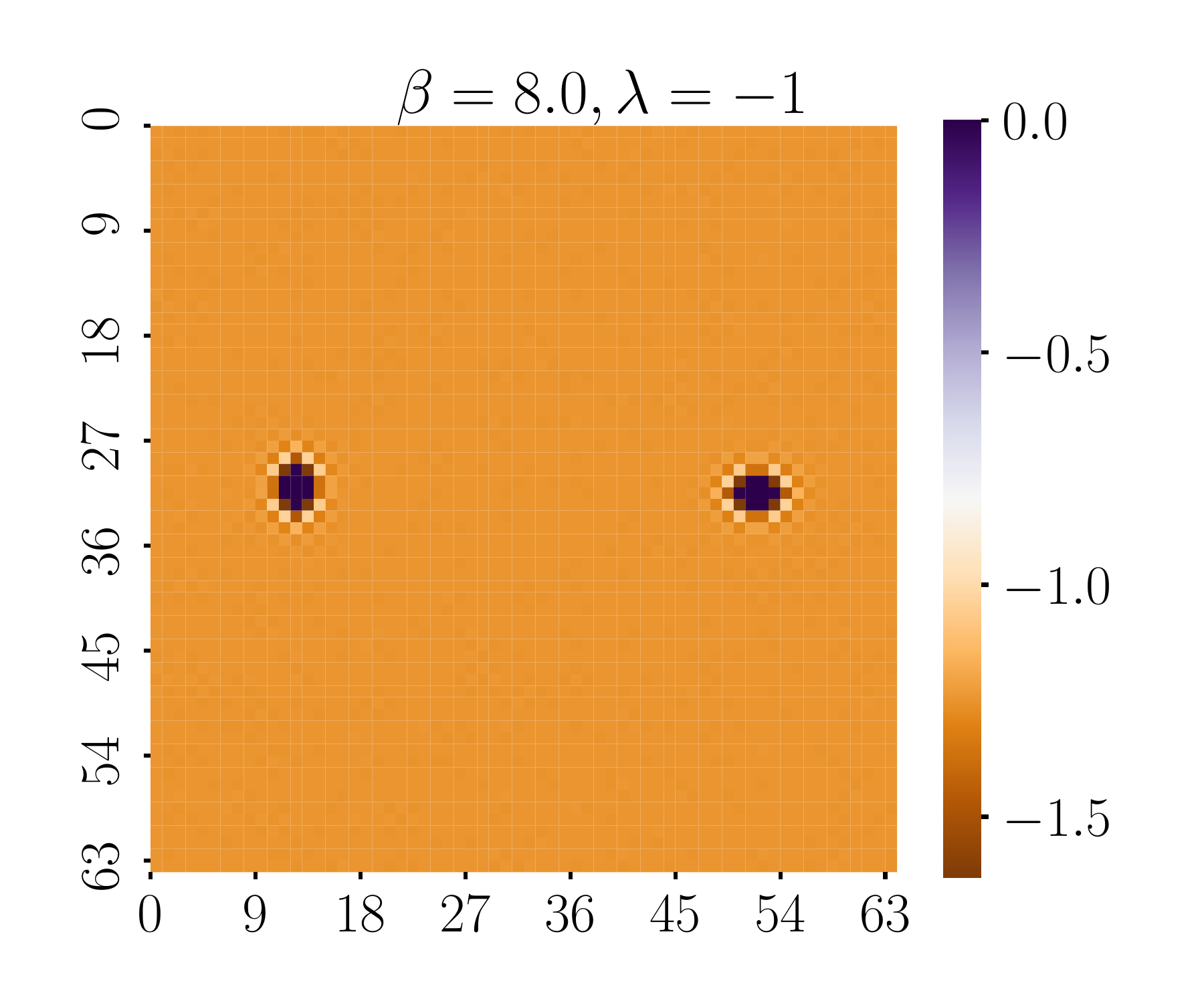}
    \includegraphics[scale=0.14]{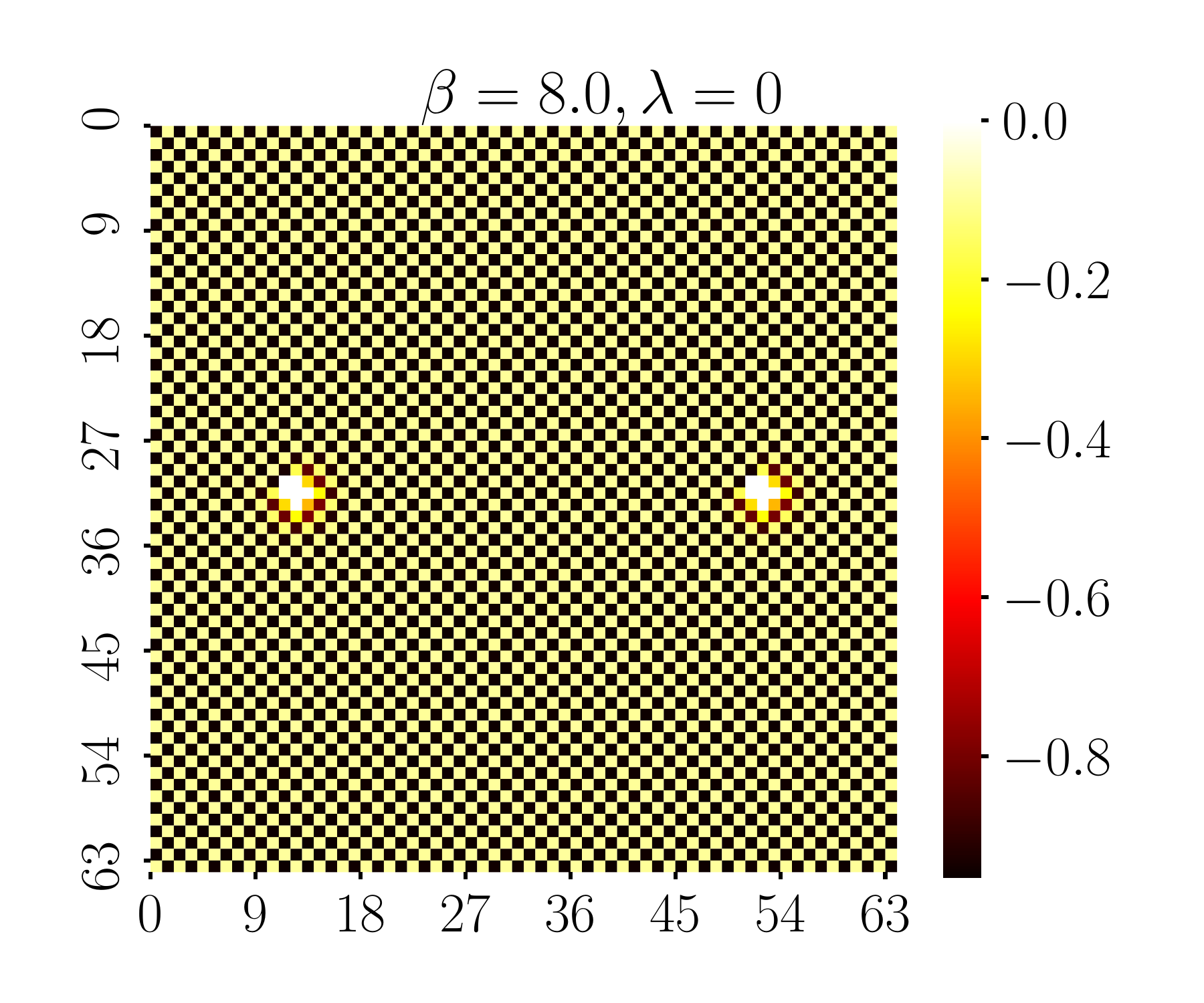}
    \caption{Total energy density with ($+2,-2$) test charge pair for pure $U(1)$ QLM with 
    $\pm2$ annealed charges at $\lambda=-1$ and $\lambda=0$ ($\beta=8.0$). $-2$ [$+2$] charge 
    is placed at $(12,32)$ $[(52,32)]$ on a $64\times64$ lattice. Screening of the static
    charges also show anisotropies analogous to the strands at low temperatures. At $\lambda=0$,
    there is a staggered pattern at the vertical ends of $Q=-2$, which shows up at the horizontal
    end for $Q=+2$. At $\lambda=0$, the charge conjugation symmetry is not broken and there is 
    no anisotropy between the screening of the two charges.} 
    \label{fig:stringZ2T0_lam}
\end{figure}

 A completely different picture is obtained when considering the response of the theory with 
annealed charges $Q_x=\pm2$ to external charges $(-2,+2)$. Figure~\ref{fig:stringZ2T0_lam} 
shows the total energy density at the low temperature $\beta=8$, for $\lambda=-1$ and $\lambda=0$. 
Instead of an extended flux tube joining the two test charges, the response is localized near 
their positions. The annealed charges rearrange themselves so as to neutralize the external 
charge-two sources. Thus, already in the low-temperature regime, a pair of external charges 
$(-2,+2)$ is completely screened and does not define an asymptotic charge-two string tension 
in the theory with annealed $Q_x=\pm2$ charges. Nevertheless, they still display the 
characteristic symmetry breaking patterns as explain in the caption of Fig.~\ref{fig:stringZ2T0_lam}.

\begin{figure}
    \centering
    \includegraphics[width=0.48\linewidth]{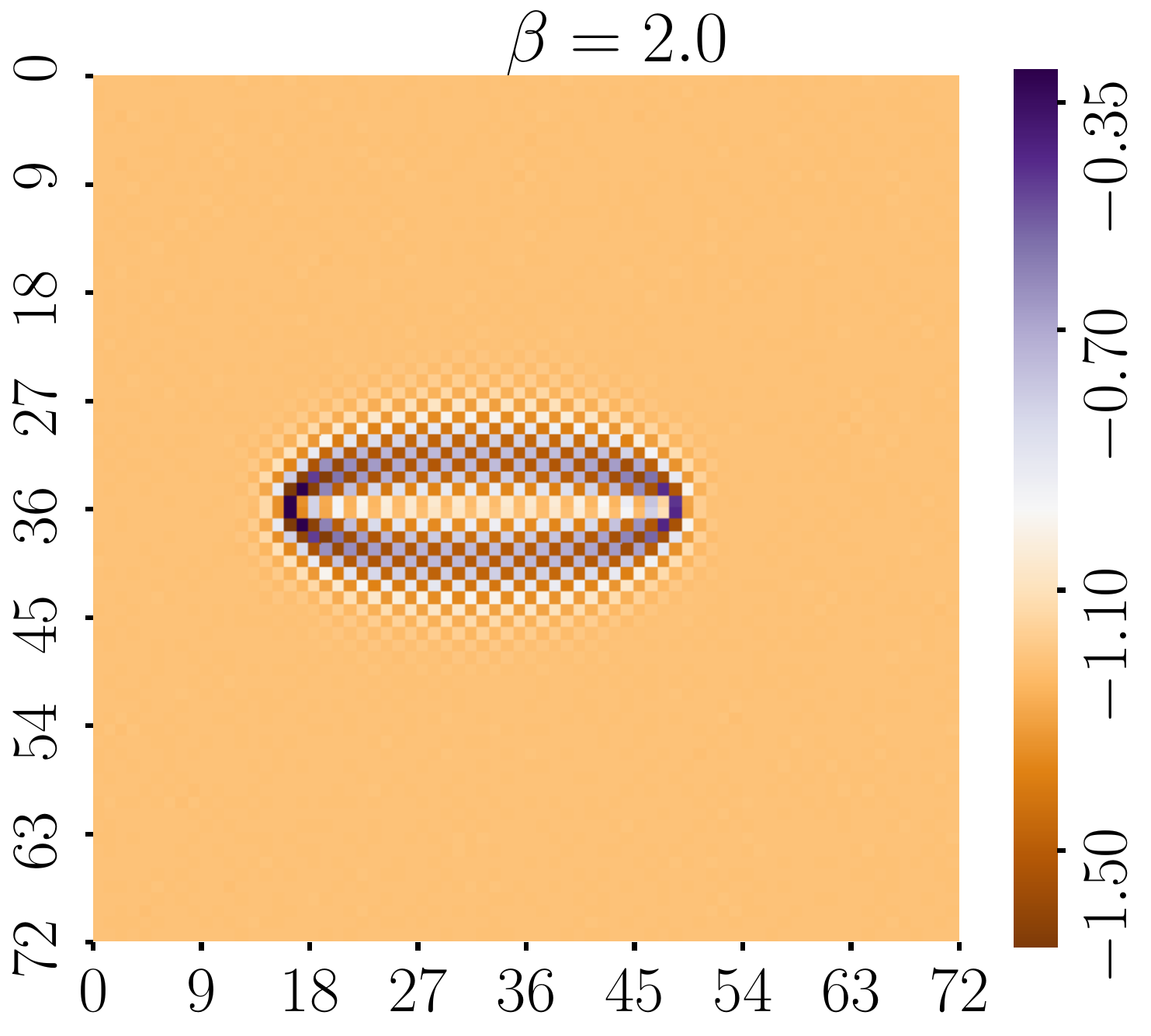}
    \includegraphics[width=0.48\linewidth]{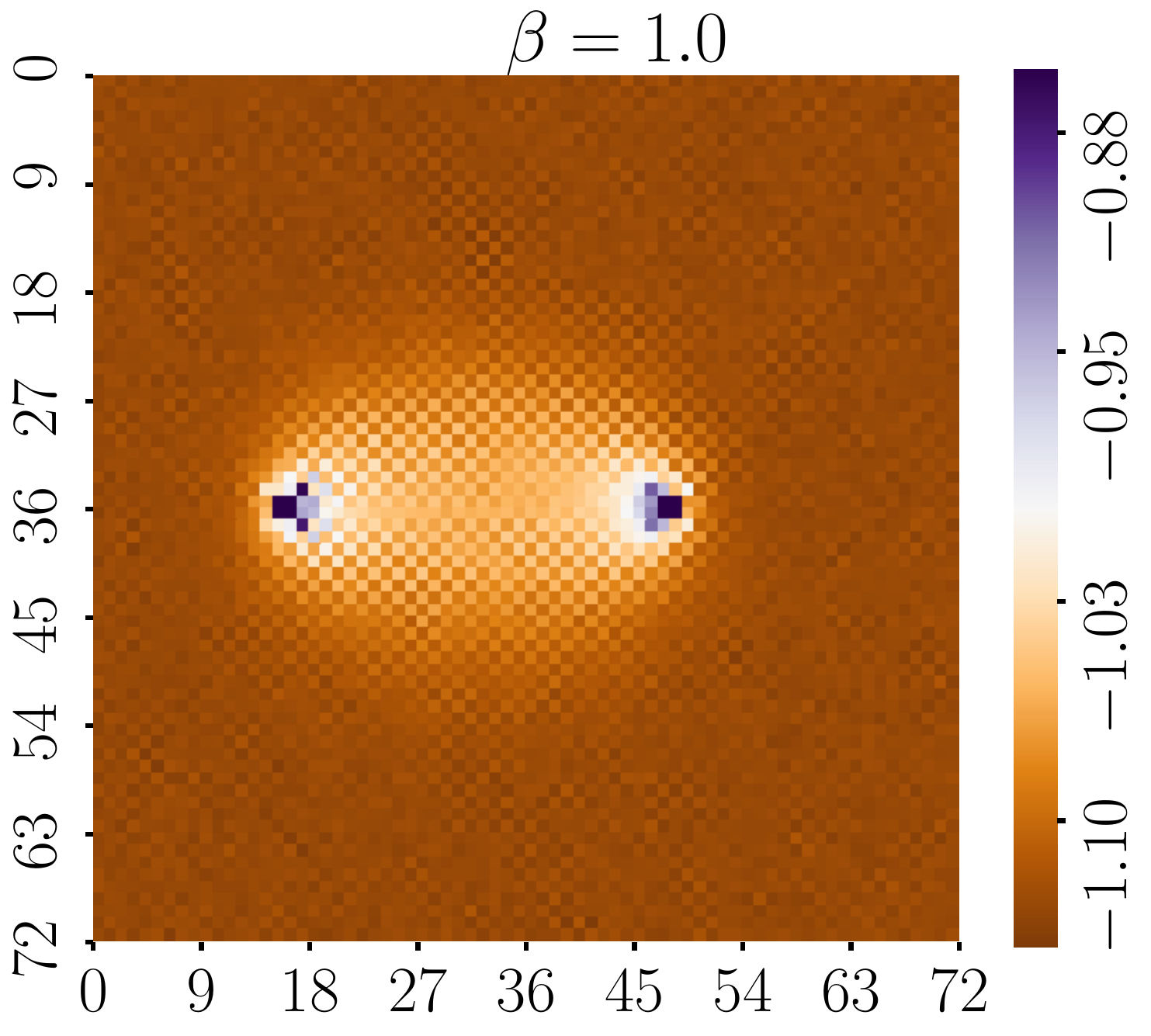}
    \includegraphics[width=0.48\linewidth]{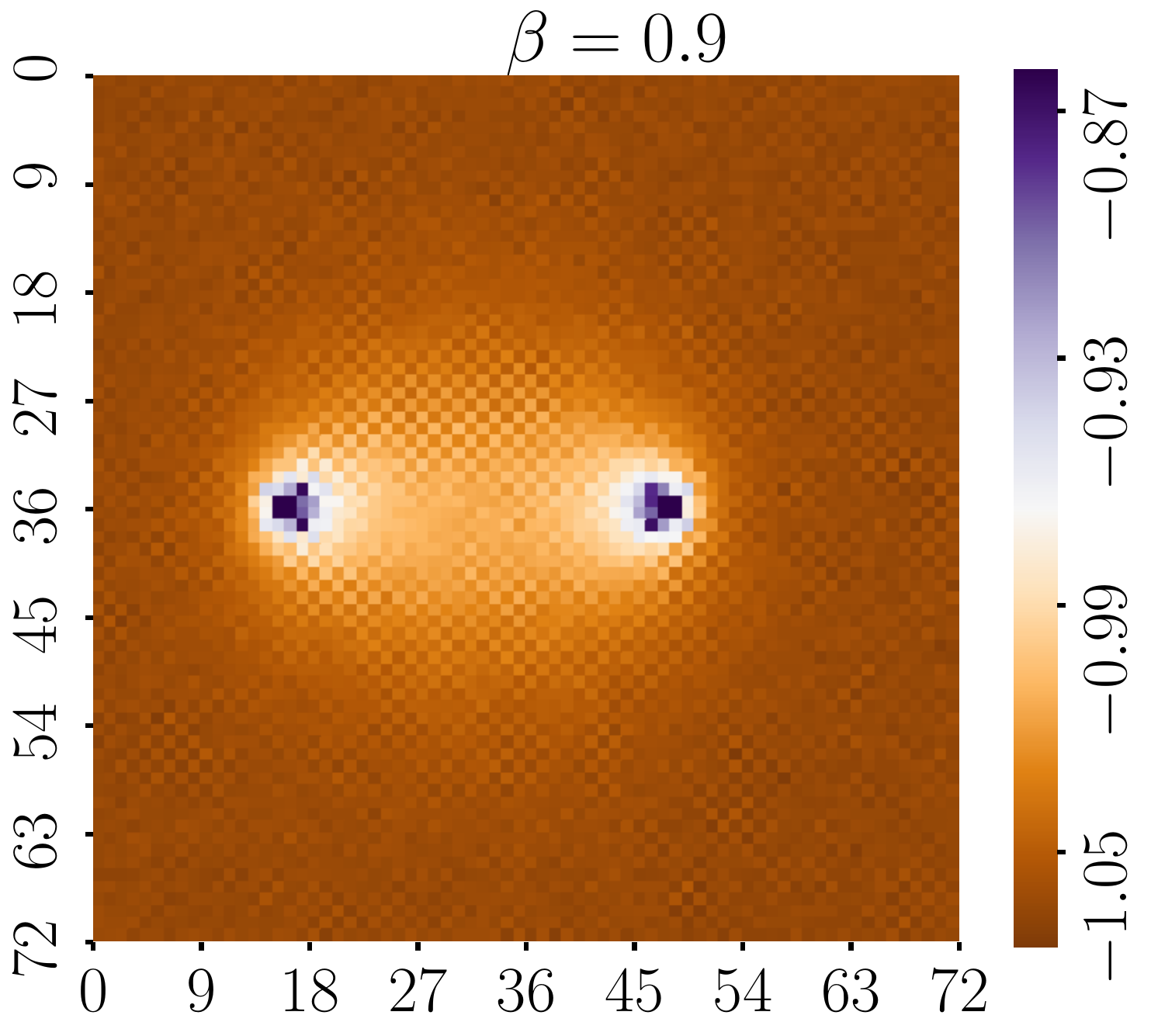}
    \includegraphics[width=0.48\linewidth]{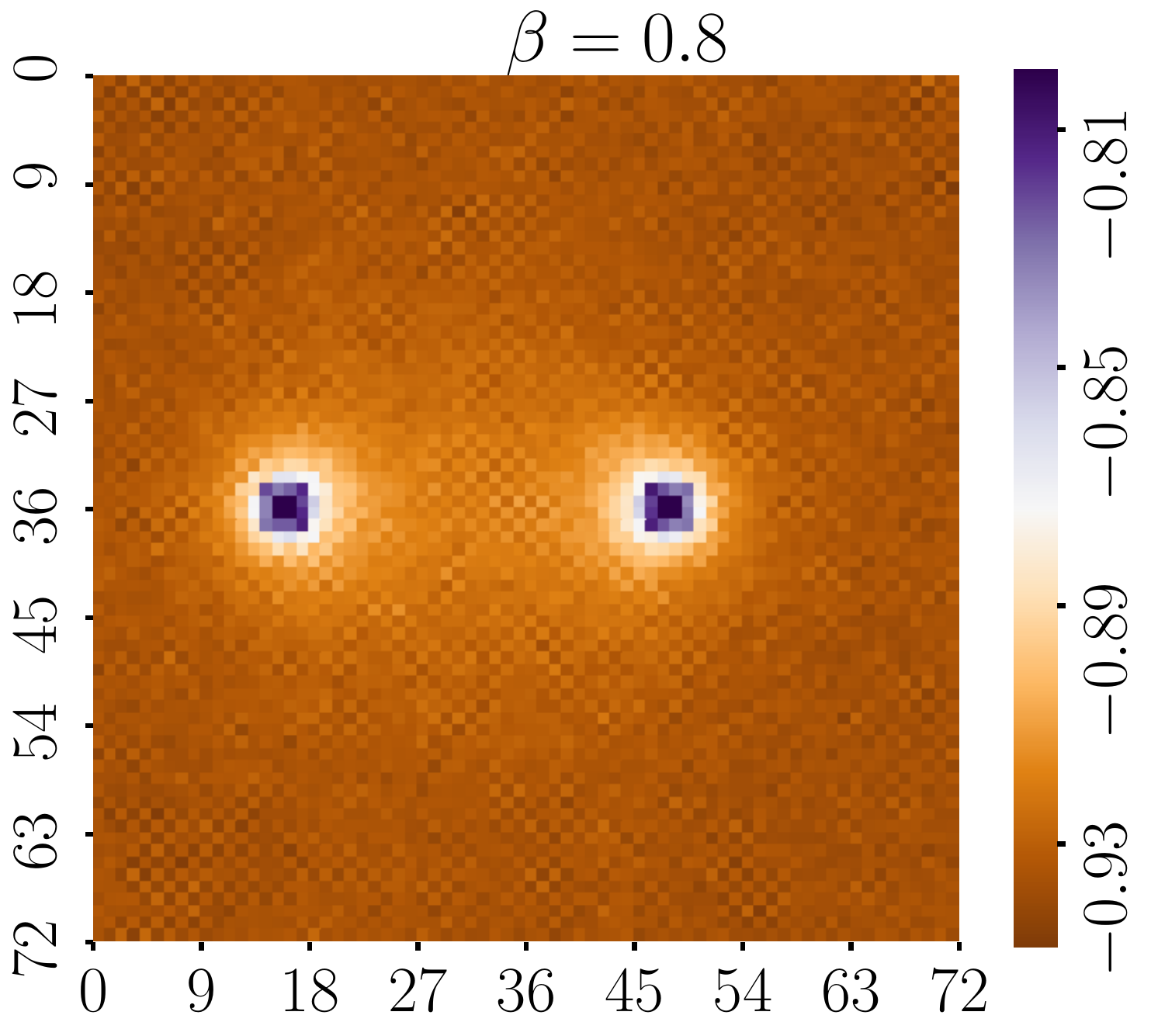}
    \caption{Total energy density with $(+1,-1)$ charge pair for $U(1)$ QLM with $\pm2$ annealed 
    charges at $\lambda=-1$ ($L=72$). $-1[+1]$ charge is placed at $(16,36)[(52,36)]$.
    The two-strand flux tube present at low temperature broadens strongly on approaching the 
    transition and is screened in the high-temperature phase.}
    \label{fig:stringZ2Q1}
\end{figure}

 In contrast, annealed charges $Q_x=\pm2$ cannot locally neutralize an external charge of odd 
magnitude. A fundamental pair $(-1,+1)$ therefore remains connected by a flux tube throughout 
the low-temperature confined phase. Its real-space structure, shown in Fig.~\ref{fig:stringZ2Q1}, 
is nevertheless strongly affected by the annealed charges. At large $\beta$, the two fractionalized 
strands remain visible. As the temperature is raised, thermally excited charge-two defects 
increasingly screen and rearrange the flux surrounding the sources. The two strands broaden rapidly 
and lose their identity as the deconfinement transition is approached. Above the transition, the 
long-distance string is screened: a short-distance response remains near each external source, but 
there is no longer a well-defined flux tube extending between them.

 It is important to emphasise the distinction in the confining properties at low temperature. 
The even external charge in Fig.~\ref{fig:stringZ2T0_lam} can be screened directly by the annealed 
$Q_x=\pm2$ charges, even in the low-temperature regime. An odd external charge cannot be neutralized 
in this way, but its confining string disappears at sufficiently high temperature because the system 
enters the deconfined phase, just as in the pure U(1) gauge theory.

\subsection{Extraction of the string tension}
 To quantify the behaviour seen in the energy-density distributions, we measure the potential 
$V_{Q\bar Q}(r)$ between two static test charges separated along a lattice axis. The negative 
charge is placed at $\mathbf{x}_{Q}=(x_Q,L/2)$ and the positive charge at $\mathbf{x}_{\bar Q}=(x_Q+r,L/2)$.
We consider both $(-1,+1)$ and $(-2,+2)$ test-charge pairs in the pure U(1) theory, and a 
fundamental $(-1,+1)$ pair in the theory with annealed $Q_x=\pm2$ charges.

 The static interaction at finite temperature can be defined in two closely related, but distinct, 
ways. Introducing partition functions $Z_{Q\bar Q}(r,\beta)$ and $Z_0(\beta)$ for the ensembles with 
and without a static charge pair, respectively, the conventional static-pair free energy is
\begin{equation}
    \Delta F_{Q\bar Q}(r,\beta) = -\frac{1}{\beta} \log\frac{Z_{Q\bar Q}(r,\beta)}{Z_0(\beta)}.
    \label{eq:static-free-energy}
\end{equation}
 This is the quantity conventionally obtained from Polyakov-loop correlators at finite temperature 
\cite{JahnPhilipsen2004}. In the present calculation, however, we measure the excess internal energy
\begin{equation}
    \Delta E_{Q\bar Q}(r,\beta) = \braket{H}_{Q\bar Q(r)} - \braket{H}_0 .
    \label{eq:static-internal-energy}
\end{equation}
 Provided that the Hamiltonian has no explicit temperature dependence, the two observables are 
related exactly by
\begin{align}
    \Delta E_{Q\bar Q}(r,\beta) &= \frac{\partial}{\partial\beta}  
      \left[ \beta\Delta F_{Q\bar Q}(r,\beta) \right] \nonumber \\
    &= \Delta F_{Q\bar Q}(r,T) - T\frac{\partial\Delta F_{Q\bar Q}(r,T)}{\partial T}.
    \label{eq:energy-free-relation}
\end{align}
Thus, $\Delta E_{Q\bar Q}$ and $\Delta F_{Q\bar Q}$ probe the same charged sector, but 
differ by the entropy associated with the flux-tube configurations. They coincide in 
the zero-temperature limit, while the distinction can become important close to the 
thermal transition.

 In particular, if the large-distance free energy is written as 
 \begin{equation}
    \Delta F_{Q\bar Q}(r,T) = \mu_F(T)+\sigma_F(T)r+\cdots ,
\end{equation}
then the coefficient of the linear term in the excess internal energy is
\begin{equation}
    \sigma_E(T) = \sigma_F(T) - T\frac{d\sigma_F(T)}{dT}
      = \frac{\partial}{\partial\beta}  \left[\beta\sigma_F(\beta)\right].
    \label{eq:energy-string-tension}
\end{equation}
Consequently, the string tension that we extract using Eq.~\eqref{eq:static-internal-energy} is 
more precisely an internal-energy string tension. It need not have the same temperature dependence 
as the free-energy string tension, and contains entropic effects of the strands. 

 Effective-string theory further clarifies the form of the subleading corrections. At low temperature 
and for $r\ll\beta$, transverse fluctuations of a long confining string produce the universal L\"usscher 
correction
\begin{equation}
    \Delta F_{Q\bar Q}(r,T) = \mu_F+\sigma_F r - \frac{\pi(D-2)}{24r} + \cdots ,
    \label{eq:luscher-term}
\end{equation}
where $D$ is the spacetime dimension \cite{LuscherWeisz2002}. In the opposite finite-temperature 
regime, $r\gg\beta$, one obtains 
\begin{equation}
    \Delta F_{Q\bar Q}(r,T) = \mu_F(T)+\sigma_F(T)r + \frac{D-2}{2}T\log(rT) + \cdots .
    \label{eq:finiteT-string-log}
\end{equation}
 The logarithmic term is the finite-temperature, long-string counterpart of the L\"uscher correction 
and has been shown to provide a universal signature of effective-string fluctuations in finite-temperature 
LGTs \cite{CaselleEtAl2011}. Although Eq.~\eqref{eq:finiteT-string-log} strictly applies to the free 
energy, we take it as a motivation to use it in the fit to extract the string tension, particularly
as we increase the temperature towards $T_c$. As we show below, the fit form with the logarithm,
when used with the leading linear large-distance behaviour in the confined phase, greatly improves the 
fit quality. We therefore consider
\begin{equation}
V_{Q\bar Q}(r) = V_0 + \sigma_Q r + c_Q\ln r ,
\label{eq:pot}
\end{equation}
 in addition to the purely linear term. 

 The improvement resulting from the logarithmic term is illustrated in Table~\ref{tab:chiStringEv} 
for an external charge-two pair at $\beta=0.8$ on a $72\times72$ lattice. The value of $\chi^2/\mathrm{dof}$
obtained from a purely linear fit remains very large unless only the largest separations are retained. 
In contrast, Eq.~\eqref{eq:pot} gives an acceptable fit already for small separations. We use variations of the fit range and the difference between the 
two fit forms to monitor the corresponding systematic uncertainty.

\begin{table}
    \centering
    \begin{tabular}{|c|c|c|}
        \hline
        Range of $r$ & $\chi^2$/dof ($f(r)$) & $\chi^2$/dof ($g(r)$) \\
        \hline
        $12-38$ & 3.5 & 1.46 \\
        $14-38$ & 3.4 & 1.41 \\
        $16-38$ & 3.45 & 1.25 \\
        $18-38$ & 1.59 & 0.99 \\
        $20-38$ & 1.53 & 1.12 \\
        $22-38$ & 1.54 & 1.29 \\
        $24-38$ & 1.42 & 1.52 \\
        \hline
    \end{tabular}
    \caption{$\chi^2$/dof using two fitting functions, $f(r)=a+\sigma r$ and 
    $g(r)=a+\sigma r + c\,\mathrm{ln}(r)$, for a range of separations $r$ at $\beta=0.8$
    on the $L=72$ lattice. The charge separation is kept even in order to
    minimize the effect of symmetry breaking.}
    \label{tab:chiStringEv}
\end{table}

\begin{figure}
    \centering
    \includegraphics[width=1\linewidth]{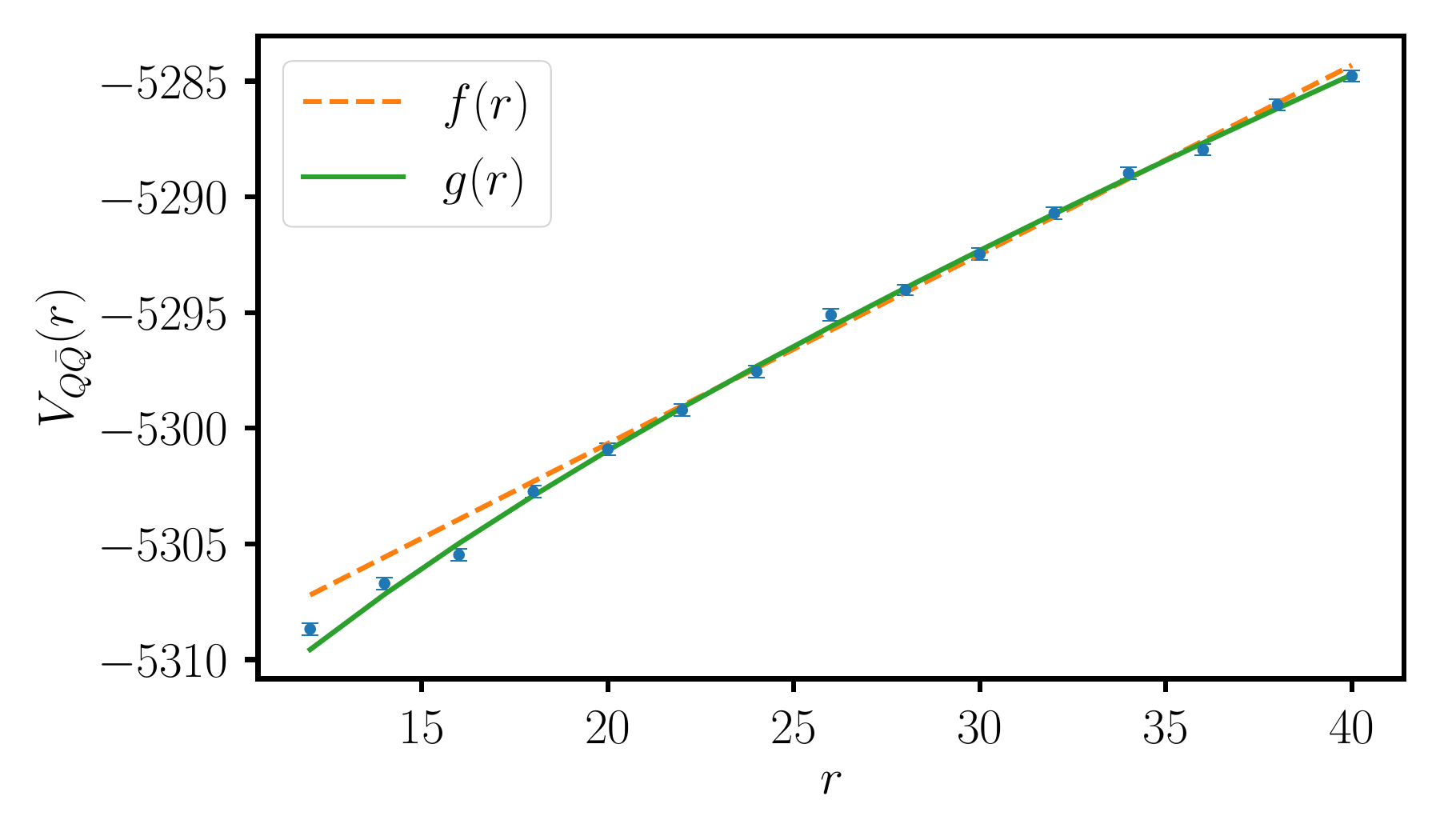}
    \caption{Potential energy of $(-1,+1)$ test charge pair as a function of separation for $\beta=0.8$ and  $\lambda=-1$ (even $r$) on a $72\times72$ lattice.}
    \label{fig:E_rEven}
\end{figure}

The string tensions obtained from linear fits on the $L=72$ lattices (using largest separations)
at $\lambda=-1$ are collected in Table~\ref{tab:sigmaQ1Q2_linear}. A typical fit for the 
charge-one case is shown in Fig.~\ref{fig:E_rEven}.  The temperature dependence of the 
string tension will be discussed in the next subsection. 

 Finally, our previous results lead us to expect that the fundamental string tension, 
$\sigma^{Z_2}_1$, in the theory with annealed charges, is close to $\sigma^{U(1)}_1$
at low temperatures. We see evidence of this in Table~\ref{tab:sigmaQ1Q2_linear} in
the fifth column. Thus, well inside the confined phase, the annealed defects have only 
a small effect on the tension of an odd-charge string.

\begin{figure}
    \centering
    \includegraphics[scale=0.27]{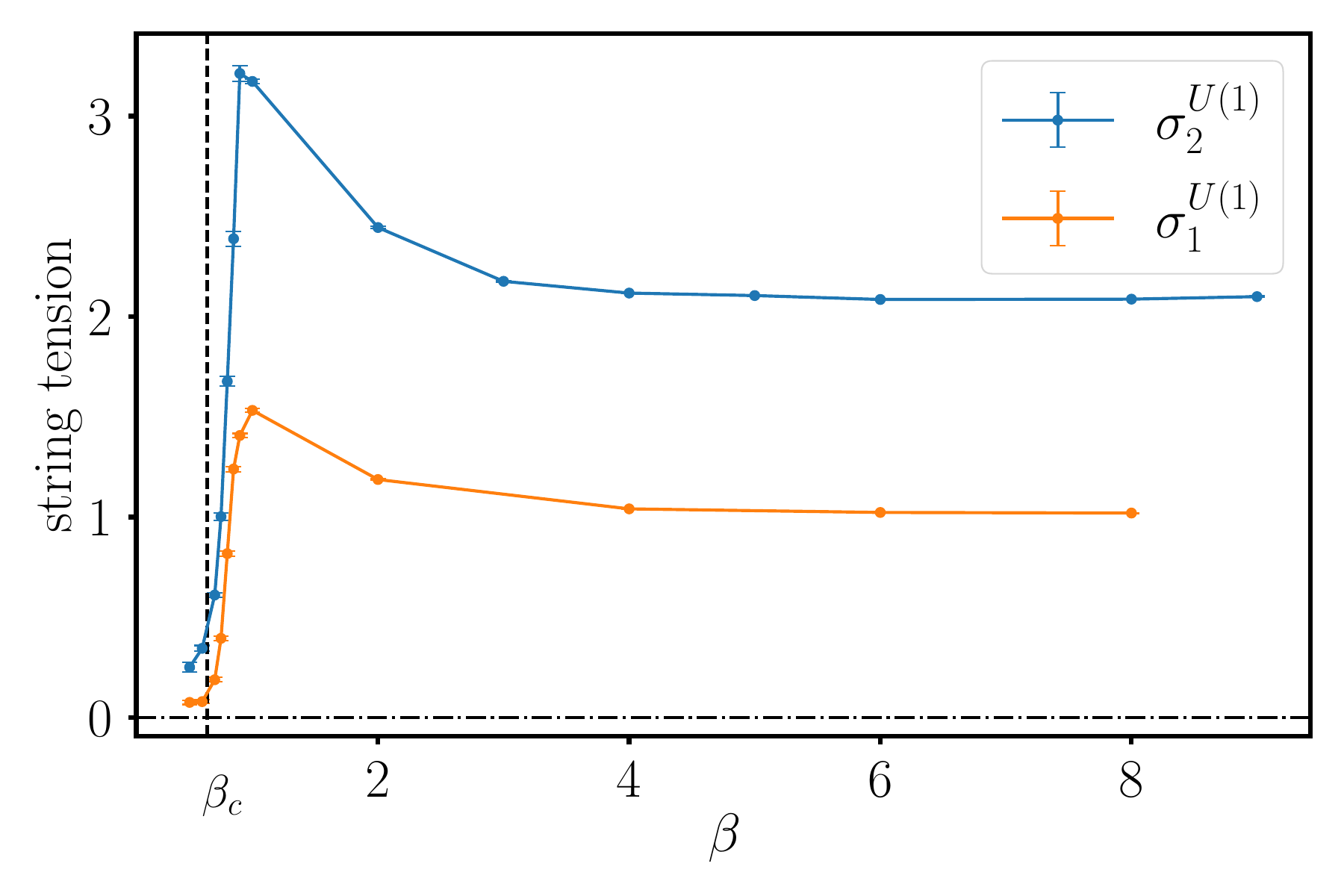}
    \caption{String tension vs $\beta$ for pure U(1) theory with $\pm1$ and 
    $\pm2$ test charges. Results are obtained for $L=72$ lattice at $\lambda=-1$.}
    \label{fig:sigVSbetaU1}
\end{figure}

\subsection{Temperature dependence}
 We conclude this section by discussing the temperature dependence of the different string 
tensions that we discussed in the last subsection. In Figure~\ref{fig:sigVSbetaU1}, we show
the temperature dependence of $\sigma^{U(1)}_1$ and $\sigma^{U(1)}_2$ in the pure theory. Deep
in the confined phase, both tensions approach nearly temperature-independent values, as in
clear from the figure. At low temperature, the charge-two tension is approximately twice the 
fundamental tension:
\begin{equation}
 \sigma^{U(1)}_2 \simeq 2\sigma^{U(1)}_1 + \Delta.
 \label{eq:sigma-relation}
\end{equation}
This approximate additivity is precisely what one expects if a fundamental string contains 
two \emph{non-interacting} flux-$\frac12$ interfaces and a charge-two string contains four. 
Since the individual strands repel each other, we paramterize the repulsion through the term
$\Delta > 0$. As one would expect, this term is temperature dependent, as is also evident
from the values in the fourth column of Table~\ref{tab:sigmaQ1Q2_linear}.

\begin{figure}
    \centering
    \includegraphics[scale=0.27]{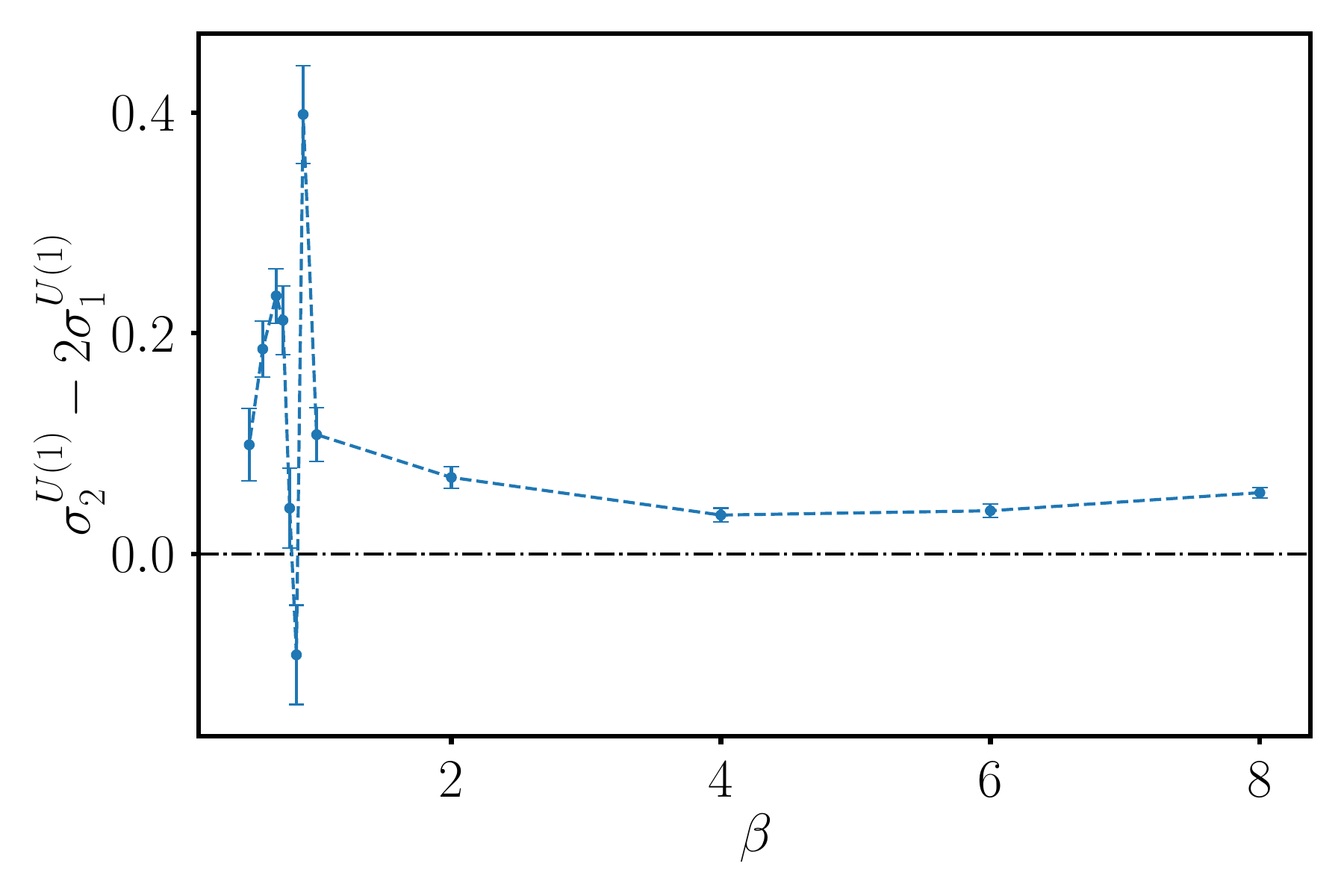}
    \caption{$\sigma_2^{U(1)}-2\sigma_1^{U(1)}$ vs $\beta$ for pure U(1) theory. The results
    are obtained for the $L=72$ lattice at $\lambda=-1$.}
    \label{fig:sig2m1VSbeta}
\end{figure}

 On heating the system from this regime, the interfaces become broader due to thermal 
fluctuations, causing an increase in mutual repulsion. This causes an increase in their 
effective string tension. Closer to the transition, the tension decreases rapidly and tends 
to zero as the system enters the deconfined phase, indicating that the string tension is 
not longer a well defined quantity. In particular, in the region between $\beta \approx 0.7-1.0$, 
as is evident from the Figs.~\ref{fig:stringU1Q1},\ref{fig:stringU1Q2} the identity of the 
flux strands as stretched domain walls become diffuse. Interestingly, the measured 
potential at very high temperatures is still mildly increasing, as shown in Fig.~\ref{fig:E_rEven}.
The nonzero coefficients obtained from finite-volume linear fits above (or even very close to) the
transition do not represent an asymptotic string tension any more, but likely reflects 
logarithmically confining nature of the Coulomb phase that replaces the confining phase 
at the high temperature regime. Fig.~\ref{fig:sig2m1VSbeta} shows the quantity 
$\Delta = \sigma_2^{U(1)}-2\sigma_1^{U(1)}$ plotted as a function of temperature: it is 
temperature independent deep in the confined phase, increases monotonically while approaching 
the phase transition, but starts displaying fluctuations even before we reach the deconfinement 
temperature indicating the onset of rapid fluctuations. This is the regime of $\beta$ where the 
notion of string tension perhaps ceases to make sense. 
 
\begin{figure}
    \centering
    \includegraphics[scale=0.27]{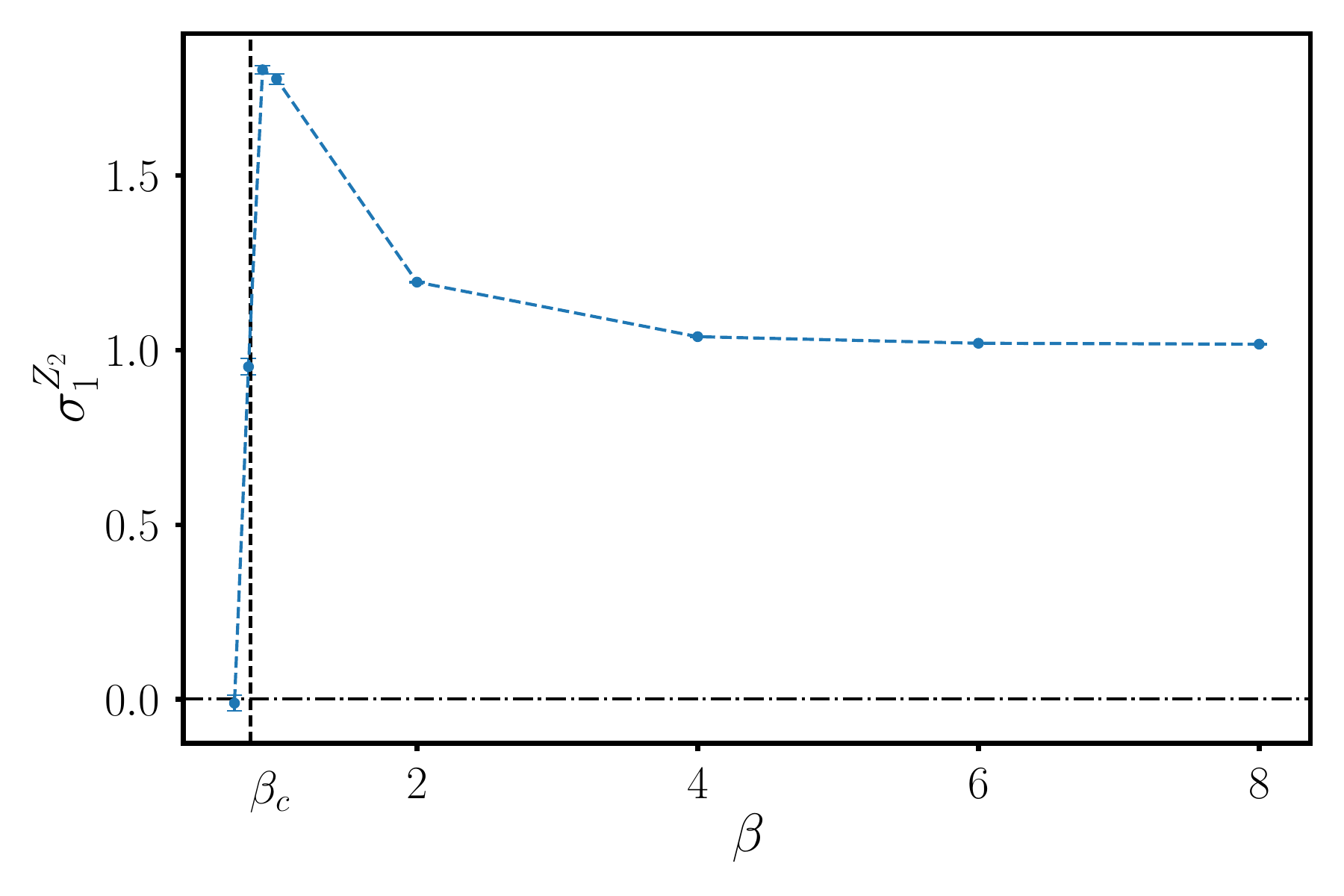}
    \caption{String tension vs $\beta$ for U(1) theory with annealed $\pm2$ charges. The 
    result is again obtained for the $L=72$ lattice at $\lambda=-1$.}
    \label{fig:sigVSbetaZ2}
\end{figure}

The fundamental string tension in the theory with annealed $Q_x=\pm2$ charges is shown in 
Fig.~\ref{fig:sigVSbetaZ2}. Its low-temperature value is again close to that of the pure 
theory. On approaching the transition from the confined side, the fitted tension increases 
before falling rapidly. The apparent nonzero tension obtained at $\beta=0.8$, slightly on the
high-temperature side of the finite-volume transition, should be interpreted as a crossover 
effect: the correlation length and the transverse width of the string are comparable to the 
system size, and a linear fit over a limited range of separations can still return a nonzero 
effective slope. By $\beta=0.7$, the fitted value is compatible with zero. This behaviour is
consistent with that observed for the pure theory in Fig.~\ref{fig:sigVSbetaU1}. 

\begin{table}
    \centering
    \begin{tabular}{|c|c|c|c|c|}
    \hline
    $\beta$  &  $\sigma_2^{U(1)}$ & $\sigma_1^{U(1)}$ & $\sigma_2^{U(1)}-2\sigma_1^{U(1)}$ & $\sigma_1^{Z_2}$ \\
    \hline
    $8.0$ & $2.0958(33)$ & 1.0202(16) & 0.0554(46) & 1.0169(14)\\
    \hline
    $6.0$ & $2.0852(31)$ & 1.0231(27) & 0.0390(62) & 1.0196(14)\\
    \hline
    $4.0$ & $2.1171(32)$ & 1.0410(27) & 0.0351(63) & 1.0384(22)\\
    \hline
    $2.0$ & $2.4440(50)$ & 1.1874(42) & 0.0692(98) & 1.1946(26)\\
    \hline
    $1.0$ & $3.1724(115)$ & 1.5322(107) & 0.1080(243) & 1.7766(145)\\
    \hline
    $0.9$ & $3.2123(394)$ & 1.4070(101) & 0.3983(443) & 1.8030(112)\\
    \hline
    $0.85$ & $2.3880(370)$ & 1.2398(128) & -0.0916(450) & $-$\\
    \hline
    $0.8$ & $1.6769(237)$ & 0.8178(136) & 0.0413(361) & 0.9523(237)\\
    \hline
    $0.75$ & $1.0025(196)$ & 0.3954(121) & 0.2117(311) & $-$\\
    \hline
    $0.7$ & $0.6114(100)$ & 0.1888(113) & 0.2338(247) & -0.0114(226)\\
    \hline
    $0.6$ & $0.3454(144)$ & 0.0799(105) & 0.1856(255) & $-$\\
    \hline
    $0.5$ & $0.2510(252)$ & 0.0761(104) & 0.0988(327) & $-$\\
    \hline
    \end{tabular}
    \caption{String tension $\sigma$ for pure $U(1)$ theory with test charge pairs 
    ($-2,+2$) and $(-1,+1)$ and $Z_2$ theory with test charge pair $(-1,+1)$ at 
    different $\beta$'s for $L=72$. The values are obtained using linear fitting
    as explained before.}
    \label{tab:sigmaQ1Q2_linear}
\end{table}

\section{$T_c/\sqrt{\sigma}$ for the two gauge theories} 
\label{sec:TcSqrtSig}
\begin{table}
    \centering
    \begin{tabular}{|c|c|c|c|}
    \hline
    $\lambda$ & $\beta_c$ & $\sigma$ & $T_c/\sqrt{\sigma}$\\
    \hline
    $-1.0$ & $0.81428(1)$ & 1.0169(14) & 1.2109(8)\\
    \hline
    $-0.9$ & $0.88529(2)$ & 0.8504(18) & 1.2227(13)\\
    \hline
    $-0.8$ & $0.96819(3)$ & 0.6978(15) & 1.2364(13)\\
    \hline
    \end{tabular}
    \caption{$T_c/\sigma$ for different $\lambda$ for $Z_2$ theory 
    with $(-1,+1)$ charge pair ($\beta=8$, $L=72$).}
    \label{tab:sigmaZ2lam}
\end{table}

 Before discussing the numerical results, let us clarify the sense in which the ratio 
$T_c/\sqrt{\sigma}$ can be used to compare different lattice gauge theories. Neither 
the critical temperature $T_c$ nor the string tension $\sigma$, expressed in lattice 
units, is separately universal. Their dimensionless ratio, however, compares the 
thermal deconfinement scale with the zero-temperature confining scale and may 
approach a characteristic value along a continuum scaling trajectory. Such a ratio 
therefore contains information that is complementary to the critical exponents of 
the finite-temperature transition.

 In making this comparison, we have to distinguish between two notions of universality.
The first concerns the critical behavior at the finite-temperature deconfinement 
transition which has been studied in this work, and concerns the long-distance 
fluctuations at $T=T_c$. The second notion concerns the zero-temperature continuum 
theory obtained, if it exists, by tuning the microscopic couplings so that a physical 
correlation length diverges in lattice units. Two lattice theories may have deconfinement 
transitions in the same thermal universality class while nevertheless possessing 
different zero-temperature spectra or different hierarchies between the glueball
and string scales. Glueballs are the elementary excitations of a non-Abelian
pure gauge theory, and in this context, we call them photoballs instead. 
 Conversely, the presence of the same residual centre symmetry 
does not by itself imply equality of all dimensionless spectral quantities. The 
ratios $T_c/\sqrt{\sigma}$ and $m_G/\sqrt{\sigma}$ (where $m_G$ is a glueball/photoball mass)
are useful precisely because they probe this additional dynamical information.

 The comparison is particularly interesting for the spin-$\tfrac12$ QLM. Its 
two-dimensional link Hilbert space and its crystalline, symmetry-breaking confined 
phases differ substantially from the conventional Wilson formulation. The confining 
strings also possess a fractionalized internal structure that has no immediate 
counterpart in the simplest Wilson theory. It is therefore not evident a priori 
whether the thermal and confining scales of the quantum link model should reproduce 
those of the corresponding Wilson LGT. At the same time, the theory with annealed 
$Q=\pm2$ charges has a natural connection with a $Z_2$ gauge theory in terms of its 
long distance confining properties, as explored here in detail. This motivates comparing 
$T_c/\sqrt{\sigma}$ in the annealed-charge theory with the corresponding ratio in the 
conventional three-dimensional $Z_2$ LGT. 

For the theory with annealed $Q=\pm2$ charges, the results in \Cref{tab:sigmaZ2lam} give
\begin{equation}
 \frac{T_c}{\sqrt{\sigma}} = 1.2109(8),~1.2227(13),~1.2364(13)
\end{equation}
at $\lambda=-1.0,~-0.9,~-0.8$, respectively. The continuum result for the conventional 
three-dimensional $Z_2$ lattice gauge theory is known to be
\begin{equation}
\left.\frac{T_c}{\sqrt{\sigma}}\right|{Z_2} = 1.235(2)
\end{equation}
obtained from high-precision studies of the dual three-dimensional Ising 
model~\cite{Caselle:2007yc}. Our values of $T_c/\sqrt{\sigma}$ vary from 
the continuum estimate for the $Z_2$ LGT by less than three percent.
This can be regarded as evidence that the deconfinement and fundamental-string 
scales are organized similarly in the two theories. 

 We should however, point out that the theory with annealed charges is not 
identical to the conventional $Z_2$ lattice gauge theory, especially at $T=0$,
where it has different confining phases and fractionalized fluxes. Moreover, 
even the finite-temperature transition exhibits weak universality rather than
a conventional second order Ising transition. 

 Next, let us consider the value of $\frac{T_c}{\sqrt{\sigma}}$ for the pure
$U(1)$ QLM, which are provided in \Cref{tab:sigmaU1lam}. The values range betweem
1.48 to 1.49 as $\lambda$ varies from -0.9 to -1.1. This value is clearly separated 
from the $Z_2$ result and suggests that the thermal and confining scales of the 
pure QLM obey a different quantitative relation.

\begin{table}
    \centering
    \begin{tabular}{|c|c|c|c|}
    \hline
    $\lambda$ & $\beta_c$ & $\sigma$ & $T_c/\sqrt{\sigma}$\\
    \hline
    $-0.9$ & $0.7273(52)$ & 0.8502(22) & 1.4912(108)\\
    \hline
    $-1.0$ & $0.6668(33)$ & 1.0202(16) & 1.4848(74)\\
    \hline
    $-1.1$ & $0.6137(38)$ & 1.1876(20) & 1.4952(93)\\
    \hline
    \end{tabular}
    \caption{$\sigma_1$ for different $\lambda$ for $U(1)$ theory with $(-1,+1)$ charge pair 
    ($\beta=8$, $L=72$).}
    \label{tab:sigmaU1lam}
\end{table}

 Let us further contrast this observation with the weak-coupling continuum limit of 
the conventional Wilson-Villain compact $U(1)$ theory in $2+1$-d. The analysis of 
G\"opfert and Mack gives the asymptotic scaling relations~\cite{Gopfert:1981er}
\begin{equation} \label{eq:GopfertMack}
a^2m_G^2 \sim c_1\beta\exp(-c_2\beta),~~a^2\sigma\sim c_3(ag^2)(am_G),
\end{equation}
where $m_G$ denotes the lightest glueball mass, or equivalently the nonperturbatively 
generated photon mass. Both $m_G/g^2$ and $\sqrt{\sigma}/g^2$ vanish exponentially as 
the Wilson continuum limit is approached ($g^2 \rightarrow 0$). Their relative scaling 
implies
\begin{equation}
  \frac{m_G}{\sqrt{\sigma}} \sim \sqrt{\frac{m_G}{g^2}} \rightarrow 0.
\label{eq:contLim}
\end{equation}
In this sense, the confining string becomes parametrically heavy when measured in units 
of the lightest glueball mass.

 Finite-temperature studies of the Wilson theory furthermore indicate that $T_c/g^2$ approaches 
a nonzero constant in the continuum limit~\cite{Athenodorou:2018sab}. Combining this result 
with the G\"opfert-Mack scaling of the string tension gives
\begin{equation}
\frac{T_c}{\sqrt{\sigma}}\rightarrow\infty. \label{eq:WilsonT}
\end{equation}
The finite value found in \Cref{tab:sigmaU1lam} therefore does not display the scale 
hierarchy characteristic of the weak-coupling Wilson theory. This observation is suggestive 
because the spin-$\frac12$ QLM constitutes a different microscopic regularization, with only 
a two-dimensional Hilbert space on each link. Although the two formulations share the same 
$U(1)$ centre symmetry and the same finite-temperature BKT universality, their zero-temperature 
confining spectra need not display the same scale hierarchy. Moreover, the approximate constancy 
of $T_c/\sqrt{\sigma}$ over the presently studied range of $\lambda$ provides evidence for a 
stable scaling regime, not necessarily a continuum limit. This qualification is relevant 
because the zero-temperature transition separating the two crystalline confining phases is 
weakly first order and hence does not automatically provide a divergent correlation length.

 A measurement of the lightest glueball mass would provide a particularly useful discriminator. 
In the conventional Wilson theory one expects $\frac{m_G}{\sqrt{\sigma}} \rightarrow 0$, whereas 
the continuum three-dimensional $Z_2$ gauge theory gives~\cite{Caselle:2007yc}
\begin{equation} \label{eq:GLmass}
 \left.\frac{m_{0^{++}}}{\sqrt{\sigma}}\right|_{Z_2} =3.037(16).
\end{equation}
 If the pure QLM possesses an order-one value of $m_G/\sqrt{\sigma}$ that remains stable 
along a trajectory of increasing correlation length, this would constitute substantially 
stronger evidence that its continuum scaling differs from the G\"opfert-Mack limit. Similarly, 
a value close to eq.~\ref{eq:GLmass} in the theory with annealed charges would strengthen the 
proposed relation between its long-distance confining sector and that of the conventional 
$Z_2$ gauge theory. The present results thus point toward a potentially different infrared 
realization of confinement in the U(1) QLM, while leaving the precise continuum interpretation 
as an important question for future investigation.

\section{Conclusions and outlook} \label{sec:concl}
In this work we have investigated the finite temperature deconfinement transition in 
spin-$\frac{1}{2}$ $U(1)$ QLM in $(2+1)$ dimensions both in the absence and presence 
of annealed $\pm2$ charges. Through extensive quantum Monte Carlo simulations we have 
characterized the critical behaviours of the two theories and studied the properties 
of confining strings across the deconfining transitions.

 We have first considered the pure $U(1)$ QLM for which the global centre symmetry is 
also $U(1)$. We find strong numerical evidence that its finite-temperature deconfinement 
transition belongs to the Berezinskii–Kosterlitz–Thouless (BKT) universality class in 
accordance with the Svetitsky-Yaffe conjecture. In particular, as would be expected, 
the entire phase below the critical coupling, $\beta_c$, demonstrates a power law 
correlation, while above $\beta_c$, the correlations decay exponentially. 

The situation changes qualitatively upon introducing thermally fluctuating (annealed) 
$\pm 2$ charges. These charges reduce the global $U(1)$ centre symmetry to its $Z_2$ 
subgroup and introduce a relevant perturbation to the effective theory describing the 
BKT transition in the pure $U(1)$ gauge theory. Our analysis of the susceptibilities, 
Binder ratios and scaling collapses provides evidence for a transition governed by weak 
universality where the critical exponent $\eta$ and the ratios of critical exponents 
remain consistent with their two-dimensional Ising values but the correlation-length 
exponent $\nu$ varies continuously with the microscopic coupling. In our previous work 
\cite{SauSenBanerjee2023}, using scaling property of connected susceptibility, we showed 
that the ratio $\frac{\gamma}{\nu}$ (and hence $\eta$) is the same as the two-dimensional 
Ising universality. In this work, using the scaling of the order parameter, we show that 
the magnetic critical exponent is also consistent with the two-dimensional Ising value. 
The value of the critical exponent $\nu$, estimated from the slope of first derivatives 
of different Binder ratios at $\beta_c$, is different from that of the two-dimensional 
Ising value and varies continuously with the microscopic coupling $\lambda$. Also, we 
have obtained consistent estimates of $\nu$ from the scaling collapse analyses of different 
Binder ratios and total susceptibility, further supporting the weak universality class. 
These observations demonstrate that the reduction of the centre symmetry to $Z_2$ does 
not necessarily imply conventional Ising criticality: the presence of some additional 
marginal coupling can lead to weak universality class which necessarily refines the 
Svetitsky-Yaffe conjecture.

 An important aspect of our work is the investigation of the confining strings and 
their evolution with temperature. In the pure $U(1)$ QLM, a $(-1,+1)$ test charge pair 
is connected by a flux tube that is fractionalized into two distinct strands. A flux 
tube connecting $(-2,+2)$ test charge pair, on the other hand, splits into four strands.  
These strands can be understood as the interfaces between distinct crystalline confined 
configurations. The energy stored inside the strand increases linearly with the distance 
between the test charge pair which is characterized by string tension. In the confined 
phase the strands have a narrow width and consequently the string tension is non-zero. 
As the temperature increases, thermal fluctuations broaden the width of the individual 
strands and progressively smoothen their microscopic structure. We have approximately 
calculated the interaction among the strands and observed that it is maximum close to 
the phase transition which is due to the maximum broadening of the strands near phase 
transition. Above the critical temperature the strands completely dissolve and the string 
tension vanishes.

 The presence of annealed $\pm2$ charges changes the scenario for $(-2,+2)$ test charge 
pair, which undergo screening. An external $\pm2$ source-charge is immediately screened 
by the thermally fluctuating $\pm2$ charges leading to non-existence of any flux tube 
between the charge pair even deep in the confined phase. In contrast, the $\pm1$ source 
charge can not be screened by thermally fluctuating $\pm2$ charges so that there exists 
a confining flux tube between $(-1,+1)$ charge pair in low temperature that is frcationalized 
into two strands like the pure $U(1)$ QLM. As the temperature increases these strands are 
broadened and ultimately lose their distinct identities as the deconfined transition is 
approached.

 We have further compared the thermal and confining energy scales through the dimensionless 
ratio $\frac{T_c}{\sqrt{\sigma}}$. For the theory with annealed charges, the ratio is 
close to the continuum value of the conventional three-dimensional $Z_2$ lattice gauge theory. 
This suggests that, despite their different microscopic formulations, the two theories exhibit 
a similar quantitative relationship between their deconfinement and fundamental-string 
scales. In contrast, the pure $U(1)$ QLM yields a different value of this ratio, 
approximately 1.49, over the range of couplings studied. This difference is particularly 
interesting because the pure QLM and the theory with annealed charges have different 
fractionalized confining structures, even though both originate from the same microscopic 
quantum link model.

 Our findings raise several interesting questions for future investigations. The key next
step would be identify the key cause of the weak-universality by investigating the finite temperature 
effective field theory quantitatively. Concretely, one would like to isolate the marginal 
couplings responsible for weak universality in the presence of annealed $\pm2$ charges.
In particular, a deformed Wilsonian LGT was also shown to have the crystalline confined phase
as the present model \cite{Banerjee:2013dda,Banerjee:2023pnb}. It is worthwhile to ask
whether doping the model with annealed even charges also gives rise to the phenomena of weak
universality. 
In terms of observables, a detailed study of the distribution of order parameter across the 
phase transition, particularly, the angular deformation of the order parameter distribution 
for pure $U(1)$ QLM might yield clues as to the nature of the relevant operators. 
It would also be useful to understand the connection of the QLM towards the modified Wilsonian
$U(1)$ theory obtained by adding monopoles \cite{Athenodorou:2026dew,Mariani:2026piy}.
The finite temperature behaviour of integer-valued quantum-spins as gauge fields provides
another distinct research direction.

\paragraph{Acknowledgements}. We acknowledge conversations with ChatGPT to make the 
construction of the effective field theory concrete. I.S. acknowledges the financial 
support received from the Anusandhan National Research Foundation (ANRF), Government 
of India, under the National Postdoctoral Fellowship (NPDF) scheme (File. No. PDF/2025/001525).
D.B. acknowledges STFC funded Particle Physics Theory Consolidated Grant ST/C004240/1 
and continued support from the Alexander von Humboldt Foundation (Germany) in the 
context of the research fellowship for experienced researchers.

\appendix 
\section{Centre symmetry in the Euclidean formulation} \label{sec:CenterS}
 The centre symmetry is cleanly demonstrated in the Euclidean formulation. In this
 case the theory is defined in terms of gauge links in Euclidean space $U_{x,\mu}$, 
 and the Polyakov loop is defined as the path-ordered product of the time-like links
 spanning $\beta = L_t \cdot a$:
 \begin{equation}
   P(\mathbf{x}) = \prod_{\tau=0}^{L_t-1} U_{(\mathbf{x},\tau),\hat 0}.
 \end{equation}
 Choosing an arbitrary Euclidean time slice $\tau_0$, the $U(1)$ centre transformation is
 \begin{equation} \label{eq:U1-centre}
    U_{(\mathbf{x},\tau_0),\hat 0} \longrightarrow z U_{(\mathbf{x},\tau_0),\hat 0},
    \qquad z=e^{i\alpha}\in U(1),
 \end{equation}
 for every spatial position $\mathbf{x}$, while all other links remain unchanged. 
 The plaquette action is invariant because every temporal plaquette contains one 
 transformed link and one oppositely oriented transformed link. The Polyakov loop, 
 however, transforms as
 \begin{equation} \label{eq:polyakov-U1}
   P(\mathbf{x}) \longrightarrow zP(\mathbf{x}) = e^{i\alpha}P(\mathbf{x}).
 \end{equation}
 Thus, in the pure gauge theory, the global centre symmetry is $U(1)$. In the presence of dynamical
 or annealed charges $Q = \pm 2$, configurations containing charge-two world-lines winding around 
 the Euclidean time direction are part of the thermal ensemble. Such a winding world-line couples 
 to the gauge field in the charge-two representation and therefore acquires the phase $z^2$
 under the centre transformation. The full partition function is consequently invariant only for 
 centre elements satisfying $z^2=1$. Thus, the $U(1)$ centre symmetry is explicitly reduced to its 
 $\mathbb{Z}_2$ subgroup, $z = \pm 1$.

 Note that even in theory with the annealed charges, the fundamental Polyakov loop remains charged 
 under this residual symmetry and transforms as
 \begin{equation}
    P(\mathbf{x})\longrightarrow -P(\mathbf{x})
 \end{equation}
 under its nontrivial element.

\section{Scaling collapse plots for $\lambda=-0.9$ and $-0.8$}
\label{app:collapse}

\begin{figure}[!tbh]
    \centering
    \includegraphics[width=0.49\linewidth]{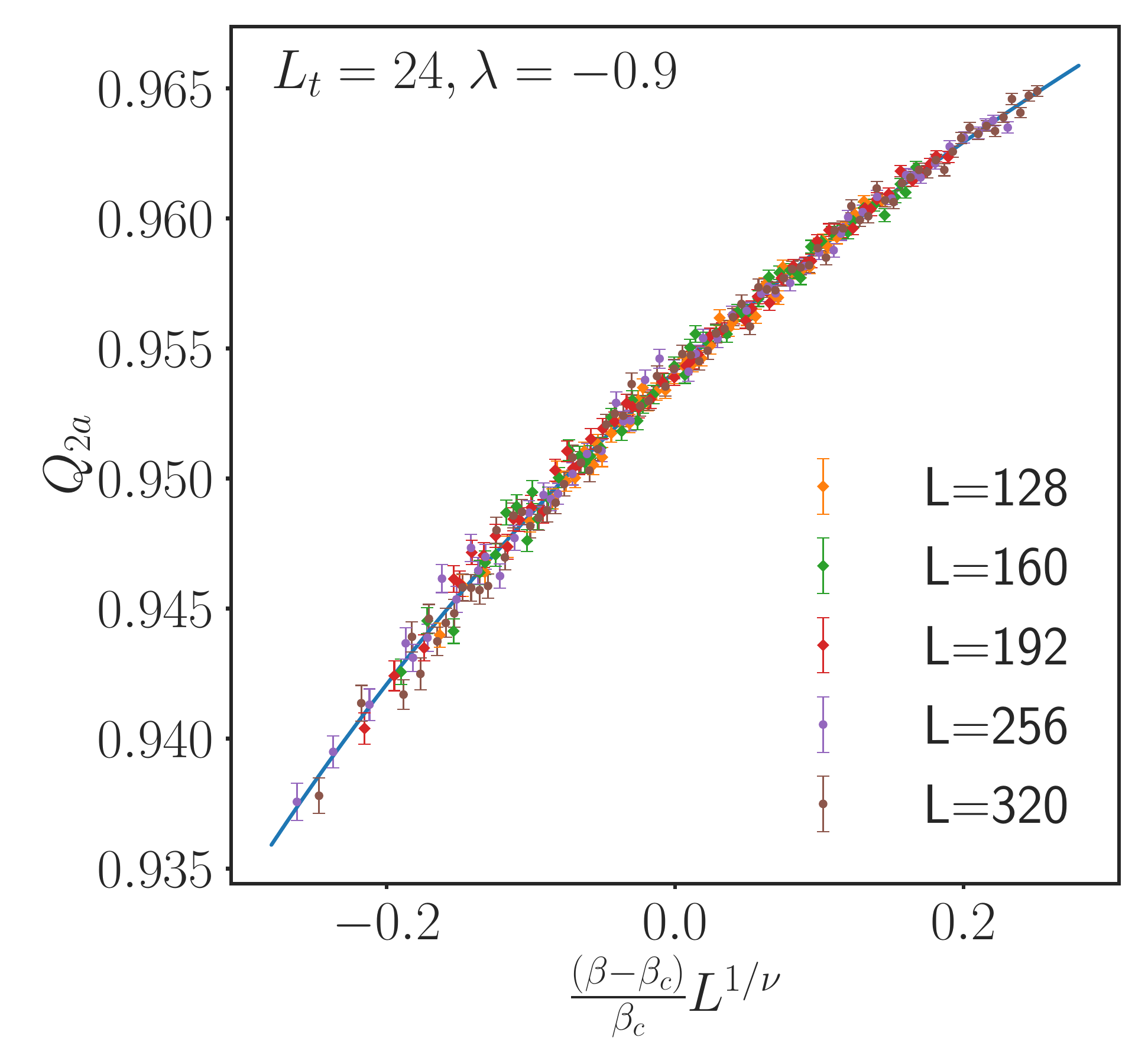}
    \includegraphics[width=0.49\linewidth]{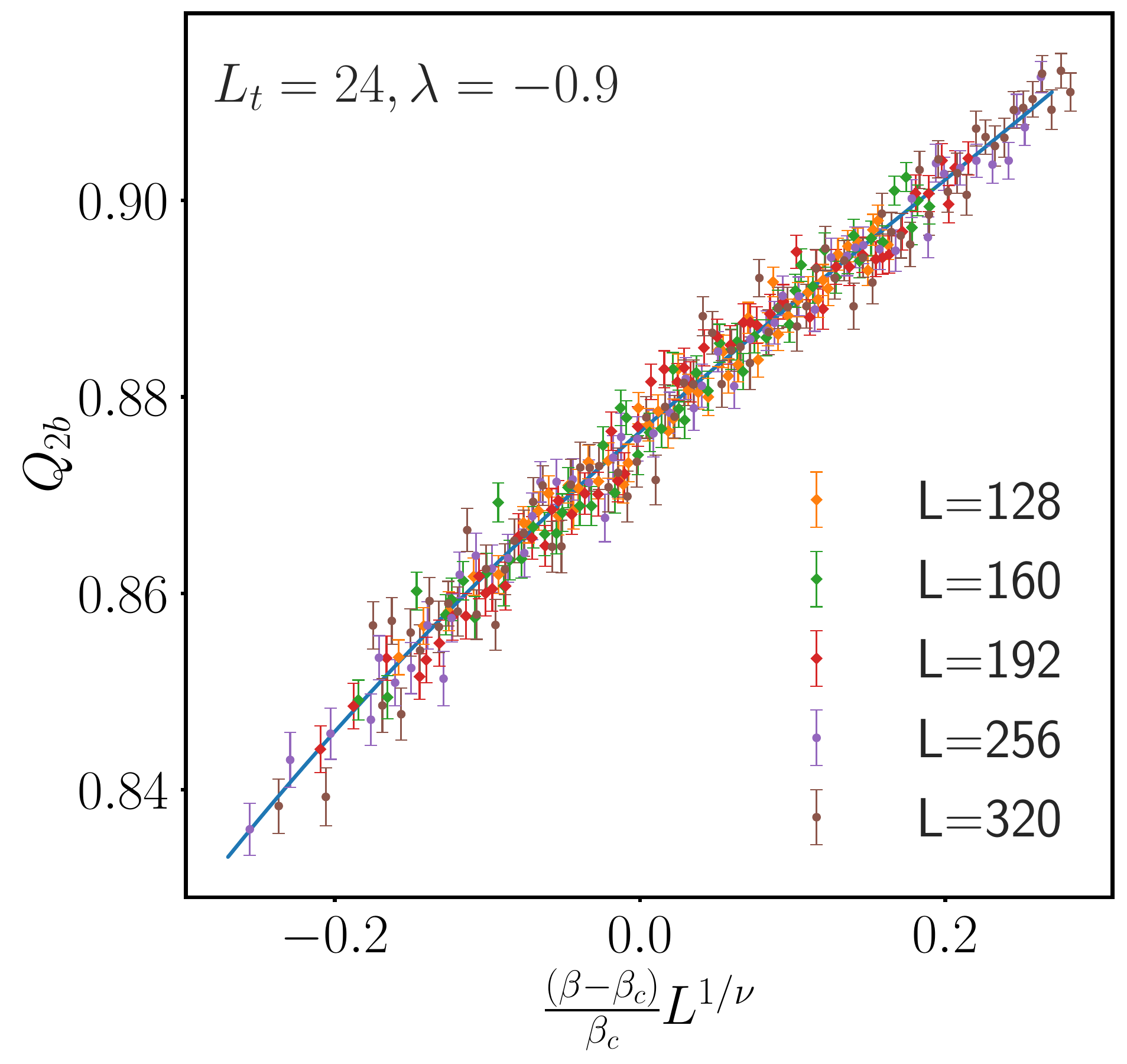}
    \includegraphics[width=0.49\linewidth]{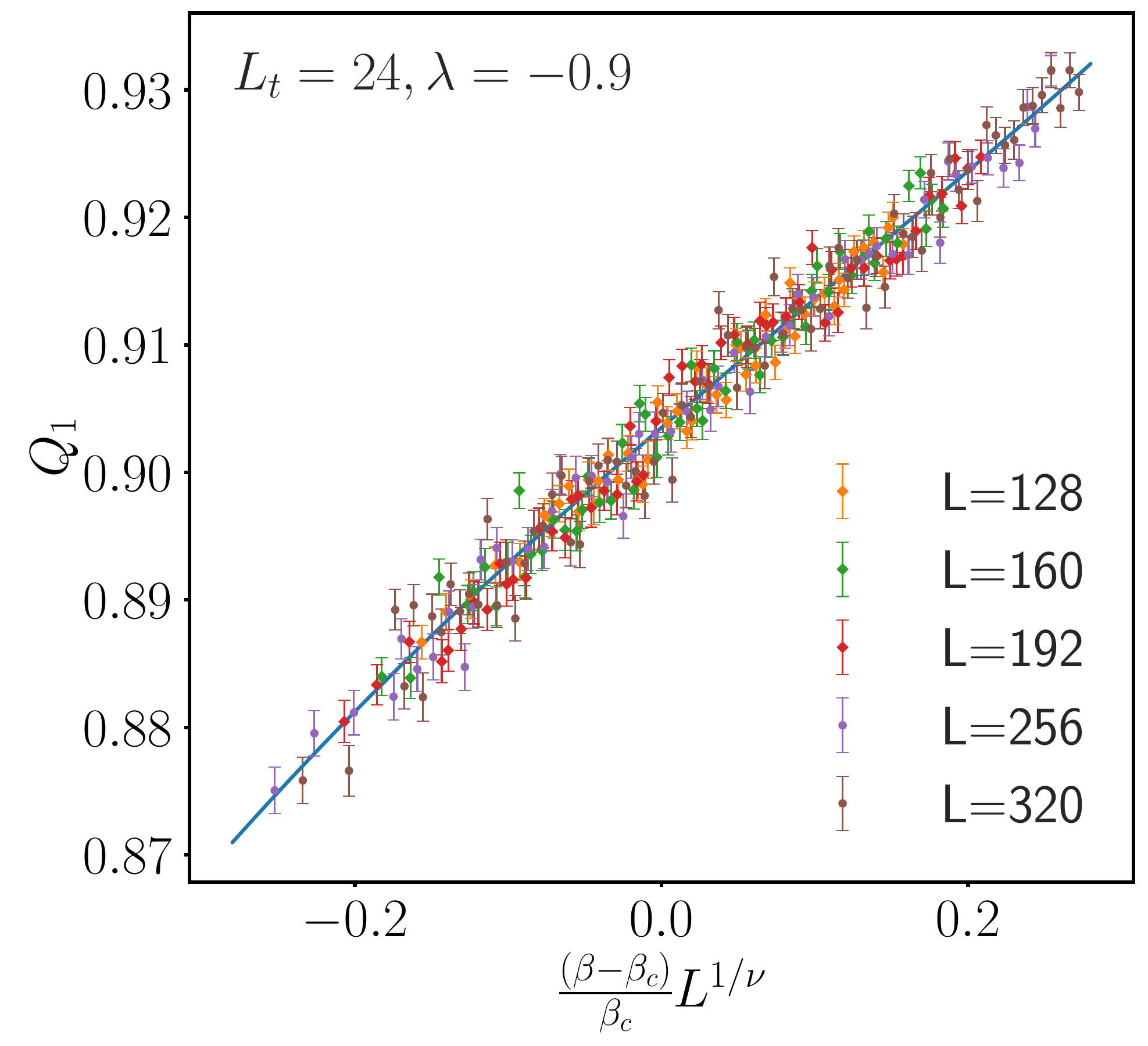}
    \includegraphics[width=0.49\linewidth]{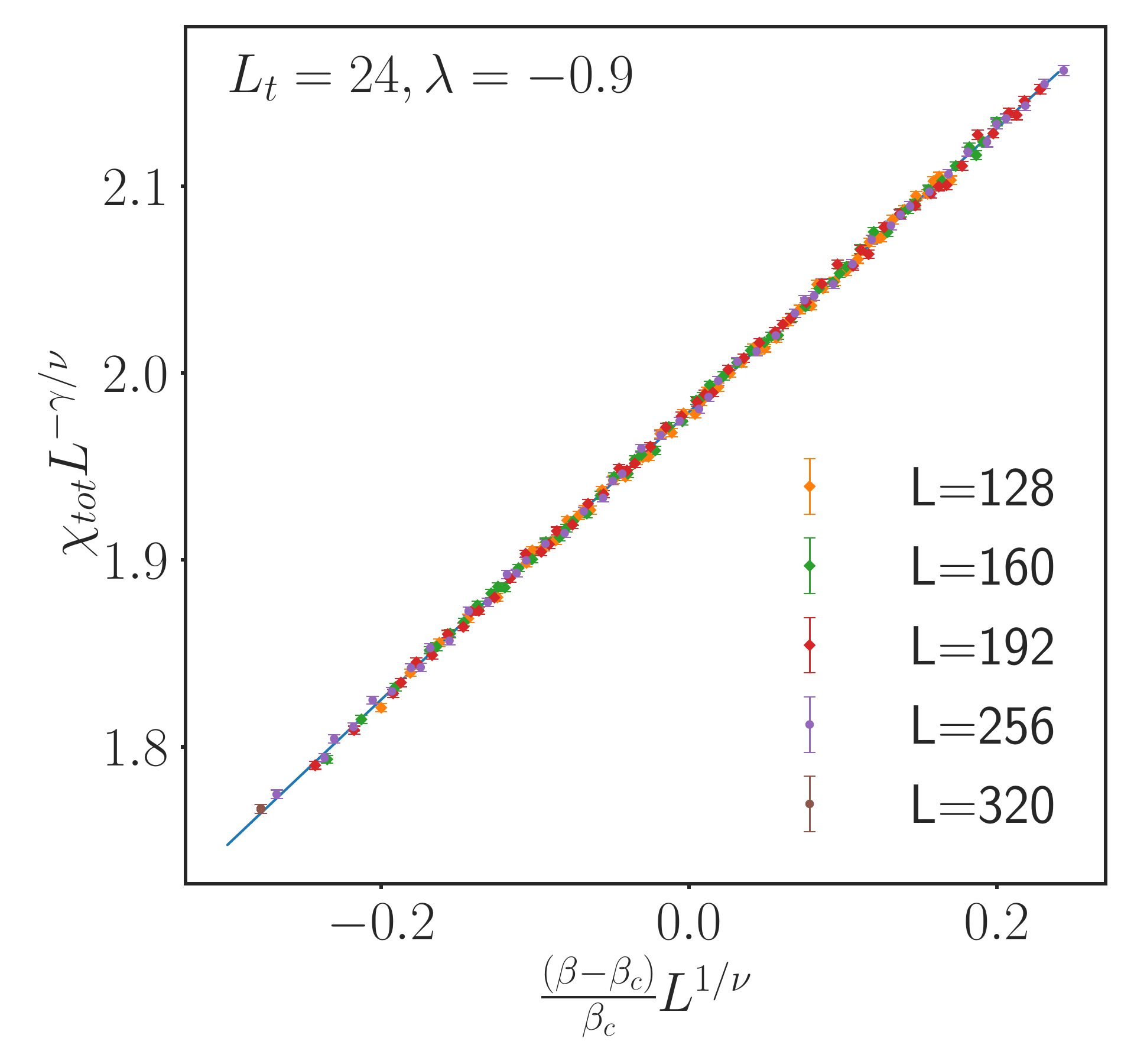}
    \caption{Scaling collapse plots at $\lambda=-0.9$ ($L_t=24$)}
    \label{fig:collapse_app09}
\end{figure}

\begin{figure}[!tbh]
    \centering
    \includegraphics[width=0.49\linewidth]{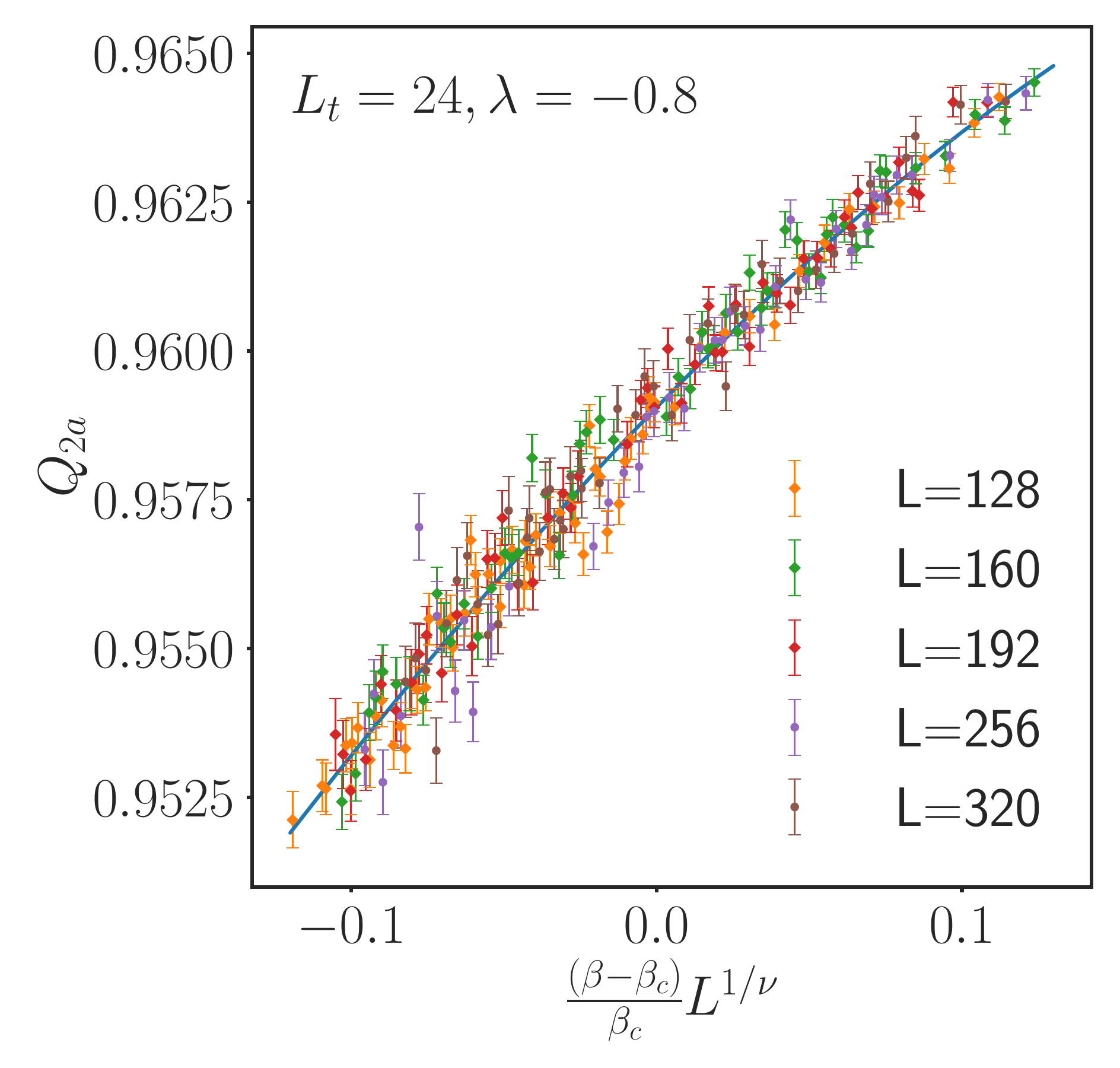}
    \includegraphics[width=0.49\linewidth]{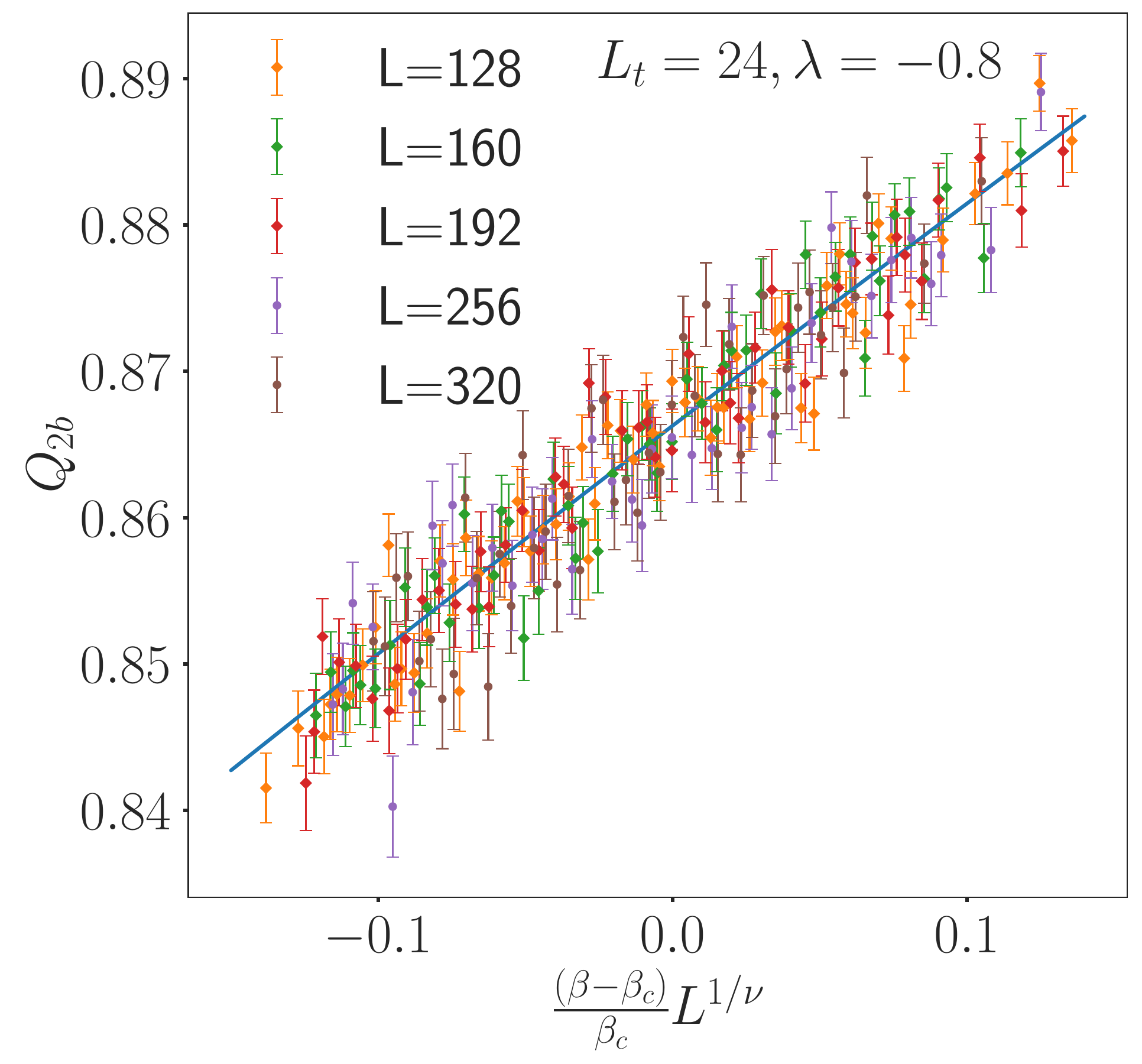}
    \includegraphics[width=0.49\linewidth]{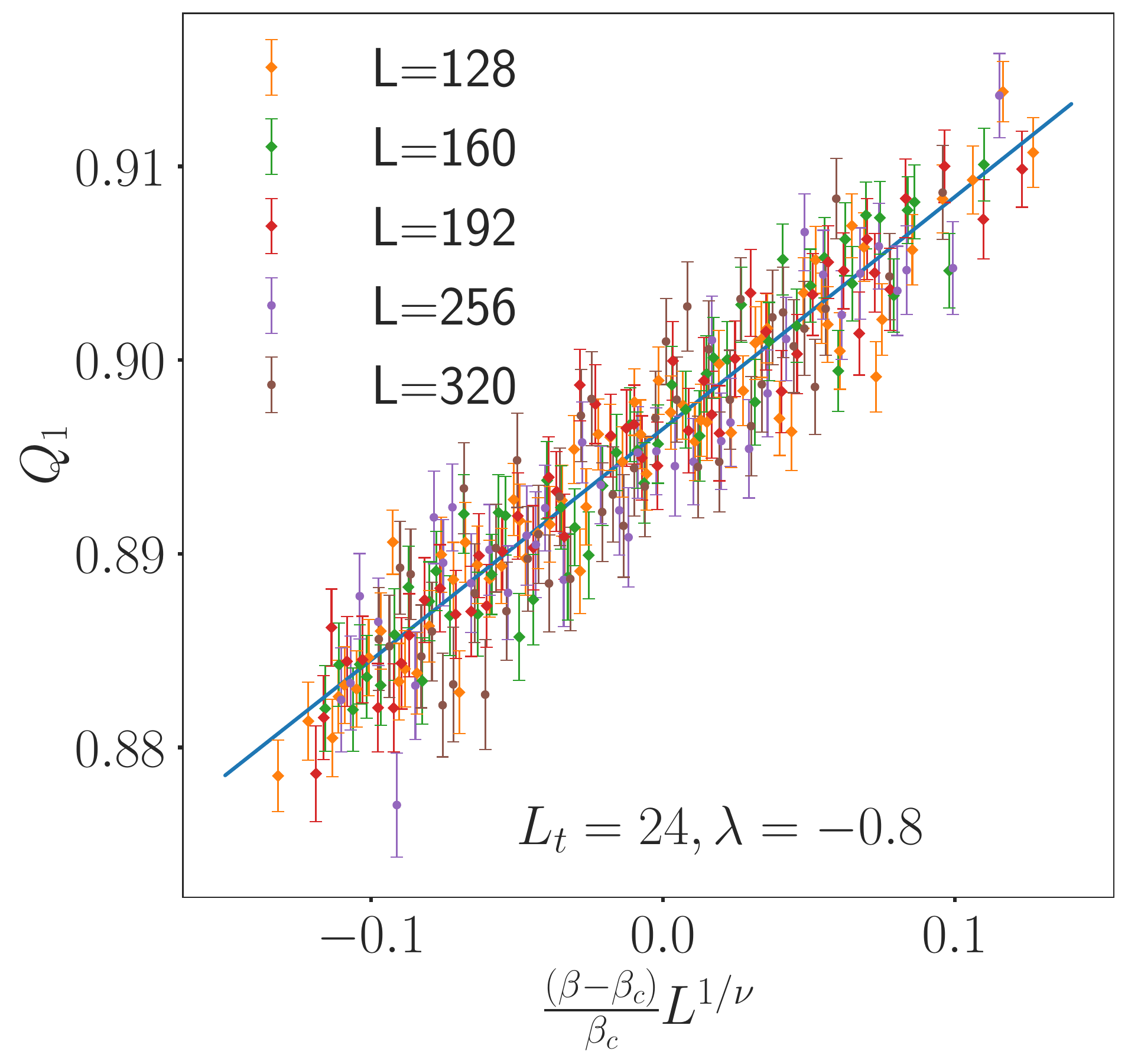}
    \includegraphics[width=0.49\linewidth]{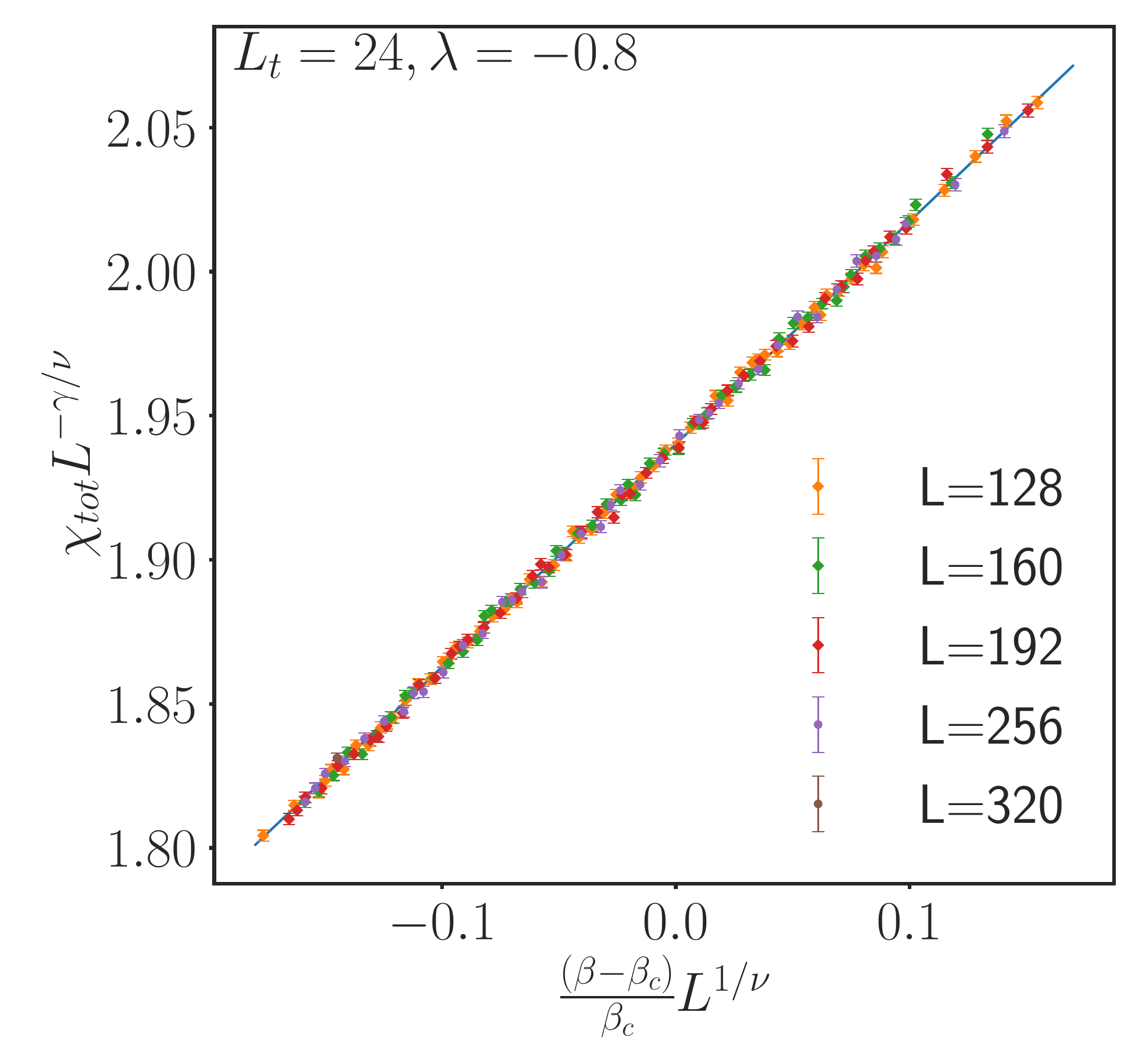}
    \caption{Scaling collapse plots at $\lambda=-0.8$ ($L_t=24$)}
    \label{fig:collapse_app08}
\end{figure}

\section{Algorithm to implement $(-1,+1)$ test charge pair}
\label{app:algQ1}
The Monte Carlo algorithm involving the test charge pair $(Q,\bar{Q})$ involves finding an initial height configuration with $Q$ $[\bar{Q}]$ charge placed at $(x_Q,y_Q)$ $[(x_Q+r,y_Q)]$. Implementing the $(-2,+2)$ test charge pair is straightforward. We start from an initial flux configuration with $-2$ $(+2)$ charge placed at $(x_Q,\frac{L}{2}-1)$ $[(x_Q+r,\frac{L}{2}-1)]$. We then find the corresponding height configuration using Eq.~\ref{eq:flux_h} in the main text. Note that the plaquettes around the $\pm2$ charges are unflippable. During Monte Carlo sweeps, we reject moves that involve flipping the height variables on these plaquettes.

Implementing the $(-1,+1)$ test charge pair requires careful consideration because the conversion rule for the flux to height variable (Eq.~\ref{eq:flux_h} in the main text) does not hold for $\pm1$ charges; one has to locally modify Eq.~\ref{eq:flux_h} to 
\begin{equation}
    E_{x,\hat{i}}= \mathcal{C}\left(\left[ h_{\tilde{x}}^X-h_{\tilde{x}+\tilde{i}-\tilde{1}-\tilde{2}}^{X^\prime} \right]
  ~~{\rm mod2} \right); ~~ {\rm X,X}^\prime \in \{{\rm A},{\rm B}\}
\end{equation}
 for any one link around the site to accommodate $-1/+1$ charge. However, this modification induces ambiguity in one of the other three links around the site that is connected by the modified link. One has to modify one of these three links around this connected site. Thus, one needs a series of modifications starting from one of the $\pm1$ charges until one reaches the other charge. It turns out that the easiest way to do that is to modify Eq.~\ref{eq:flux_h} for the links on the horizontal line connecting $\pm1$ charges. (This line connecting $-1$ and $+1$ charges is called `Dirac string'.) Unlike $\pm2$ charges, two of the four plaquettes around $\pm1$ charge can be flippable, in principle. We have to take care while flipping the heights around $\pm1$ charges during Monte Carlo sweeps. We discuss it below.

\begin{figure}
    \centering
    \includegraphics[scale=0.3]{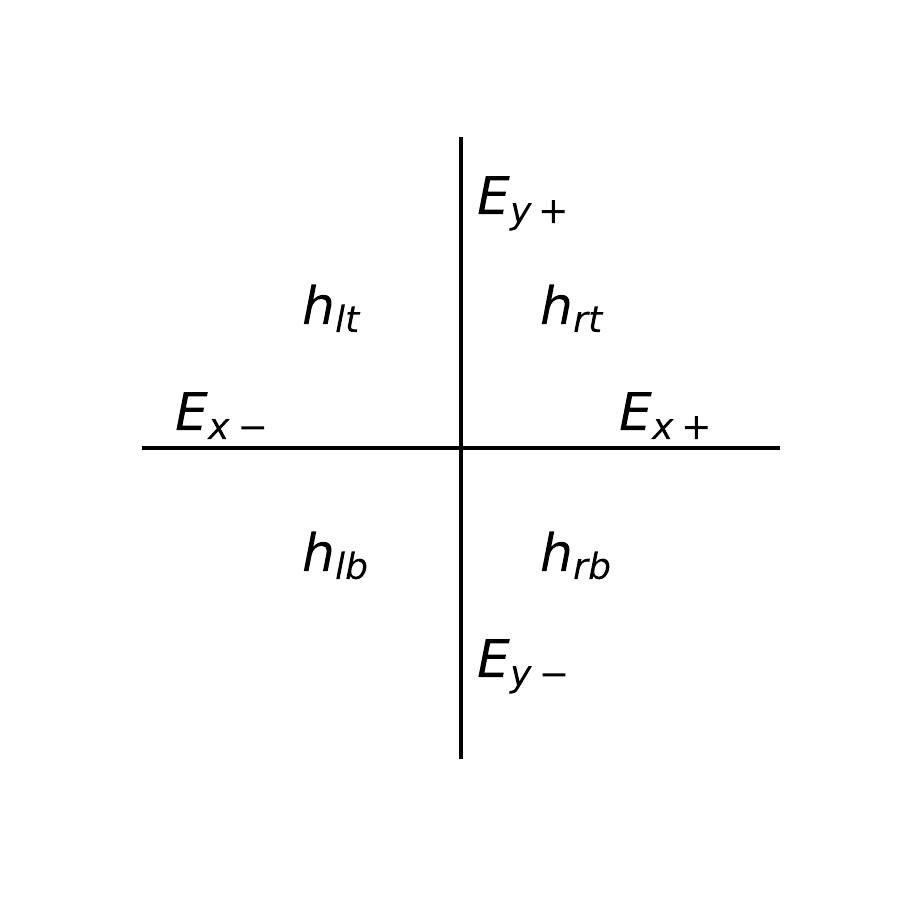}
    \caption{Schematic depiction of fluxes and heights around $\pm1$ charge.}
    \label{fig:appC}
\end{figure}

 Let's denote by $E_{x+}$, $E_{x-}$, $E_{y+}$ and $E_{y-}$ the flux on the right, left, top and bottom link of a site, respectively. Also denote by $h_{rt}$, $h_{rb}$, $h_{lt}$ and $h_{lb}$ the height variables on the right-top, right-bottom, left-top and left-bottom respectively (see Fig.~\ref{fig:appC}). In computation we set the height variables $\pm1$. Then the fluxes around $\pm1$ charges can be written as
 \begin{align}\label{appEq:h_to_f}
     2E_{x+} = -s_R h_{rt}h_{rb},~2E_{x-}=-s_L h_{lt}h_{lb}\\
     2E_{y+} = -h_{lt}h_{rt},~2E_{y-} = -h_{lb}h_{rb},
 \end{align}
 where $s_R=1(-1)$ implies the absence (presence) of Dirac string on the right link. Similarly for $s_L$ on the left link. The charge, in terms of heights, at this site is given by
 \begin{align}\label{appEq:Q}
     Q &= E_{x+}+E_{y+}-E_{x-}-E_{y-}\\
       &= \frac{1}{2}[h_{lb}(s_L h_{lt}+h_{rb}) - h_{rt}(s_R h_{rb}+h_{lt})]\\
       &= \frac{1}{2}[h_{lt}(s_L h_{lb}-h_{rt})-h_{rb}(s_R h_{rt}-h_{lb})].
 \end{align}
 Now if $h_{lb}$ flips (i.e. $h_{lb}\rightarrow -h_{lb}$), the change in the charge is $dQ=h_{lb}(s_L h_{lt}+h_{rb})$. So if we want $Q$ to be unchanged and at the same time flip $h_{lb}$, we have to satisfy $dQ=0$. Similarly to flip $h_{rt}$ we need to check $(s_R h_{rb}+h_{lt})$, to flip $h_{lt}$ we need to check $(s_L h_{lb}-h_{rt})$ and to flip $h_{rb}$ we need to check $(s_R h_{rt}-h_{lb})$.

\bibliography{ref-DeConf.bib}

@article{Zohar2016,
  author        = {Zohar, Erez and Cirac, J. Ignacio and Reznik, Benni},
  title         = {Quantum Simulations of Lattice Gauge Theories Using
                   Ultracold Atoms in Optical Lattices},
  journal       = {Rep. Prog. Phys.},
  volume        = {79},
  number        = {1},
  pages         = {014401},
  year          = {2016},
  doi           = {10.1088/0034-4885/79/1/014401},
  eprint        = {1503.02312},
  archivePrefix = {arXiv},
  primaryClass  = {quant-ph}
}

@article{Banuls2020,
  author        = {Ba{\~n}uls, Mari Carmen and others},
  title         = {Simulating Lattice Gauge Theories within Quantum
                   Technologies},
  journal       = {Eur. Phys. J. D},
  volume        = {74},
  number        = {8},
  pages         = {165},
  year          = {2020},
  doi           = {10.1140/epjd/e2020-100571-8},
  eprint        = {1911.00003},
  archivePrefix = {arXiv},
  primaryClass  = {quant-ph}
}

@article{Bauer2023,
  author        = {Bauer, Christian W. and others},
  title         = {Quantum Simulation for High-Energy Physics},
  journal       = {PRX Quantum},
  volume        = {4},
  number        = {2},
  pages         = {027001},
  year          = {2023},
  doi           = {10.1103/PRXQuantum.4.027001},
  eprint        = {2204.03381},
  archivePrefix = {arXiv},
  primaryClass  = {quant-ph}
}

@article{Martinez2016,
  author  = {Martinez, Esteban A. and Muschik, Christine A. and
             Schindler, Philipp and Nigg, Daniel and Erhard, Alexander and
             Heyl, Markus and Hauke, Philipp and Dalmonte, Marcello and
             Monz, Thomas and Zoller, Peter and Blatt, Rainer},
  title   = {Real-Time Dynamics of Lattice Gauge Theories with a
             Few-Qubit Quantum Computer},
  journal = {Nature},
  volume  = {534},
  number  = {7608},
  pages   = {516--519},
  year    = {2016},
  doi     = {10.1038/nature18318}
}

@article{Mil2020,
  author  = {Mil, Alexander and Zache, Torsten V. and Hegde, Apoorva and
             Xia, Andy and Bhatt, Rohit P. and Oberthaler, Markus K. and
             Hauke, Philipp and Berges, J{\"u}rgen and
             Jendrzejewski, Fred},
  title   = {A Scalable Realization of Local {$U(1)$} Gauge Invariance
             in Cold Atomic Mixtures},
  journal = {Science},
  volume  = {367},
  number  = {6482},
  pages   = {1128--1130},
  year    = {2020},
  doi     = {10.1126/science.aaz5312}
}

@article{Yang2020,
  author  = {Yang, Bing and Sun, Hui and Ott, Robert and Wang, Han-Yi and
             Zache, Torsten V. and Halimeh, Jad C. and Yuan, Zhen-Sheng and
             Hauke, Philipp and Pan, Jian-Wei},
  title   = {Observation of Gauge Invariance in a 71-Site
             Bose--Hubbard Quantum Simulator},
  journal = {Nature},
  volume  = {587},
  number  = {7834},
  pages   = {392--396},
  year    = {2020},
  doi     = {10.1038/s41586-020-2910-8}
}

@article{Wilson1974,
  author  = {Wilson, Kenneth G.},
  title   = {Confinement of Quarks},
  journal = {Phys. Rev. D},
  volume  = {10},
  number  = {8},
  pages   = {2445--2459},
  year    = {1974},
  doi     = {10.1103/PhysRevD.10.2445}
}

@article{Kogut1979,
  author  = {Kogut, John B.},
  title   = {An Introduction to Lattice Gauge Theory and Spin Systems},
  journal = {Rev. Mod. Phys.},
  volume  = {51},
  number  = {4},
  pages   = {659--713},
  year    = {1979},
  doi     = {10.1103/RevModPhys.51.659}
}

@article{LeeNagaosaWen2006,
  author        = {Lee, Patrick A. and Nagaosa, Naoto and Wen, Xiao-Gang},
  title         = {Doping a Mott Insulator: Physics of High-Temperature
                   Superconductivity},
  journal       = {Rev. Mod. Phys.},
  volume        = {78},
  number        = {1},
  pages         = {17--85},
  year          = {2006},
  doi           = {10.1103/RevModPhys.78.17},
  eprint        = {cond-mat/0410445},
  archivePrefix = {arXiv}
}

@article{Hermele2004,
  author        = {Hermele, Michael and Fisher, Matthew P. A. and
                   Balents, Leon},
  title         = {Pyrochlore Photons: The {$U(1)$} Spin Liquid in a
                   {$S=1/2$} Three-Dimensional Frustrated Magnet},
  journal       = {Phys. Rev. B},
  volume        = {69},
  number        = {6},
  pages         = {064404},
  year          = {2004},
  doi           = {10.1103/PhysRevB.69.064404},
  eprint        = {cond-mat/0305401},
  archivePrefix = {arXiv}
}

@article{SavaryBalents2017,
  author        = {Savary, Lucile and Balents, Leon},
  title         = {Quantum Spin Liquids: A Review},
  journal       = {Rep. Prog. Phys.},
  volume        = {80},
  number        = {1},
  pages         = {016502},
  year          = {2017},
  doi           = {10.1088/0034-4885/80/1/016502},
  eprint        = {1601.03742},
  archivePrefix = {arXiv},
  primaryClass  = {cond-mat.str-el}
}

@article{Castelnovo2008,
  author        = {Castelnovo, Claudio and Moessner, Roderich and
                   Sondhi, Shivaji L.},
  title         = {Magnetic Monopoles in Spin Ice},
  journal       = {Nature},
  volume        = {451},
  number        = {7174},
  pages         = {42--45},
  year          = {2008},
  doi           = {10.1038/nature06433},
  eprint        = {0710.5515},
  archivePrefix = {arXiv},
  primaryClass  = {cond-mat.str-el}
}

@article{SavaryBalents2013,
  author        = {Savary, Lucile and Balents, Leon},
  title         = {Spin Liquid Regimes at Nonzero Temperature in
                   Quantum Spin Ice},
  journal       = {Phys. Rev. B},
  volume        = {87},
  number        = {20},
  pages         = {205130},
  year          = {2013},
  doi           = {10.1103/PhysRevB.87.205130},
  eprint        = {1301.5643},
  archivePrefix = {arXiv},
  primaryClass  = {cond-mat.str-el}
}

@article{Polyakov1977,
  author  = {Polyakov, Alexander M.},
  title   = {Quark Confinement and Topology of Gauge Groups},
  journal = {Nucl. Phys. B},
  volume  = {120},
  pages   = {429--458},
  year    = {1977},
  doi     = {10.1016/0550-3213(77)90086-4}
}

@article{Polyakov1978,
  author  = {Polyakov, Alexander M.},
  title   = {Thermal Properties of Gauge Fields and Quark Liberation},
  journal = {Phys. Lett. B},
  volume  = {72},
  pages   = {477--480},
  year    = {1978},
  doi     = {10.1016/0370-2693(78)90737-2}
}

@article{Susskind1979,
  author  = {Susskind, Leonard},
  title   = {Lattice Models of Quark Confinement at High Temperature},
  journal = {Phys. Rev. D},
  volume  = {20},
  pages   = {2610--2618},
  year    = {1979},
  doi     = {10.1103/PhysRevD.20.2610}
}

@article{McLerranSvetitsky1981,
  author  = {McLerran, Larry D. and Svetitsky, Benjamin},
  title   = {Quark Liberation at High Temperature:
             A Monte Carlo Study of {$SU(2)$} Gauge Theory},
  journal = {Phys. Rev. D},
  volume  = {24},
  pages   = {450--460},
  year    = {1981},
  doi     = {10.1103/PhysRevD.24.450}
}

@article{SvetitskyYaffe1982,
  author  = {Svetitsky, Benjamin and Yaffe, Laurence G.},
  title   = {Critical Behavior at Finite-Temperature
             Confinement Transitions},
  journal = {Nucl. Phys. B},
  volume  = {210},
  pages   = {423--447},
  year    = {1982},
  doi     = {10.1016/0550-3213(82)90172-9}
}

@article{CaselleHasenbusch1996,
  author        = {Caselle, Michele and Hasenbusch, Martin},
  title         = {Deconfinement Transition and Dimensional Crossover
                   in the Three-Dimensional Gauge Ising Model},
  journal       = {Nucl. Phys. B},
  volume        = {470},
  pages         = {435--453},
  year          = {1996},
  doi           = {10.1016/0550-3213(96)00161-7},
  eprint        = {hep-lat/9511015},
  archivePrefix = {arXiv}
}

@article{Balog:2000ra,
    author = "Balog, Janos",
    title = "{Kosterlitz-Thouless theory and lattice artifacts}",
    eprint = "hep-lat/0011078",
    archivePrefix = "arXiv",
    reportNumber = "MPI-PHT-2000-44",
    doi = "10.1088/0305-4470/34/25/306",
    journal = "J. Phys. A",
    volume = "34",
    pages = "5237--5250",
    year = "2001"
}

@article{Hasenbusch:2005xm,
    author = "Hasenbusch, Martin",
    title = "{The Two dimensional XY model at the transition temperature: A High precision Monte Carlo study}",
    eprint = "cond-mat/0502556",
    archivePrefix = "arXiv",
    doi = "10.1088/0305-4470/38/26/003",
    journal = "J. Phys. A",
    volume = "38",
    pages = "5869--5884",
    year = "2005"
}

@article{KogutSusskind1975,
  author  = {Kogut, John and Susskind, Leonard},
  title   = {Hamiltonian Formulation of Wilson's Lattice Gauge Theories},
  journal = {Phys. Rev. D},
  volume  = {11},
  pages   = {395--408},
  year    = {1975},
  doi     = {10.1103/PhysRevD.11.395}
}

@article{Banerjee2013,
    author = "Banerjee, D. and Jiang, F. -J. and Widmer, P. and Wiese, U. -J.",
    title = "{The (2 + 1)-d U(1) quantum link model masquerading as deconfined criticality}",
    eprint = "1303.6858",
    archivePrefix = "arXiv",
    primaryClass = "cond-mat.str-el",
    doi = "10.1088/1742-5468/2013/12/P12010",
    journal = "J. Stat. Mech.",
    volume = "1312",
    pages = "P12010",
    year = "2013"
}

@article{Shannon2004,
  title = {Cyclic exchange, isolated states, and spinon deconfinement in an $XXZ$ Heisenberg model on the checkerboard lattice},
  author = {Shannon, Nic and Misguich, Gr\'egoire and Penc, Karlo},
  journal = {Phys. Rev. B},
  volume = {69},
  issue = {22},
  pages = {220403(R)},
  numpages = {4},
  year = {2004},
  month = {Jun},
  publisher = {American Physical Society},
  doi = {10.1103/PhysRevB.69.220403},
  url = {https://link.aps.org/doi/10.1103/PhysRevB.69.220403}
}

@article{ZoharBurrello2015,
  author        = {Zohar, Erez and Burrello, Michele},
  title         = {A Formulation of Lattice Gauge Theories
                   for Quantum Simulations},
  journal       = {Phys. Rev. D},
  volume        = {91},
  pages         = {054506},
  year          = {2015},
  doi           = {10.1103/PhysRevD.91.054506},
  eprint        = {1409.3085},
  archivePrefix = {arXiv},
  primaryClass  = {quant-ph}
}

@article{Hackett2019,
  author        = {Hackett, Daniel C. and Howe, Kiel and Hughes, Ciaran
                   and Jay, William and Neil, Ethan T. and Simone, James N.},
  title         = {Digitizing Gauge Fields: Lattice Monte Carlo Results
                   for Future Quantum Computers},
  journal       = {Phys. Rev. A},
  volume        = {99},
  pages         = {062341},
  year          = {2019},
  doi           = {10.1103/PhysRevA.99.062341},
  eprint        = {1811.03629},
  archivePrefix = {arXiv},
  primaryClass  = {quant-ph}
}

@article{ChandrasekharanWiese1997,
  author        = {Chandrasekharan, Shailesh and Wiese, Uwe-Jens},
  title         = {Quantum Link Models:
                   A Discrete Approach to Gauge Theories},
  journal       = {Nucl. Phys. B},
  volume        = {492},
  pages         = {455--474},
  year          = {1997},
  doi           = {10.1016/S0550-3213(97)80041-7},
  eprint        = {hep-lat/9609042},
  archivePrefix = {arXiv}
}

@article{Wiese2022,
  author        = {Wiese, Uwe-Jens},
  title         = {From Quantum Link Models to {D}-Theory:
                   A Resource-Efficient Framework for the Quantum
                   Simulation and Computation of Gauge Theories},
  journal       = {Philos. Trans. R. Soc. A},
  volume        = {380},
  pages         = {20210068},
  year          = {2022},
  doi           = {10.1098/rsta.2021.0068},
  eprint        = {2107.09335},
  archivePrefix = {arXiv},
  primaryClass  = {hep-lat}
}

@article{LiuChandrasekharan2022,
  author        = {Liu, Hanqing and Chandrasekharan, Shailesh},
  title         = {Qubit Regularization and Qubit Embedding Algebras},
  journal       = {Symmetry},
  volume        = {14},
  number        = {2},
  pages         = {305},
  year          = {2022},
  doi           = {10.3390/sym14020305},
  eprint        = {2112.02090},
  archivePrefix = {arXiv},
  primaryClass  = {hep-lat}
}

@article{Zache2023,
  author        = {Zache, Torsten V. and Gonz{\'a}lez-Cuadra, Daniel
                   and Zoller, Peter},
  title         = {Quantum and Classical Spin-Network Algorithms for
                   {$q$}-Deformed Kogut--Susskind Gauge Theories},
  journal       = {Phys. Rev. Lett.},
  volume        = {131},
  pages         = {171902},
  year          = {2023},
  doi           = {10.1103/PhysRevLett.131.171902},
  eprint        = {2304.02527},
  archivePrefix = {arXiv},
  primaryClass  = {quant-ph}
}

@article{Banerjee:2021tfo,
    author = "Banerjee, D. and Caspar, S. and Jiang, F. -J. and Peng, J. -H. and Wiese, U. -J.",
    title = "{Nematic confined phases in the U(1) quantum link model on a triangular lattice: Near-term quantum computations of string dynamics on a chip}",
    eprint = "2107.01283",
    archivePrefix = "arXiv",
    primaryClass = "hep-lat",
    doi = "10.1103/PhysRevResearch.4.023176",
    journal = "Phys. Rev. Res.",
    volume = "4",
    number = "2",
    pages = "023176",
    year = "2022"
}

@article{Berezinskii1971,
  author  = {Berezinskii, V. L.},
  title   = {Destruction of Long-Range Order in One-Dimensional
             and Two-Dimensional Systems Having a Continuous
             Symmetry Group. I. Classical Systems},
  journal = {Sov. Phys. JETP},
  volume  = {32},
  pages   = {493--500},
  year    = {1971}
}

@article{KosterlitzThouless1973,
  author  = {Kosterlitz, J. Michael and Thouless, David J.},
  title   = {Ordering, Metastability and Phase Transitions in
             Two-Dimensional Systems},
  journal = {J. Phys. C},
  volume  = {6},
  pages   = {1181--1203},
  year    = {1973},
  doi     = {10.1088/0022-3719/6/7/010}
}

@article{Kosterlitz1974,
  author  = {Kosterlitz, J. Michael},
  title   = {The Critical Properties of the Two-Dimensional
             {$XY$} Model},
  journal = {J. Phys. C},
  volume  = {7},
  pages   = {1046--1060},
  year    = {1974},
  doi     = {10.1088/0022-3719/7/6/005}
}

@article{Coddington1986,
  author  = {Coddington, P. D. and Hey, A. J. G. and
             Middleton, A. A. and Townsend, J. S.},
  title   = {The Deconfining Transition for Finite-Temperature
             {$U(1)$} Lattice Gauge Theory in
             {$(2+1)$} Dimensions},
  journal = {Phys. Lett. B},
  volume  = {175},
  pages   = {64--68},
  year    = {1986}
}

@article{Borisenko2015,
  author        = {Borisenko, Oleg and Chelnokov, Volodymyr and
                   Gravina, Mario and Papa, Alessandro},
  title         = {Deconfinement and Universality in the Three-Dimensional
                   {$U(1)$} Lattice Gauge Theory at Finite Temperature:
                   Study in the Dual Formulation},
  journal       = {JHEP},
  volume        = {09},
  pages         = {062},
  year          = {2015},
  doi           = {10.1007/JHEP09(2015)062},
  eprint        = {1507.00833},
  archivePrefix = {arXiv},
  primaryClass  = {hep-lat}
}

@article{SauSenBanerjee2023,
  author        = {Sau, Indrajit and Sen, Arnab and Banerjee, Debasish},
  title         = {Weak Universality Induced by {$Q=\pm2e$} Charges
                   at the Deconfinement Transition of a
                   {$(2+1)$}-Dimensional {$U(1)$} Lattice Gauge Theory},
  journal       = {Phys. Rev. Lett.},
  volume        = {130},
  pages         = {071901},
  year          = {2023},
  doi           = {10.1103/PhysRevLett.130.071901},
  eprint        = {2209.14606},
  archivePrefix = {arXiv},
  primaryClass  = {hep-lat}
}

@article{Ashkin1943,
  author  = {Ashkin, Julius and Teller, Edward},
  title   = {Statistics of Two-Dimensional Lattices
             with Four Components},
  journal = {Phys. Rev.},
  volume  = {64},
  pages   = {178--184},
  year    = {1943},
  doi     = {10.1103/PhysRev.64.178}
}

@article{Baxter1971,
       author = {{Baxter}, R.~J.},
        title = "{Eight-Vertex Model in Lattice Statistics}",
      journal = {Physical Review Letters},
         year = 1971,
        month = apr,
       volume = {26},
       number = {14},
        pages = {832-833},
          doi = {10.1103/PhysRevLett.26.832},
       adsurl = {https://ui.adsabs.harvard.edu/abs/1971PhRvL..26..832B}
}

@article{Suzuki1974,
       author = {{Suzuki}, M.},
        title = "{New Universality of Critical Exponents}",
      journal = {Progress of Theoretical Physics},
         year = 1974,
        month = jun,
       volume = {51},
       number = {6},
        pages = {1992-1993},
          doi = {10.1143/PTP.51.1992},
       adsurl = {https://ui.adsabs.harvard.edu/abs/1974PThPh..51.1992S}
}

@article{Jin2012,
       author = {{Jin}, Songbo and {Sen}, Arnab and {Sandvik}, Anders W.},
        title = "{Ashkin-Teller Criticality and Pseudo-First-Order Behavior in a Frustrated Ising Model on the Square Lattice}",
      journal = {Physical Review Letters},
         year = 2012,
        month = jan,
       volume = {108},
       number = {4},
          eid = {045702},
        pages = {045702},
          doi = {10.1103/PhysRevLett.108.045702},
archivePrefix = {arXiv},
       eprint = {1110.5874},
 primaryClass = {cond-mat.stat-mech},
       adsurl = {https://ui.adsabs.harvard.edu/abs/2012PhRvL.108d5702J}
}

@article{Jin2013,
  title = {Phase transitions in the frustrated Ising model on the square lattice},
  author = {Jin, Songbo and Sen, Arnab and Guo, Wenan and Sandvik, Anders W.},
  journal = {Phys. Rev. B},
  volume = {87},
  issue = {14},
  pages = {144406},
  numpages = {12},
  year = {2013},
  month = {Apr},
  publisher = {American Physical Society},
  doi = {10.1103/PhysRevB.87.144406},
  url = {https://link.aps.org/doi/10.1103/PhysRevB.87.144406}
}

@article{Queiroz2011,
       author = {{de Queiroz}, S.~L.~A.},
        title = "{Scaling behavior of a square-lattice Ising model with competing interactions in a uniform field}",
      journal = {Physical Review E},
         year = 2011,
        month = sep,
       volume = {84},
       number = {3},
          eid = {031132},
        pages = {031132},
          doi = {10.1103/PhysRevE.84.031132},
archivePrefix = {arXiv},
       eprint = {1107.6022},
 primaryClass = {cond-mat.stat-mech},
       adsurl = {https://ui.adsabs.harvard.edu/abs/2011PhRvE..84c1132D}
}

@article{Alet2005,
       author = {{Alet}, Fabien and {Jacobsen}, Jesper Lykke and {Misguich}, Gr{\'e}goire and {Pasquier}, Vincent and {Mila}, Fr{\'e}d{\'e}ric and {Troyer}, Matthias},
        title = "{Interacting Classical Dimers on the Square Lattice}",
      journal = {Physical Review Letters},
         year = 2005,
        month = jun,
       volume = {94},
       number = {23},
          eid = {235702},
        pages = {235702},
          doi = {10.1103/PhysRevLett.94.235702},
archivePrefix = {arXiv},
       eprint = {cond-mat/0501241},
 primaryClass = {cond-mat.stat-mech},
       adsurl = {https://ui.adsabs.harvard.edu/abs/2005PhRvL..94w5702A}
}

@article{Pearce1987,
       author = {{Pearce}, P.~A. and {Kim}, D.},
        title = "{Continuously varying exponents in magnetic hard squares}",
      journal = {Journal of Physics A Mathematical General},
         year = 1987,
        month = dec,
       volume = {20},
       number = {18},
        pages = {6471-6485},
          doi = {10.1088/0305-4470/20/18/044},
       adsurl = {https://ui.adsabs.harvard.edu/abs/1987JPhA...20.6471P}
}

@article{Malakis2009,
       author = {{Malakis}, A. and {Berker}, A. Nihat and {Hadjiagapiou}, I.~A. and {Fytas}, N.~G.},
        title = "{Strong violation of critical phenomena universality: Wang-Landau study of the two-dimensional Blume-Capel model under bond randomness}",
      journal = {Physical Review E},
         year = 2009,
        month = jan,
       volume = {79},
       number = {1},
          eid = {011125},
        pages = {011125},
          doi = {10.1103/PhysRevE.79.011125},
archivePrefix = {arXiv},
       eprint = {0809.4241},
 primaryClass = {cond-mat.dis-nn},
       adsurl = {https://ui.adsabs.harvard.edu/abs/2009PhRvE..79a1125M}
}

@article{Suzuki2015,
       author = {{Suzuki}, Takafumi and {Harada}, Kenji and {Matsuo}, Haruhiko and {Todo}, Synge and {Kawashima}, Naoki},
        title = "{Thermal phase transition of generalized Heisenberg models for SU (N ) spins on square and honeycomb lattices}",
      journal = {Physical Review B},
         year = 2015,
        month = mar,
       volume = {91},
       number = {9},
          eid = {094414},
        pages = {094414},
          doi = {10.1103/PhysRevB.91.094414},
archivePrefix = {arXiv},
       eprint = {1505.06273},
 primaryClass = {cond-mat.stat-mech},
       adsurl = {https://ui.adsabs.harvard.edu/abs/2015PhRvB..91i4414S}
}

@article{Jose1977,
  author  = {Jos{\'e}, Jorge V. and Kadanoff, Leo P. and
             Kirkpatrick, Scott and Nelson, David R.},
  title   = {Renormalization, Vortices, and Symmetry-Breaking
             Perturbations in the Two-Dimensional Planar Model},
  journal = {Phys. Rev. B},
  volume  = {16},
  pages   = {1217--1241},
  year    = {1977},
  doi     = {10.1103/PhysRevB.16.1217}
}

@article{Penc2023,
  title = {Crystalline phases and devil's staircase in qubit spin ice},
  author = {Kond\'akor, M\'ark and Penc, Karlo},
  journal = {Phys. Rev. Res.},
  volume = {5},
  issue = {4},
  pages = {043172},
  numpages = {28},
  year = {2023},
  month = {Nov},
  publisher = {American Physical Society},
  doi = {10.1103/PhysRevResearch.5.043172},
  url = {https://link.aps.org/doi/10.1103/PhysRevResearch.5.043172}
}

@article{JahnPhilipsen2004,
    author        = {Jahn, Oliver and Philipsen, Owe},
    title         = {The Polyakov Loop and its Relation to Static Quark Potentials and Free Energies},
    journal       = {Phys. Rev. D},
    volume        = {70},
    pages         = {074504},
    year          = {2004},
    doi           = {10.1103/PhysRevD.70.074504},
    eprint        = {hep-lat/0407042},
    archivePrefix = {arXiv}
}

@article{LuscherWeisz2002,
    author        = {L{\"u}scher, Martin and Weisz, Peter},
    title         = {Quark Confinement and the Bosonic String},
    journal       = {JHEP},
    volume        = {07},
    pages         = {049},
    year          = {2002},
    doi           = {10.1088/1126-6708/2002/07/049},
    eprint        = {hep-lat/0207003},
    archivePrefix = {arXiv}
}

@article{CaselleEtAl2011,
    author        = {Caselle, Michele and Feo, Alessandra and
                     Panero, Marco and Pellegrini, Roberto},
    title         = {Universal Signatures of the Effective String in
                     Finite Temperature Lattice Gauge Theories},
    journal       = {JHEP},
    volume        = {04},
    pages         = {020},
    year          = {2011},
    doi           = {10.1007/JHEP04(2011)020},
    eprint        = {1102.0723},
    archivePrefix = {arXiv},
    primaryClass  = {hep-lat}
}

@article{Caselle:2007yc,
  author        = {Caselle, Michele and Hasenbusch, Martin
                   and Panero, Marco},
  title         = {The interface free energy: Comparison of accurate
                   Monte Carlo results for the 3D Ising model with
                   effective interface models},
  journal       = {JHEP},
  volume        = {09},
  pages         = {117},
  year          = {2007},
  eprint        = {0707.0055},
  archivePrefix = {arXiv},
  primaryClass  = {hep-lat},
  doi           = {10.1088/1126-6708/2007/09/117}
}

@article{Gopfert:1981er,
    author = "Gopfert, Markus and Mack, Gerhard",
    title = "{Proof of Confinement of Static Quarks in Three-Dimensional U(1) Lattice Gauge Theory for All Values of the Coupling Constant}",
    reportNumber = "DESY-81-036",
    doi = "10.1007/BF01961240",
    journal = "Commun. Math. Phys.",
    volume = "82",
    pages = "545",
    year = "1981"
}

@article{Athenodorou:2018sab,
    author = "Athenodorou, Andreas and Teper, Michael",
    title = "{On the spectrum and string tension of U(1) lattice gauge theory in 2 + 1 dimensions}",
    eprint = "1811.06280",
    archivePrefix = "arXiv",
    primaryClass = "hep-lat",
    doi = "10.1007/JHEP01(2019)063",
    journal = "JHEP",
    volume = "01",
    pages = "063",
    year = "2019"
}

@article{Athenodorou:2026dew,
    author = "Athenodorou, Andreas and Bonati, Claudio and Soler Calero, Ivan",
    title = "{A different kind of continuum limit for the three-dimensional U(1) gauge theory}",
    eprint = "2608.27976",
    archivePrefix = "arXiv",
    primaryClass = "hep-lat",
    month = "8",
    year = "2026"
}

@article{Mariani:2026piy,
    author = "Mariani, Alessandro",
    title = "{Monopole crystal and effective theory for a monopole-enhanced 3D U(1) gauge theory}",
    eprint = "2611.15922",
    archivePrefix = "arXiv",
    primaryClass = "hep-lat",
    month = "11",
    year = "2026"
}

@article{Caselle:2025vhx,
    author = "Caselle, M. and Mariani, A.",
    title = "{The finite temperature ground state energy of the confining string in three-dimensional U(1) gauge theory}",
    eprint = "2503.05354",
    archivePrefix = "arXiv",
    primaryClass = "hep-lat",
    doi = "10.1007/JHEP06(2025)010",
    journal = "JHEP",
    volume = "06",
    pages = "010",
    year = "2025"
}

@article{Caselle:2025bgu,
    author = "Caselle, Michele and Mariani, Alessandro and Panero, Marco and Smecca, Antonio",
    title = "{On the equation of state of U(1) lattice gauge theory in three dimensions}",
    eprint = "2501.16185",
    archivePrefix = "arXiv",
    primaryClass = "hep-lat",
    reportNumber = "HIP-2025-02/TH",
    doi = "10.1007/JHEP03(2025)130",
    journal = "JHEP",
    volume = "03",
    pages = "130",
    year = "2025"
}

@article{Banerjee:2023pnb,
    author = "Banerjee, A. and Banerjee, D. and Kanwar, G. and Mariani, A. and Rindlisbacher, T. and Wiese, U. -J.",
    title = "{Broken symmetry and fractionalized flux strings in a staggered U(1) pure gauge theory}",
    eprint = "2309.17109",
    archivePrefix = "arXiv",
    primaryClass = "hep-lat",
    doi = "10.1103/PhysRevD.109.014506",
    journal = "Phys. Rev. D",
    volume = "109",
    number = "1",
    pages = "014506",
    year = "2024"
}

@article{Banerjee:2013dda,
    author = "Banerjee, D. and Jiang, F. -J. and Widmer, P. and Wiese, U. -J.",
    title = "{The (2 + 1)-d U(1) quantum link model masquerading as deconfined criticality}",
    eprint = "1303.6858",
    archivePrefix = "arXiv",
    primaryClass = "cond-mat.str-el",
    doi = "10.1088/1742-5468/2013/12/P12010",
    journal = "J. Stat. Mech.",
    volume = "1312",
    pages = "P12010",
    year = "2013"
}
\end{document}